\documentclass[a4paper,11pt]{article}
\pdfoutput=1

\usepackage{setup}
\usepackage[utf8]{inputenc}
\usepackage{jheppub}

\preprint{IPPP/19/6, Nikhef-2019-003}

\title{\boldmath Asymmetric heavy-quark hadroproduction at LHCb: \\ Predictions and applications}
    
\author[a]{Rhorry Gauld,}
\author[b]{Ulrich Haisch,}
\author[c]{and Benjamin D. Pecjak}

\affiliation[a]{Nikhef Theory Group, Science Park 105, 1098 XG Amsterdam, The Netherlands}
\affiliation[b]{Max Planck Institute for Physics, F{\"o}hringer Ring 6,  80805 M{\"u}nchen, Germany}
\affiliation[e]{Institute for Particle Physics Phenomenology, Durham University,  Durham DH1 3LE, UK}

\emailAdd{rgauld@nikhef.nl}
\emailAdd{haisch@mpp.mpg.de}
\emailAdd{ben.pecjak@durham.ac.uk}

\abstract{We present a phenomenological analysis of asymmetric bottom- and charm-quark production within the LHCb acceptance relevant for $pp$ collisions at $\sqrt{s} = 13 \, {\rm TeV}$. Predictions are provided for both anti-$k_t$ bottom- and charm-jet pairs, which are kept differentially with respect to the invariant mass of the jet pair. It is quantified how data in this region can provide sensitivity to the couplings of the $Z$ boson to heavy quarks, and we investigate what precision is needed to compete with LEP. We also discuss how asymmetry and rate measurements can provide constraints on a particular class of new-physics models, which contains gauge bosons with small/moderate couplings to light/heavy quarks and masses of the order of $100 \, {\rm GeV}$. Predictions are obtained including all relevant QCD and QED/weak contributions up to next-to-leading order, which have been implemented in a Fortran code which allows to directly compute the asymmetric cross sections. We provide all relevant analytic formulas for our computations.}


\def \sh{\hat{s}}
\def\bm#1{\mbox{\boldmath$#1$\unboldmath}}
\def\Aone{A}
\def\Atwo{A}
\def\Athree{B}

\begin{document} 
\maketitle
\flushbottom

\section{Introduction} 
\label{sec:introduction}

The production of  bottom and charm quarks at high-energy colliders is a topic of considerable interest. While not directly observed, these quarks fragment into unstable bottom and charm  hadrons with a typical mean lifetime of $10^{-12} \, {\rm s}$. As a consequence of the short but finite lifetime, bottom and charm  hadrons decay within the detector at a location which is displaced from the primary collision point. This distinct experimental signature can be used to associate the production of a particle jet in the collision with that originating from a heavy quark, or to improve the efficiency for exclusively reconstructing the heavy-flavour hadron, which in turn has allowed detailed studies of heavy-quark production. 

A relevant example is the pair-production of  bottom- and charm-quarks in $e^+e^-$ collisions in the vicinity of the $Z$ pole, as studied at both LEP and SLC. Precision measurements of both the production rates and the asymmetries in angular distributions of the produced heavy-quarks has allowed to perform precision tests of the Standard Model~(SM), and has led to the most stringent constraints on the coupling structure of the $Z$ boson to all quarks but the top quark~\cite{ALEPH:2005ab}. Similar studies of the angular asymmetries in heavy-quark production  have also been carried out at hadron colliders. In $p \bar p$ collisions at the Tevatron, a measurement of the asymmetry in $b$-quark pair production has been performed for $B$-hadrons by the D\O\ collaboration~\cite{Abazov:2014ysa}, and also for bottom-quark jet ($b$-jet) pairs by the CDF collaboration~\cite{Aaltonen:2016azt}. A measurement of the $b$-jet pair asymmetry has also been achieved by the LHCb collaboration in $pp$ collisions at the LHC~\cite{Aaij:2014ywa}.

The asymmetric hadroproduction of heavy-quarks provides important information as compared to what is accessible in $e^+e^-$ collisions. First, the production mechanisms are entirely different in these collisions, and therefore unique information is provided in hadron collisions. In addition, a measurement of the asymmetry can be performed differentially in the invariant mass of the $b \bar b$ system across a large range of values. This information allows to test a number of new-physics scenarios which are not accessible in $e^+e^-$ collisions, and there have been a number of relevant phenomenological studies both in the SM and beyond~(cf.~\cite{Bai:2011ed,Kahawala:2011sm,Saha:2011wr,Manohar:2012rs,Drobnak:2012cz,Delaunay:2012kf,Ipek:2013zi,Grinstein:2013iws,Murphy:2015cha,Gauld:2015qha} for instance).  It is, however, important to note that the prediction and measurement of heavy-quark asymmetries at hadron colliders also come with a number of challenges. Experimentally it is necessary to account for the effects of pile-up, and to suppress the extremely large background contributions from light-flavour jet production. In addition, the absolute value of the predicted asymmetry is typically quite small. This is mainly a consequence of the large suppression introduced by the symmetric gluon-fusion subprocess for heavy-quark pair production. On the theoretical side, the evaluation of QCD corrections (which are dominant) to  heavy-quark production are  more complicated at hadron colliders because all external particles are coloured. Obtaining predictions are furthermore computationally more intensive, as the partonic cross sections have to be convoluted with parton distribution functions (PDFs). 

The purpose of this work is to provide robust predictions for both bottom- and charm-quark jet-pair production in $pp$ collisions at the $13 \, {\rm TeV}$ LHC in the forward direction. There are at least two  motivations for focussing on this specific kinematic region. First, the forward regime provides unique opportunities to measure heavy-quark asymmetries at the LHC, because of the increased asymmetry between $q$  and $\bar q$ PDFs present  when these partons carry large  energy fractions, and the reduced dilution of the symmetric gluon-fusion contribution. Second, the LHCb experiment is a forward detector~\cite{Alves:2008zz} and able  to perform both charge- and flavour-tagging of heavy-quark jets~\cite{Aaij:2014ywa,Aaij:2015yqa}. In fact, the recent LHCb measurement of the $Z \to b \bar b$ production cross section~\cite{Aaij:2017eru}  indicates that finely binned heavy-quark asymmetry measurements in the $Z$-pole region  should be possible as well. As~we will show in this article,  the latter point is of relevance as there is a long-standing tension between the measured and the SM value of the $b \bar b$ forward-backward asymmetry in~$e^+e^-$ collisions~\cite{ALEPH:2005ab}. It is also discussed how measurements of bottom- and charm-quark pair production can provide constraints on new-physics models, which contain gauge bosons with masses of around $100 \, {\rm GeV}$ and small/moderate couplings to light/heavy quarks. \\

The remainder of the paper is laid out as follows. In Section~\ref{setup} we provide details of the theoretical set-up that we use to obtain our numerical results. The SM predictions for the cross sections and the asymmetries are given in Section~\ref{sigma} and Section~\ref{asym}, respectively. Two applications of our results are presented in Section~\ref{sec:applications} and conclude  our article. The technical details of our calculations and their numerical implementation can be found in Appendix~\ref{calculation}.

\section{Theoretical framework} 
\label{setup}

As discussed in the introduction, the goal of the experimental analysis~\cite{Aaij:2014ywa} is to measure an asymmetry in the rapidity distributions of $b$- and $\bar{b}$-quarks produced in $pp$ collisions. Experimentally, this has been achieved by requiring the presence of two anti-$k_t$ jets~\cite{Cacciari:2008gp}  which are both charge- (in the presence of a semi-leptonic $B$ decay) and flavour-tagged. This procedure allows to differentiate between $b$- and $\bar{b}$-quark jets and to construct asymmetric observables. Practically, the asymmetry  is measured differentially with respect to the invariant mass  of the $b$- and $\bar{b}$-jet pair system.

The corresponding theoretical predictions for the inclusive process $pp \to Q\bar Q X$ with $Q$ referring to either a bottom or charm quark in this work are obtained assuming a standard factorisation theorem~\cite{Collins:1989gx}, whereby the hadron-level cross section can be computed by convoluting the individual partonic cross sections with the relevant PDFs. Theoretical predictions for heavy-quark production can be characterised in terms of the perturbative accuracy of the partonic cross sections according to
\begin{align} \label{eq:expansion}
\rd \hat{\sigma} = \sum_{n,\,m} \alpha^n \hspace{0.25mm} \alpha_s^m \hspace{0.5mm} \rd \hat{\sigma}^{(n,\,m)} \,,
\end{align}
where $\rd\hat{\sigma}^{(n,\,m)}$ denotes the coupling-stripped differential partonic cross section and $\alpha$ ($\alpha_s$) is the QED (QCD) coupling.  The leading order (LO) contributions to (\ref{eq:expansion}) correspond to $n+m = 2$, while the next-to-leading order (NLO) contributions have $n+m = 3$, and so forth. In this work, we include all numerically relevant NLO corrections to the distributions.\footnote{The tiny $\mathcal{O}(\alpha^3)$ corrections are neglected in our predictions.} The technical details of the calculation and implementation of the various contributions to the partonic cross sections are discussed in Appendix~\ref{calculation}. The techniques to obtain NLO corrections to $2 \to 2$ processes are by now standard, and in the case of $pp \to Q\bar Q X$  all relevant NLO contributions are known since some time~\cite{Nason:1987xz,Nason:1989zy,Mangano:1991jk,Beenakker:1990maa,Beenakker:1988bq,Kuhn:2005it,Kuhn:2006vh,Bernreuther:2006vg,Hollik:2007sw,Kuhn:2009nf}  (see also references therein for partial results). We therefore refrain from giving NLO expressions for~(\ref{eq:expansion})  in the main text. Instead we provide  an overview of the numerical implementation of our calculations in the following, and discuss the details of the various inputs and scheme choices, which are relevant to the numerical predictions provided in this article. 

\subsection{General set-up}

The numerical predictions in this paper are obtained by means of a private Fortran code, which is linked to a number of external libraries: the evaluation of the input PDFs is performed with {\tt LHAPDF}~\cite{Buckley:2014ana}, the numerical integration algorithms of~{\tt CUBA}~\cite{Hahn:2004fe} are used, and one-loop scalar integrals are evaluated with {\tt OneLOop}~\cite{vanHameren:2009dr,vanHameren:2010cp}. An important aspect of the implementation of our analytic calculations is that we separate the contributions to the partonic cross sections into parts that are symmetric and asymmetric under interchange of the final-state heavy quarks. The numerical integration of the symmetric and asymmetric contributions can therefore be performed independently. This approach significantly improves the stability of the numerical integration of the asymmetric cross sections, as only the asymmetric contributions of the partonic cross sections are integrated and the numerical adaption of the integration is specifically optimised for these contributions.

The mass of the considered heavy quark is included in our calculations, and we therefore work in a scheme with $N_F = 4 \, (3)$ massless quarks for the bottom-quark (charm-quark) predictions. The $\mathcal{O}(\alpha_s^3)$ corrections to the symmetric cross sections are obtained with the matrix elements~\cite{Nason:1987xz} implemented in~{\tt POWHEG~BOX}~\cite{Frixione:2007nw}. The calculation of the weak box-diagram corrections of $\mathcal{O}(\alpha\alpha_s^2)$ have instead been obtained using {\tt MadLoop}~\cite{Hirschi:2011pa} as part of the loop-induced module~\cite{Hirschi:2015iia} available in {\tt MadGraph5\_aMC@NLO}~\cite{Alwall:2014hca}.\footnote{Details on this set-up are available on the wiki page~\cite{mg5online}. We thank Valentin~Hirschi for his assistance with this part of the calculation.} All other contributions have been computed with the aid of~{\tt FeynArts}~\cite{Hahn:2000kx} and~{\tt FormCalc}~\cite{Hahn:1998yk}, and the relevant analytic formulas for the asymmetric contributions to the partonic cross section are collected in Appendix~\ref{calculation}. In these cases, we have used the technique of phase-space slicing~\cite{Harris:2001sx} or dipole subtraction~\cite{Catani:1996vz} to regulate the explicit (implicit) divergences present in the virtual~(real) phase-space.

We add that differential $\mathcal{O}(\alpha_s^4)$ results have been first presented for top-quark production in~\cite{Czakon:2014xsa}, as well as for massless partons to leading colour in~\cite{Currie:2016bfm}. At present, a calculation of $b$-tagged jets (either massive or massless) is  instead not available. However, it can be expected that such predictions will become available in the future when issues related to numerical stability or flavour-tagging of subtraction terms in the next-to-next-to-leading order~(NNLO) QCD calculations have been resolved.

\subsection{Observables}

To match the experimental definition of jet observables, we construct anti-$k_t$ jets with a radius parameter $R = 0.5$ and  tag them as a $Q$-jet  ($\bar Q$-jet) if they contain a $Q$~$(\bar Q)$, with $Q$ being the heavy quark. Throughout this work, all observables are computed in terms of these flavour-tagged jets. The label ``jet'' will however be suppressed, meaning for example that   the invariant mass of a $b$- and $\bar b$-jet pair will be simply called $m_{b \bar b}$. If not stated otherwise, we will always place the following kinematic cuts on the flavour-tagged heavy-quark jets
\beq \label{eq:fiducial}
p_{T,Q \, (\bar Q)} > 20.0 \, {\rm GeV} \,, \qquad \eta_{Q \, (\bar Q)} \in [2.2, 4.2] \,,\qquad \phi_{Q \bar Q} > 2.6 \,.
\eeq
referring to these selections as ``LHCb kinematic cuts". The requirements on the transverse momentum ($p_T$) and the pseudorapidity $(\eta)$ ensure that the jets are reconstructed according to the flavour tagging algorithm in use at LHCb~\cite{Aaij:2015yqa}. The cut on the angular separation~($\phi$) between the two flavour-tagged jets in the azimuthal plane  ensures that the two  heavy-quark jets are well separated. This cut therefore avoids configuration which can appear for instance at $\mathcal{O}(\alpha_s^3)$ where both the $Q$ and $\bar Q$ are contained within a single jet (at LO this cannot occur as the heavy quarks are produced back-to-back). The  impact of the choice of the angular  cut on the predicted asymmetries is discussed in Section~\ref{asym}.

The two primary observables of interest are the heavy-quark production cross sections and the corresponding asymmetries. The cross sections are computed differentially in  $m_{Q\bar Q}$ within the LHCb fiducial region~(\ref{eq:fiducial}). The asymmetries are also computed differentially in  $m_{Q\bar Q}$, and defined according to
\beq \label{eq:asym}
\frac{\rd A}{\rd m_{Q\bar Q}} = \left ( \frac{\rd \sigma_S}{\rd m_{Q\bar Q} } \right )^{-1} \left ( \frac{ \rd \sigma_A}{ \rd m_{Q\bar Q}} \,  \Bigg |_{\Delta y > 0} -  \frac{ \rd \sigma_A}{ \rd m_{Q\bar Q}} \, \Bigg |_{\Delta y < 0} \right )  \,.
\eeq
Here $\rd \sigma_{S \, (A)}$ refers to the convolution integral of the differential (a)symmetric partonic cross sections with the relevant PDFs, and $\Delta y = y_{Q} - y_{\bar{Q}}$ is the difference between the rapidities of the $Q$ and the $\bar Q$. 

\subsection{Heavy-quark mass effects}

\begin{figure}
\centering
\includegraphics[width=.49\linewidth]{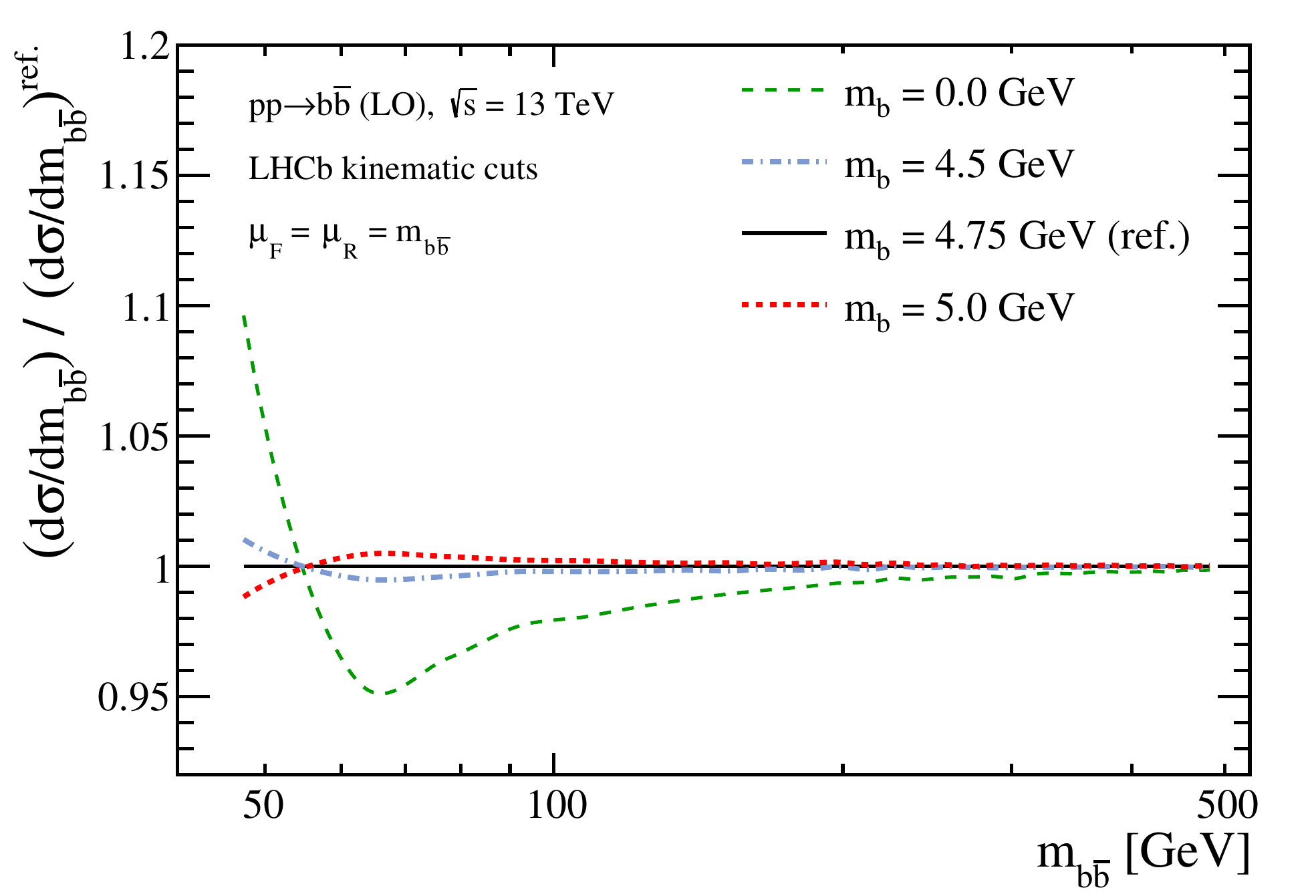} \hfill
\includegraphics[width=.49\linewidth]{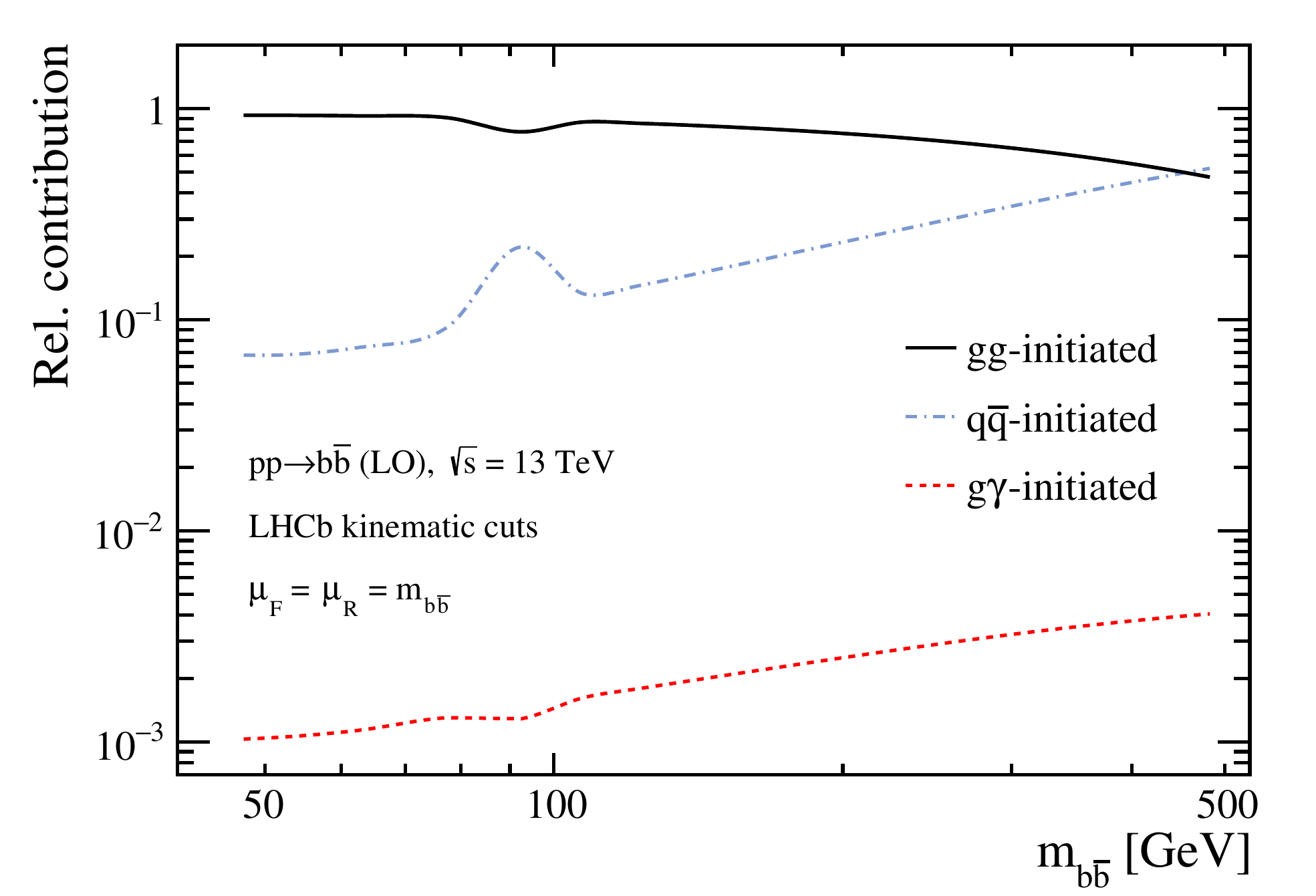} \hfill
\vspace{2mm}
\caption{
Left: Differential cross sections for producing $b$-jet pairs  for different choices of $m_b$, normalised to the result obtained with $m_b = 4.75 \, {\rm GeV}$. Right: Relative contributions of each partonic channel to the differential cross section for the choice $m_b = 4.75 \, {\rm GeV}$. The shown results are LO accurate and correspond to $pp$ collisions at $\sqrt{s} = 13 \, {\rm TeV}$, employing the selections~(\ref{eq:fiducial}).}
\label{fig:LOmass}
\end{figure}

As mentioned above, we retain the effects of the heavy-quark mass throughout our calculations. The following choices for the heavy-quark masses in the on-shell scheme are adopted
\beq \label{eq:mbmc}
m_b = 4.75 \, {\rm GeV}\,,  \qquad  m_c = 1.5 \, {\rm GeV} \,.
\eeq
These values are broadly consistent with the recommendations of the LHC Higgs Cross Section Working Group~\cite{deFlorian:2016spz}. When we provide predictions for either cross sections or asymmetries, we do not consider the uncertainties associated to~(\ref{eq:mbmc}). The motivation for this is that the mass corrections within the considered fiducial region are typically small (although not negligible), and the resulting ambiguities are small compared to the scale uncertainties. This statement is corroborated in Figure~\ref{fig:LOmass} (left), which shows LO differential $b \bar b$ cross sections within the LHCb fiducial region~(\ref{eq:fiducial}) for different choices of~$m_b$. These distributions are obtained with the {\tt LUXqed15}~\cite{Manohar:2016nzj} central PDF set member with  factorisation ($\mu_F$) and renormalisation ($\mu_R$) scales set dynamically to $m_{b\bar b}$, and the distributions have been normalised to the result obtained with $m_b = 4.75 \, {\rm GeV}$. As can be seen from the distribution obtained with $m_b = 0$ the mass corrections amount to $3\%$ to $10\%$ within  $m_{b\bar b} \in [40,100]$~GeV. On the other hand, a variation of $m_b$ in the range $m_b \in [4.5,5.0] \, {\rm GeV}$ results in cross-section changes below the percent level. We note that the inclusion of mass effects lead to a positive correction to the cross section within the LHCb fiducial region, while the inclusive cross section within the same invariant mass region receives negative corrections.

%------------------------------------------------
%------------------------------------------------

While to achieve precision predictions in the region of $m_{b\bar b} \in [40,100] \, {\rm GeV}$ including mass corrections is clearly important, at larger values of $m_{b\bar b}$ one could alternatively perform the calculation taking the heavy quarks to be massless. Employing a massless scheme would have the advantage that logarithmic mass corrections could be resummed, but also has some weaknesses. In this context it is important to recall that the measurement is performed by requiring the presence of two well separated flavour-tagged anti-$k_t$ jets according to~\eqref{eq:fiducial}. With such kinematic requirements, the phase-space regions where the NLO fixed-order calculation receives large logarithmic corrections (for example, due to the presence of $g\to b\bar{b}$ collinear enhancements) are avoided, and the effects of resumming these types of contributions is therefore typically small.\footnote{It was checked that for the LHCb kinematic cuts~(\ref{eq:fiducial}), the contribution from the subprocess $b\bar b\to b\bar b$ in the massless scheme accounts for only around $1\%$ of the total cross section.} Another consideration is that the prediction of flavour-tagged anti-$k_t$ jets is only infrared (IR)-safe for the massive calculation. Due to the presence of wide-angle $g\to Q\bar{Q}$ splittings of soft gluons, the massless calculation is IR-unsafe~\cite{Banfi:2006hf}. We therefore provide our predictions including the full mass effects up to~NLO, such that it is possible that numerical predictions computed with the massive NNLO calculation~\cite{Czakon:2013goa} can be added consistently at a later date. Alternatively one could consider a flavour-tagging algorithm which is IR-safe and achievable experimentally. 

\subsection{PDFs,  input parameters and scale variation}
\label{sec:PDFsEWSCAL}

As a baseline PDF set in this work, we use the variant of {\tt NNPDF31\_nlo\_as\_0118}~\cite{Ball:2017nwa} where the charm-quark PDF is generated purely perturbatively. This is a variable flavour number scheme set with $N_F = 5$, which is therefore evolved (both PDFs and $\alpha_s$) with five active flavours above the $b$-quark mass threshold. As discussed throughout this section, we deliver predictions for bottom- and charm-quark pair production at NLO including both QCD and QED/weak  corrections. For consistency, these predictions should be obtained by convoluting the partonic cross sections with PDFs which have been extracted from a PDF fit including both QCD and QED effects. There are two important points related to the choice of PDFs which we describe below.

First, when calculating $b \bar b$ ($c \bar c$) production using $N_F = 4 \, (3)$ active flavours, there is a  mis-match at $\mathcal{O}(\alpha_s^3)$ between the  perturbative cross-section calculation and our input~PDF set which uses $N_F =5$. To account for this we include the relevant compensation terms following~\cite{Cacciari:1998it}. Second, our baseline PDF does not include a photon~PDF or the effects of a joint QCD-QED evolution~\cite{deFlorian:2015ujt}. There has recently been quite some activity in precisely determining the photon PDF~\cite{Manohar:2016nzj,Harland-Lang:2016apc,Harland-Lang:2016kog,Manohar:2017eqh,Bertone:2017bme} and also a number of studies of electroweak~(EW) effects in top-quark pair production have been presented~\cite{Pagani:2016caq,Czakon:2017wor,Czakon:2017lgo}. We have studied the impact of the latter two types of contributions for bottom- and charm-quark pair production, and found these effects of very limited importance. This is demonstrated in Figure~\ref{fig:LOmass}~(right) where the contributions of gluon-fusion, quark-annihilation, and gluon-photon scattering  to the $b \bar b$ cross section are shown. The given predictions are obtained at LO with the {\tt LUXqed15} PDF set using $\mu_F = \mu_{R}= m_{b\bar b}$ and employing the reference cuts~\eqref{eq:fiducial}  at the 13~TeV LHC. In~the considered invariant mass range, we find that the photon-induced contributions lead to effects at the permille level. Compared to the uncertainty of the total cross section (which is around $10\%$), these effects are thus entirely negligible.\footnote{The only exception is the ratio of bottom- and charm-quark jet rates which is discussed in Section~\ref{sigma}.} We have therefore chosen to use a~PDF set based on NLO QCD which does not include a photon~PDF. A~consequence of ignoring the mixed QCD-QED evolution effects is to slightly overestimate the uncertainty due to $\mu_F$ variation used to assess the theoretical uncertainty of our predictions.

In this work, we use the following input parameters: $m_h = 125 \, {\rm GeV}$, $m_t = 173 \, {\rm GeV}$, $M_W = 80.385 \, {\rm GeV}$, $\Gamma_W = 2.085 \, {\rm GeV}$, $M_Z = 91.1876 \, {\rm GeV}$, $\Gamma_Z = 2.4952 \, {\rm GeV}$ as well as  $G_{F} = 1.16638 \cdot 10^{-5} \, {\rm GeV}^{-2}$. Employing this input and including the dominant one- and two-loop universal corrections to the $\rho \hspace{0.25mm}$-parameter~\cite{Fleischer:1993ub}, we have derived  the following values for the square of the sine of the weak mixing angle $\sin^2 \theta_{w} = 0.2293$ and the electromagnetic coupling $\alpha = 1/128.55$. For the values of the Cabibbo-Kobayashi-Maskawa (CKM) matrix, we take $|V_{us}| = |V_{cd}| = \sqrt{1-|V_{ud}|^2} =  \sqrt{1-|V_{cs}|^2} = 0.23$, $|V_{tb}| = 1$, while all other elements are set to zero. For the evaluation of the ${\cal O } (\alpha \alpha_s^2)$ corrections we use a complex mass-scheme~\cite{Denner:2005fg}, accounting for the width effects of the $Z$ boson. In the latter case, the (complex) value of the weak mixing angle is derived from the complex $W$- and $Z$-boson masses.

To assess the uncertainty due to missing higher-order corrections, scale variations are performed by changing both $\mu_F$ and $\mu_R$ independently by a factor of two around a reference scale $\mu_0$, with the constraint that $1/2 < \mu_F / \mu_R < 2$. Predictions are obtained for the two following choices of the reference scale
\beq \label{eq:scalechoices}
\mu_0 = m_{Q\bar Q} \,, \qquad \mu_0  = \overline{E}_{T,Q} = \frac{E_{T,Q}+E_{T,\bar{Q}}}{2}\,,
\eeq
corresponding to the invariant mass of the heavy-quark jet pair and the mean transverse energy of the heavy-quark jets, respectively. When observables such as the asymmetry defined in~\eqref{eq:asym} or a cross-section ratio between heavy quarks are considered, the scale variations are computed by correlating the scales between numerator and denominator.

\section{Cross-section predictions}
\label{sigma}

The main goal of this work is to provide reliable predictions for asymmetric heavy-quark production in the fiducial region~(\ref{eq:fiducial}). Having a clear understanding of the associated  cross sections is, however, an important ingredient of this analysis as well. From the theoretical point of view, it is important to validate the absolute heavy-quark jet rates as well as the shape of the invariant mass distributions, in particular in the region around the $Z$ pole. Experimentally, measurements of the cross sections may give handles on the~(charged) flavour-tagging efficiency and the mis-tag rates, as well as  providing an important validation of the jet-energy scale and resolution corrections. The differential heavy-quark cross sections may also lead to constraints on new-physics models which contain light gauge bosons, a point we will return to in Section~\ref{sec:applications}. The remainder of the current Section is dedicated to the study of the symmetric distributions. 

\subsection{Cross sections}

\begin{figure}[t!]
\centering
\includegraphics[width=.49\linewidth]{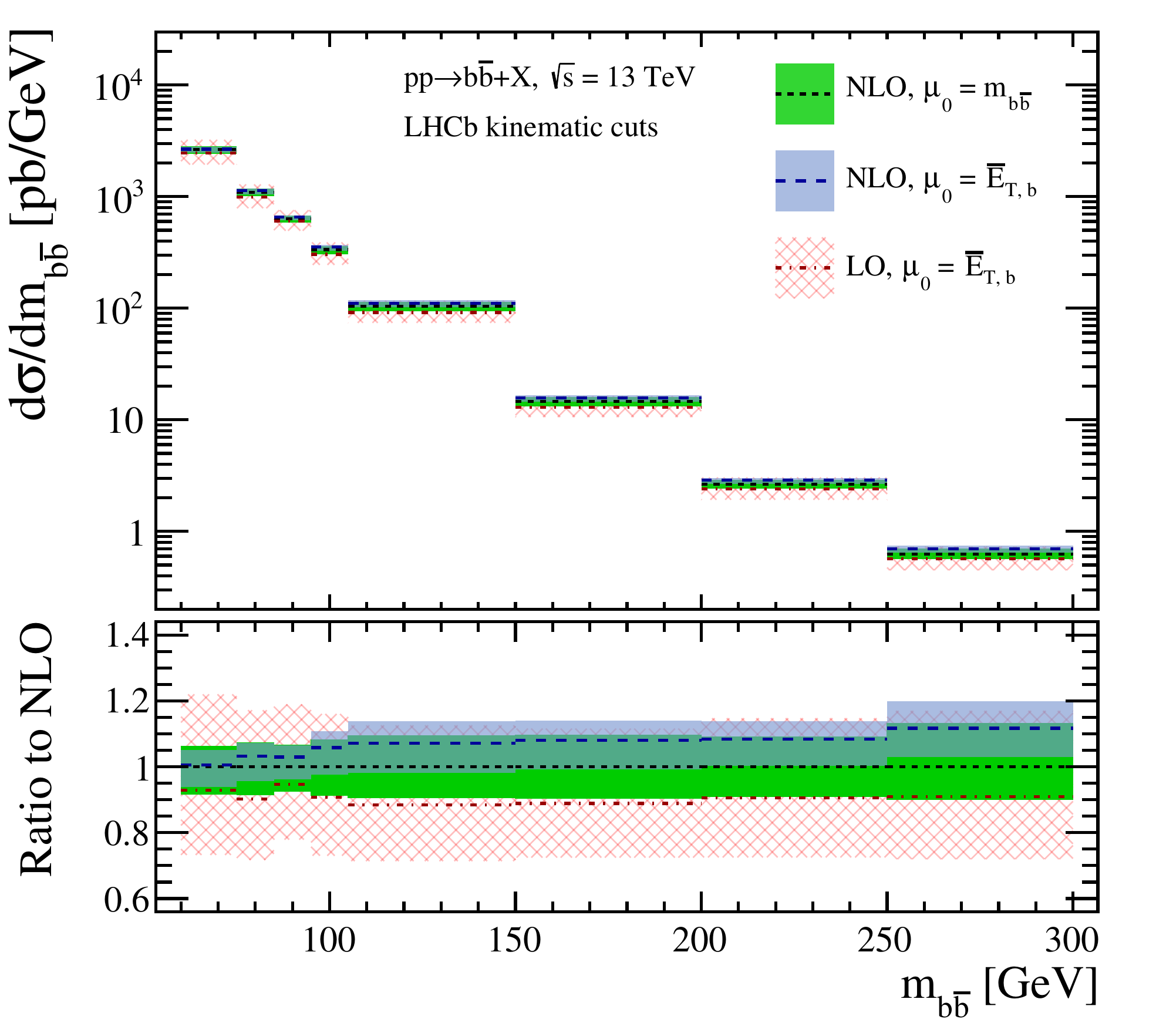} \hfill
\includegraphics[width=.49\linewidth]{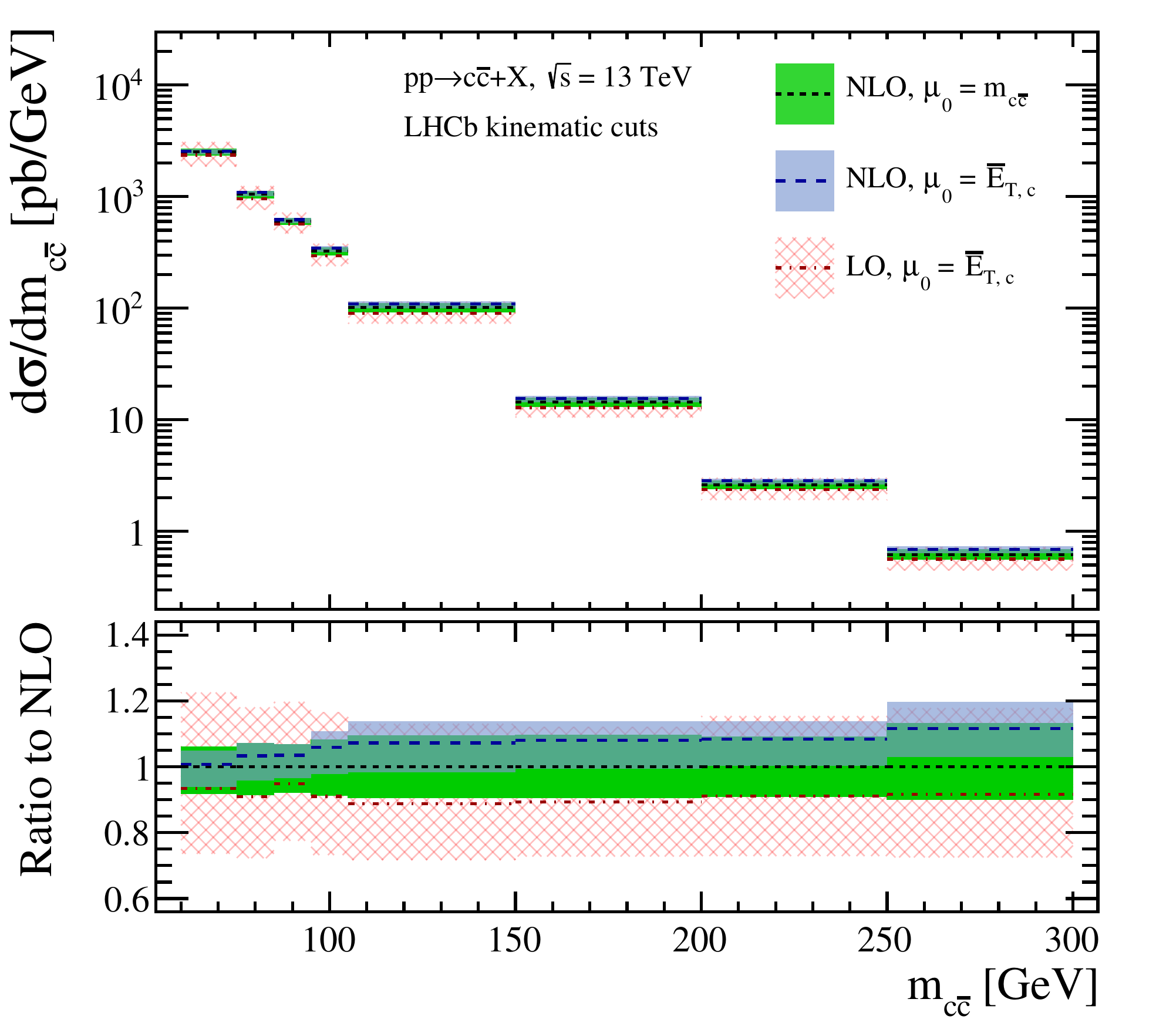} 
\vspace{2mm}
\caption{
Differential $b \bar b$ (left) and $c \bar c$ (right) cross sections for $pp$ collisions at $\sqrt{s} = 13 \, {\rm TeV}$  in the fiducial region~\eqref{eq:fiducial}. The NLO distributions employ the two scale choices  $\mu_0 = m_{Q\bar{Q}}$ and $\mu_0 = \overline{E}_{T,Q}$, and LO distributions obtained with $\mu_0 = \overline{E}_{T,Q}$ are also shown. The lower part of each panel displays the distributions normalised  to the central NLO prediction obtained with $\mu_0 = m_{Q\bar{Q}}$.
}
\label{fig:dsigdmqq}
\end{figure}

Figure~\ref{fig:dsigdmqq} gives our $\sqrt{s} = 13 \, {\rm TeV}$ predictions for the heavy-quark jet rates within the LHCb acceptance~(\ref{eq:fiducial}). The results have been obtained at NLO for both $b$-~(left) and $c$-jet~(right) pairs for the two dynamical references scales~(\ref{eq:scalechoices}) with the corresponding scale uncertainties shown as error bands. In order to allow to assess  the perturbative stability of the predictions, the LO predictions obtained with $\mu_0 = \overline{E}_{T,Q}$ are also displayed.  In both cases,  the lower panel of the plots shows the distributions normalised to the central NLO prediction obtained with $\mu_0 = m_{Q\bar{Q}}$. 

%------------------------------------------------
%------------------------------------------------

In the figure we have focussed on the region of $m_{Q\bar{Q}} \in [60,300] \, {\rm GeV}$, where the differential cross sections span several orders of magnitude. The scale uncertainties of the~NLO distributions are about $10\%$, which represents a marked improvement  with respect to the LO results. We also find that the NLO distributions corresponding to the two scale choices~(\ref{eq:scalechoices}) lead to consistent results, and tend to lie within the uncertainty bands of the LO distributions. In the region of $m_{Q\bar{Q}} \ge 100 \, {\rm GeV}$ the cross sections are entirely dominated by the QCD contributions, and there is a $5\%$ to $10\%$ difference between the central values of the NLO results obtained with $\mu_0 = m_{Q\bar{Q}}$ or $\mu_0 = \overline{E}_{T,Q}$. An improvement in the perturbative stability of the predictions in this region would require the inclusion of $\mathcal{O}(\alpha_s^4)$ corrections, either through a fixed-order calculation or by performing resummation (see for example~\cite{Czakon:2018nun}). The fiducial cross sections within the invariant mass bin $m_{Q\bar{Q}} \in [250,300] \, {\rm GeV}$ are approximately $30 \, {\rm pb}$. Assuming an integrated luminosity of~$5 \, {\rm fb}^{-1}$ and a signal efficiency $\epsilon_{Q\bar Q} = 0.6\%$, these numbers imply that a relative statistical uncertainty of about $1\%$ may be achievable with future LHCb data.  

%------------------------------------------------
\begin{figure}[t!]
\centering
\includegraphics[width=.49\linewidth]{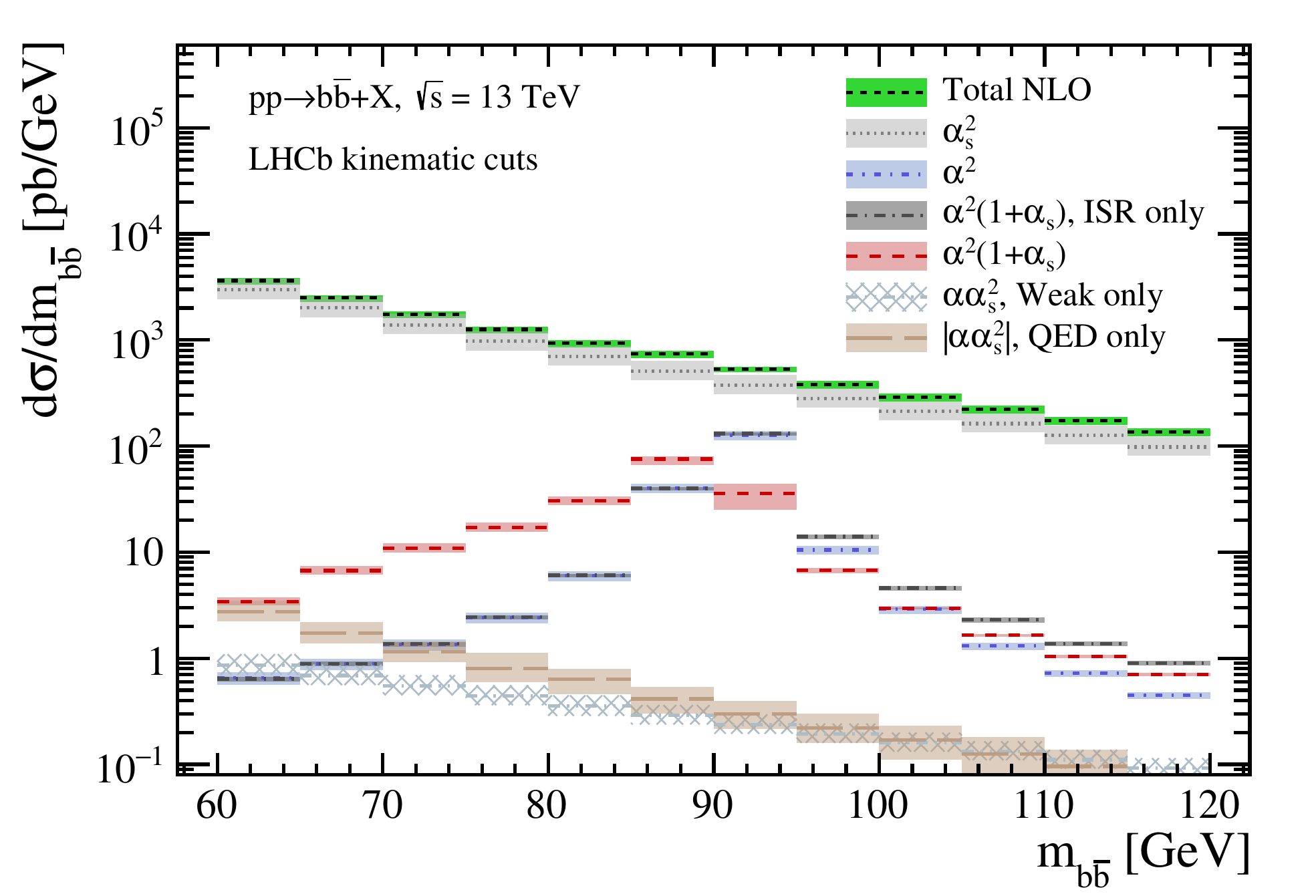} 
\includegraphics[width=.49\linewidth]{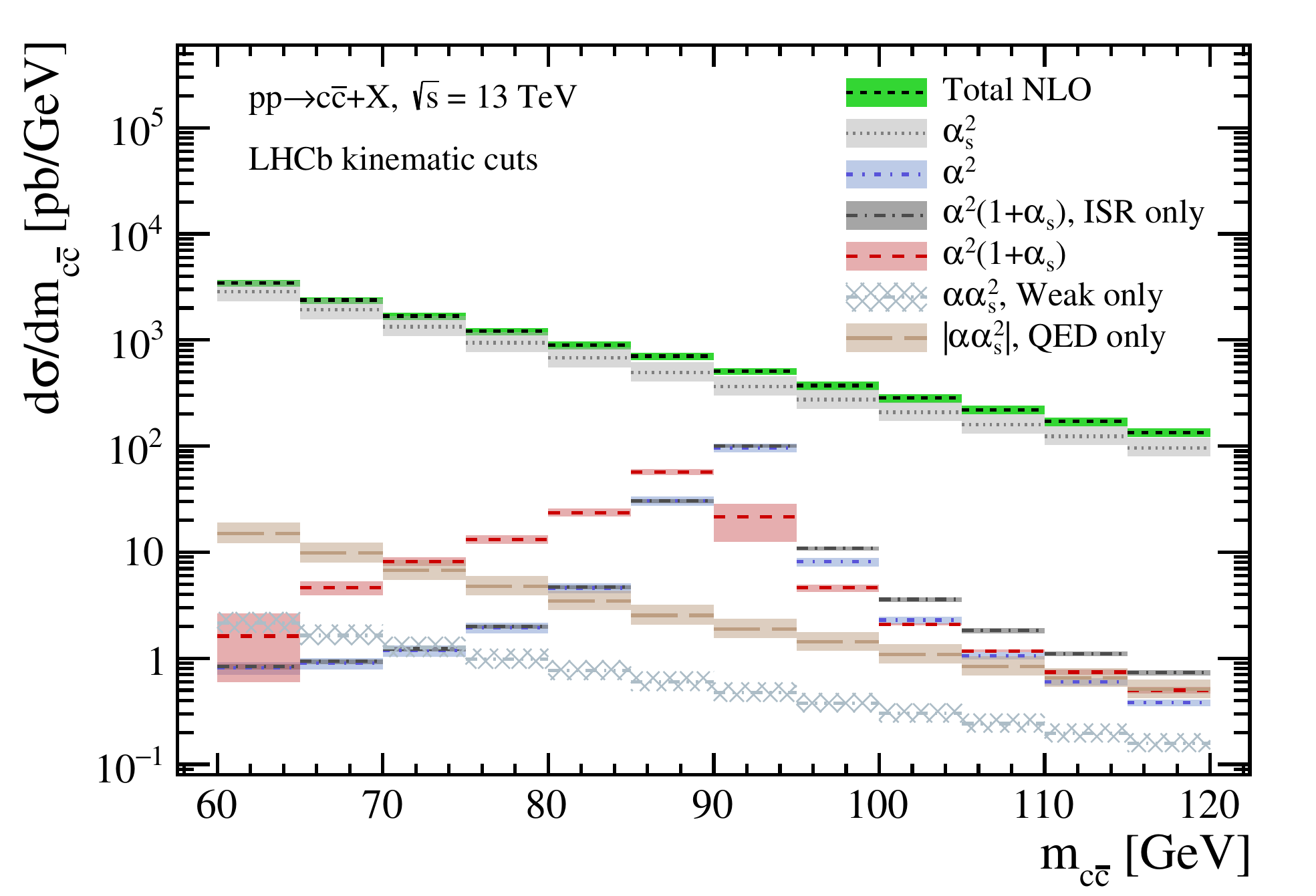} 
\vspace{2mm}
\caption{
Same as Figure~\ref{fig:dsigdmqq} but restricted to invariant masses $m_{Q\bar Q} \in [60,120] \, {\rm GeV}$. Besides the full NLO results various individual contributions are shown. See text for further details. 
}
\label{fig:dsigdmqqZpole}
\end{figure}
%------------------------------------------------

LHCb has recently performed a measurement of the process $pp  \to Z \to b \bar{b} +{\rm jets}$ at~$\sqrt{s} = 8 \, {\rm TeV}$~\cite{Aaij:2017eru}. This measurement is performed differentially with respect to $m_{b \bar b}$ in bins of width $4 \, {\rm GeV}$ in the $Z$-pole region, suggesting that future measurements of inclusive heavy-quark pair production will also be possible with similarly fine binning. In~Figure~\ref{fig:dsigdmqqZpole} we provide predictions for both bottom (left) and charm (right) jet-pair production, focussing on the invariant-mass region of $m_{Q\bar Q} \in [60,120] \, {\rm GeV}$. Besides the total rates also spectra of the various subprocesses are shown. At LO we display the purely QCD ($\alpha_s^2$) and EW~($\alpha^2$) contributions, while at NLO we have chosen to depict various combinations of mixed QCD-EW corrections. When considering the EW corrections to the LO QCD processes ($\alpha\alpha_s^2$) only the values of the NLO coefficient is displayed, where we have separated the impact of the QED and weak corrections. The QED corrections in this case are negative, and thus the absolute values of the NLO coefficient are shown. In the case of the QCD corrections to the LO EW processes ($\alpha^2\alpha_s$) the sums of the LO and NLO coefficients are given, where we have also displayed the result when including only initial-state radiation (ISR)  from~QCD (labelled as ISR only).

The LO QCD contributions are by far dominant, while the LO EW contribution only becomes relevant (reaching roughly $10\%$) in the region of $m_{Q\bar Q} \in [85,95] \, {\rm GeV}$. The QED corrections to the LO QCD process are negative and more important in the case of charm-quark production,\footnote{This contribution is dominated by the QED correction to the $gg\to Q\bar Q$ subprocess, which is proportional to the squared electric charge of the heavy quark.} where these effects amount to half a percent of the total cross section. For both bottom- and charm-quark production, the weak corrections are negligibly small. The QCD corrections to the LO EW process have a more important impact on the obtained results. This is primarily due to the contribution of hard QCD corrections to the heavy-quark lines, where an emitted gluon is not reconstructed as part of the heavy-flavour jet. A consequence of such resonantly enhanced events which lose energy via the emission of such a hard gluon results in a shift of the $Z$ peak to lower $m_{Q\bar Q}$ values.  For $m_{Q\bar Q} \in [85,95] \, {\rm GeV}$, the corresponding numerical effect amounts to several percent at the cross-section level. 

As a final comment to Figure~\ref{fig:dsigdmqqZpole} (right), we note that the scale variation present for the~$c \bar c$ prediction of $\mathcal{O}(\alpha^2\alpha_s)$ in the bin $m_{c\bar c} \in [60,65] \, {\rm GeV}$ is a genuine effect in our calculation. It  originates from the NLO contributions associated to $qg \to Q\bar{Q}q$ (and the corresponding collinear counterterm) with purely photon exchange --- this term can be considered as part of the QED correction to the LO transition $g\gamma\to Q\bar{Q}$. The scale dependence of the NLO corrections would normally compensate that of the LO contribution, which is however absent in our computation due to the missing photon PDF. This increased scale dependence has a permille effect on the total cross section.

\begin{figure}[t!]
\centering
\includegraphics[width=.49\linewidth]{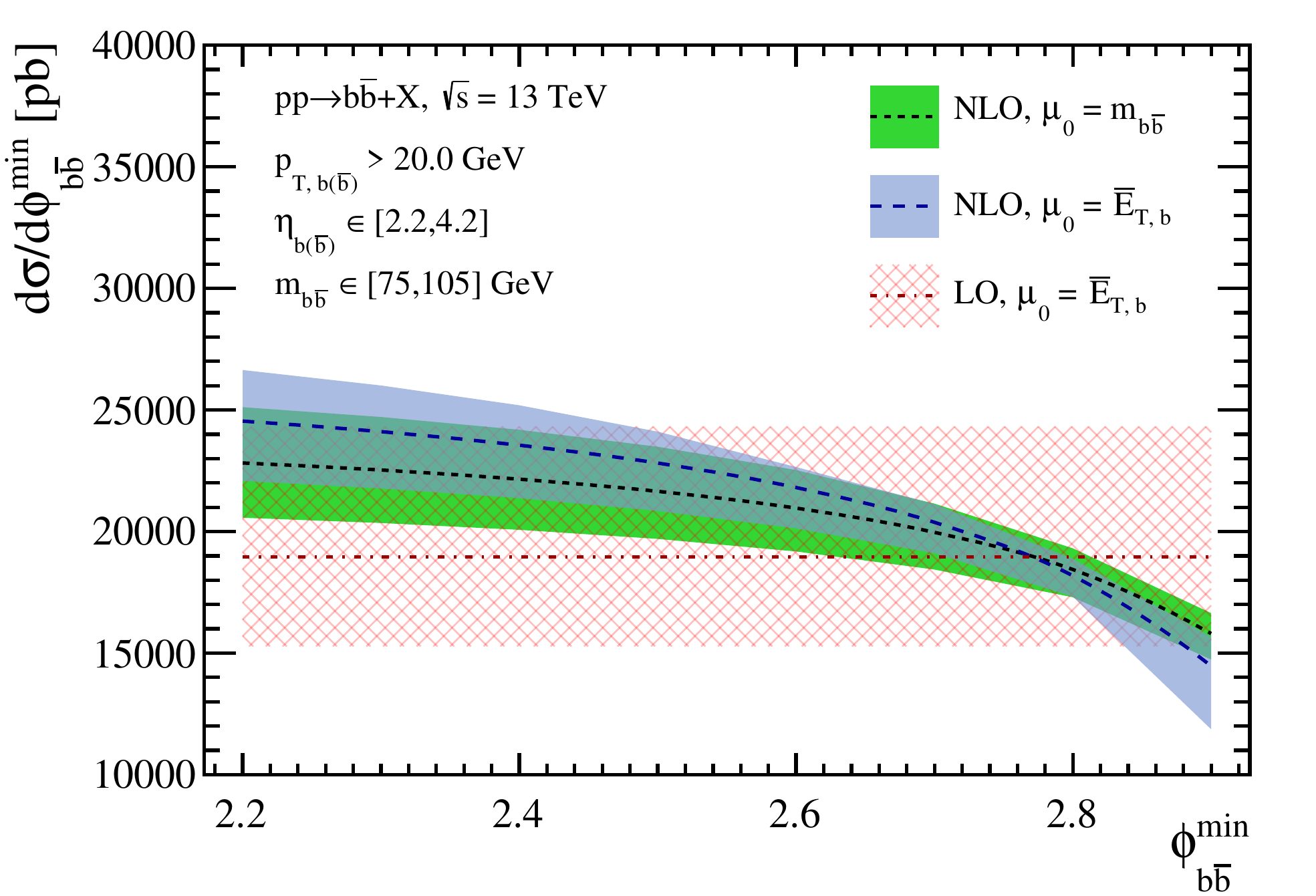} \hfill
\includegraphics[width=.49\linewidth]{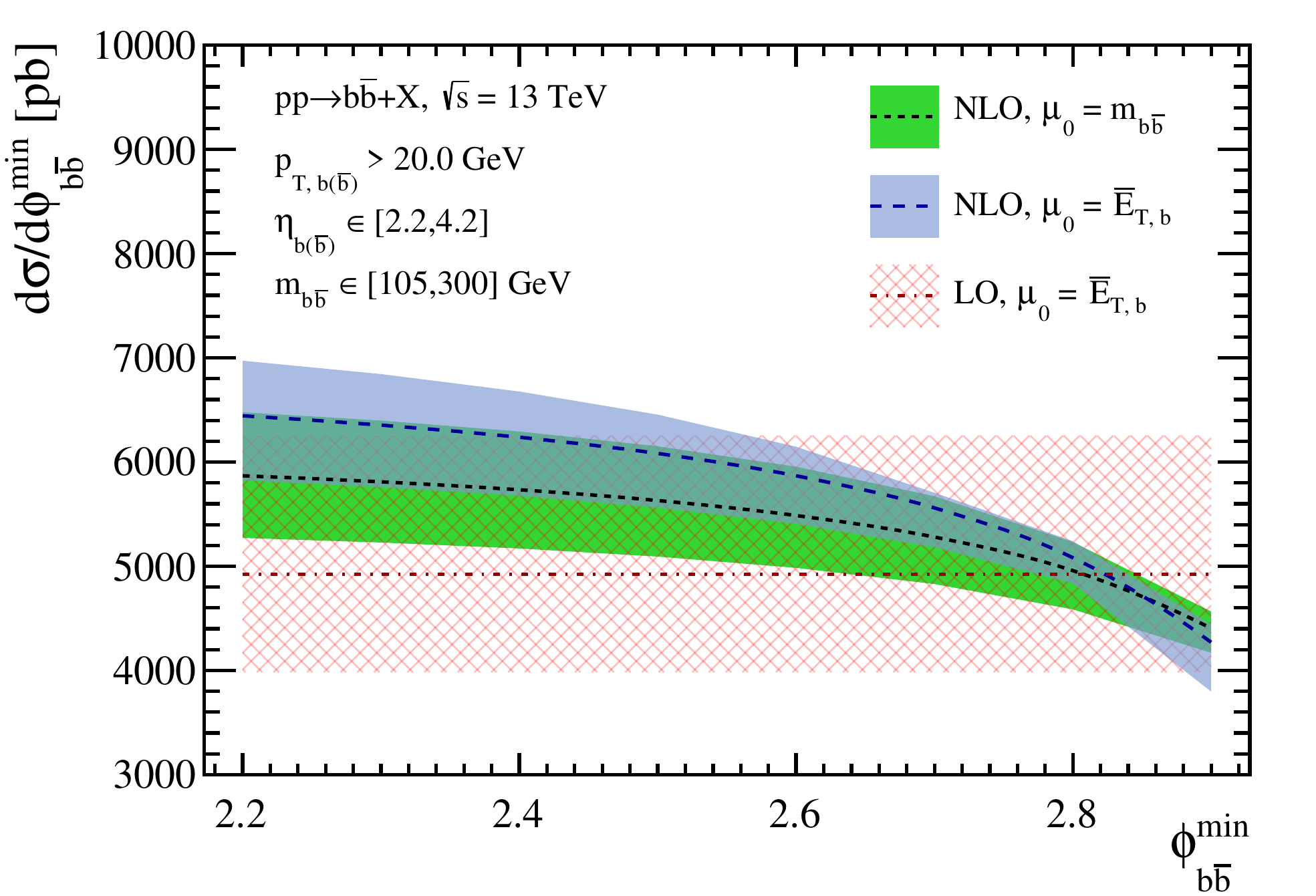} 
\vspace{2mm}
\caption{
Dependence of the $b \bar b$ cross section on the minimum angular separation $\phi_{b\bar b}^{\rm min}$ between the $b$-jets. The two plots correspond to the invariant-mass regions $m_{b\bar b} \in [75,105] \, {\rm GeV}$~(left) and $m_{b\bar b} \in [105,300] \, {\rm GeV}$~(right). The NLO results are obtained for $\mu_0 = m_{Q\bar{Q}}$ and $\mu_0 = \overline{E}_{T,Q}$, while the LO distributions employ $\mu_0 = \overline{E}_{T,Q}$ and  only shown for comparison. 
}
\label{fig:dsigdphi}
\end{figure}

To conclude the discussion of the differential cross sections, we present predictions for $b$-jet pair production within the LHCb acceptance as a function of the minimum angular separation $\phi_{b\bar b}^{\rm min}$ between the $b$-jets. In the LHCb analysis~\cite{Aaij:2014ywa}, a cut $\phi_{b\bar b} > 2.6$ is imposed, and said to provide improved sensitivity to the asymmetry of the signal process by enhancing ``non-$gg$ production mechanisms". In Section~\ref{asym} we will show that the value of this cut is not so important for the signal process, however it is still likely to be relevant for reducing the background contribution from light-flavoured jets. An experimental study of this distribution may therefore be useful when studying/reducing the contamination of background events. The relevant predictions for $pp$ collisions at $\sqrt{s} = 13 \, {\rm TeV}$ are presented in Figure~\ref{fig:dsigdphi} for the invariant mass bins of $m_{b\bar b} \in [75,105] \, {\rm GeV}$ (left) and $m_{b\bar b} \in [105,300] \, {\rm GeV}$ (right). The kinematic requirements on the $b$-jets are indicated in the plots, and are consistent with the standard cuts introduced in~\eqref{eq:fiducial}. 

%------------------------------------------------
%------------------------------------------------

\subsection{Cross-section ratios}
\label{sec:ratio}

In addition to the measurements of the $b \bar b$ and $c \bar c$ cross sections discussed in the last section, it is also of interest to perform  measurements of the cross-section ratio between the different heavy-quark types. As the theoretical predictions for the cross-section ratios are very precise, these measurements will provide an important experimental benchmark for testing and validating the (charged) flavour-tagging efficiency and mis-tag rates.

\begin{figure}[t!]
\centering
\includegraphics[width=.49\linewidth]{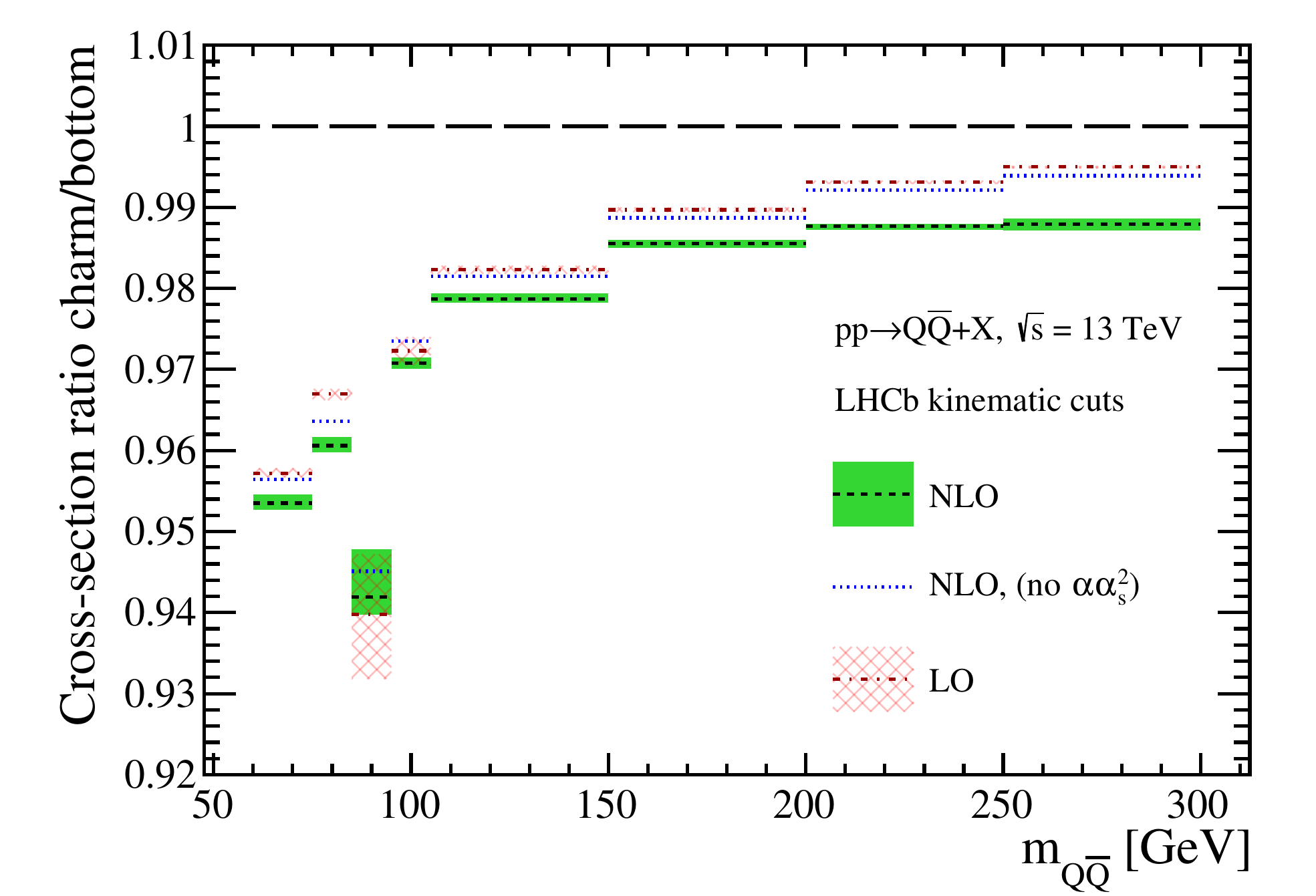} 
\includegraphics[width=.49\linewidth]{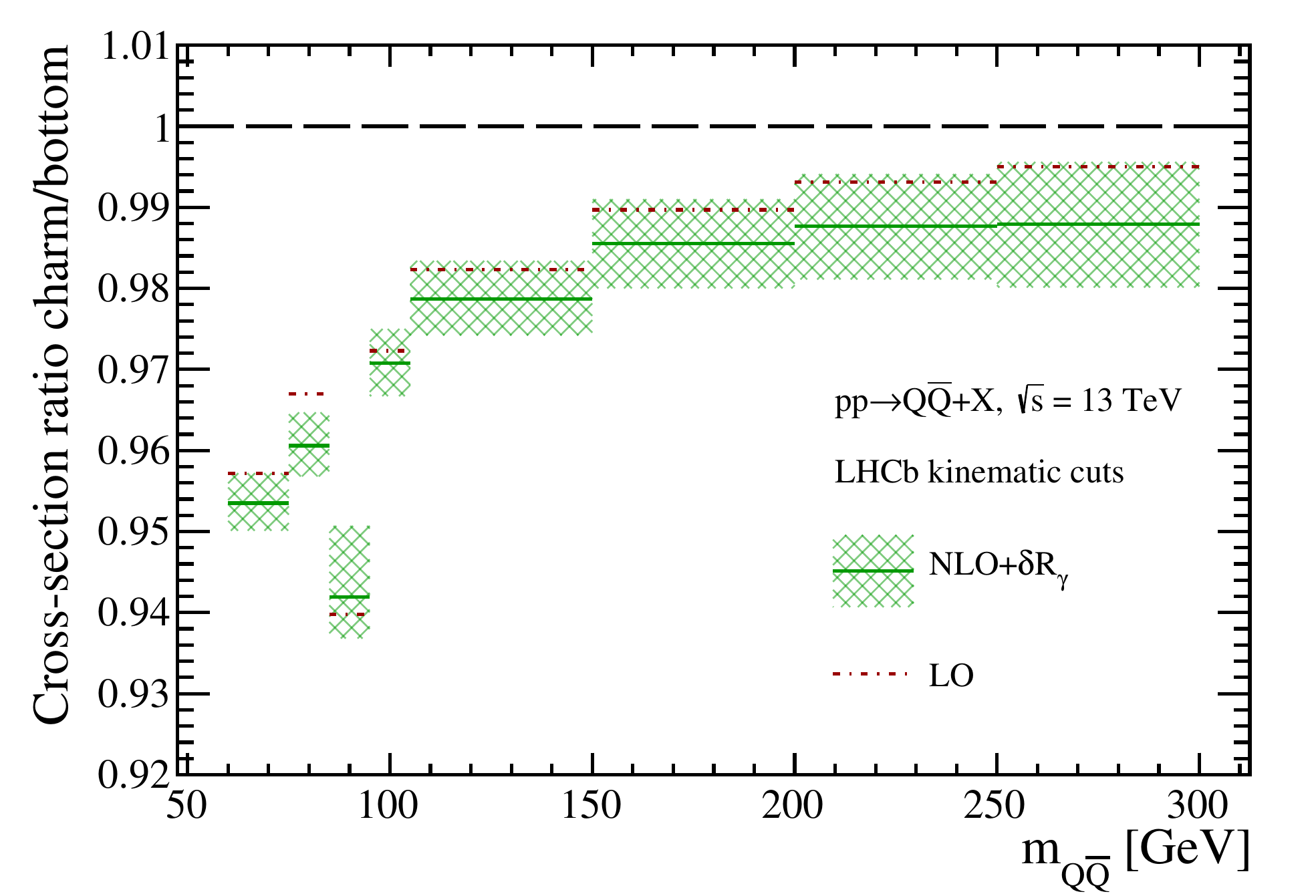} 
\vspace{2mm}
\caption{
Ratio between the differential cross section of $c$- and $b$-jet pair production at the 13~TeV~LHC.  Left: LO and NLO distributions for $\mu_0 = m_{Q\bar Q}$, as well as the NLO distribution obtained without the $\mathcal{O}(\alpha \alpha_s^2)$ corrections. Right: NLO distribution obtained with $\mu_0 = m_{Q\bar Q}$, where an additional uncertainty due to photon-induced contributions has been included as explained in the main text. The central value of the LO distribution is also shown for reference.
}
\label{fig:ratio}
\end{figure}

Our predictions for the cross-section ratio of $c$- and $b$-jet pairs at the 13~TeV LHC in the phase-space region~(\ref{eq:fiducial}) are shown in Figure~\ref{fig:ratio}. These distributions are obtained with $\mu_0 = m_{Q\bar{Q}}$, and the uncertainty has been evaluated by correlating the scale variations between the charm- and bottom-quark predictions. In the considered $m_{Q \bar Q}$ range, the  ratio between the $c \bar c$ and $b \bar b$ cross sections is below 1. The observed deviation of the ratio from~1~can be attributed to the mass dependence of the LO cross section --- see~Figure~\ref{fig:LOmass}~(left) --- and also to the different EW charges of up- and down-type quarks which affects the ratio both close to and away from the $Z$ peak. In Figure~\ref{fig:ratio} (left), the NLO ratio is also displayed for the case that  the $\mathcal{O}(\alpha\alpha_s^2)$ corrections have been removed.  These corrections arise dominantly in the form of  QED corrections to the $gg\to Q\bar Q$ subprocess. They are negative  and amount to effects of the order of $e_Q^2 \cdot 1\%$ on the spectra, where $e_Q$ denotes the electric charge of the heavy quark $Q$. While the $\mathcal{O}(\alpha\alpha_s^2)$  contributions thus have a negligible impact at the level of the cross sections, they have a visible effect on the ratio of the symmetric rates.

%------------------------------------------------
%------------------------------------------------

As discussed in Section~\ref{setup}, we have chosen to use PDFs that do not include a photon PDF, and as a result photon-initiated contributions are not included in our computations. To assess the potential uncertainty due to these missing contributions, we have recomputed the ratio  of the $c \bar c$ and $b \bar b$ spectra  at LO with the {\tt LUXqed15} PDF set. An uncertainty is then calculated  according to
\begin{align}
\delta R_{\gamma} = \frac{\sigma ({pp \to c\bar c})}{\sigma ({pp \to b\bar b})} - \left( \frac{\sigma ({pp \to c\bar c})}{\sigma ({pp \to b\bar b})}  \right)_{\text{no photon PDF}} \,,
\end{align} 
where the second ratio is computed excluding all photon-initiated contributions. The modulus of this uncertainty is then added linearly to the scale uncertainty, both in the positive and negative directions. The corresponding results are computed at NLO accuracy with $\mu_0 = m_{Q\bar{Q}}$ and shown in Figure~\ref{fig:ratio} (right). Computing the uncertainty this way is likely to overestimate the total theoretical uncertainties. However, to our knowledge, the only publicly available PDF sets based on the precise {\tt LUXqed15} photon PDF determination are NNLO QCD accurate or have been determined at NLO QCD accuracy while fitting an intrinsic charm-quark PDF. Using either of these PDF sets for the current predictions would introduce some level of inconsistency. At present, we therefore recommend to use the conservative uncertainty including $\delta R_{\gamma}$ when comparing our results to future data. 

\section{Asymmetry predictions}
\label{asym}

In this Section we provide differential predictions for the asymmetries as defined in~\eqref{eq:asym}. These predictions are obtained  by separately computing both the numerator and denominator of this expression at NLO,~i.e.~including terms up to $n+m = 3$ in the expansion defined in~\eqref{eq:expansion}. The corresponding LO results are obtained by including terms up to $n+m = 2$. The exception is that we also take into account the $\mathcal{O}(\alpha_s^3)$ contribution to the numerator when evaluating the asymmetry at LO. This procedure is  motivated by the well-known fact that the $\mathcal{O}(\alpha_s^2)$ corrections do not generate an asymmetry in the SM, so that  the $\mathcal{O}(\alpha_s^3)$ terms should  be considered the leading contribution to the asymmetry. In fact, these terms are numerically dominant except for $m_{Q \bar Q}$ values close to the $Z$ pole. 

\begin{figure}[t!]
\centering
\includegraphics[width=.49\linewidth]{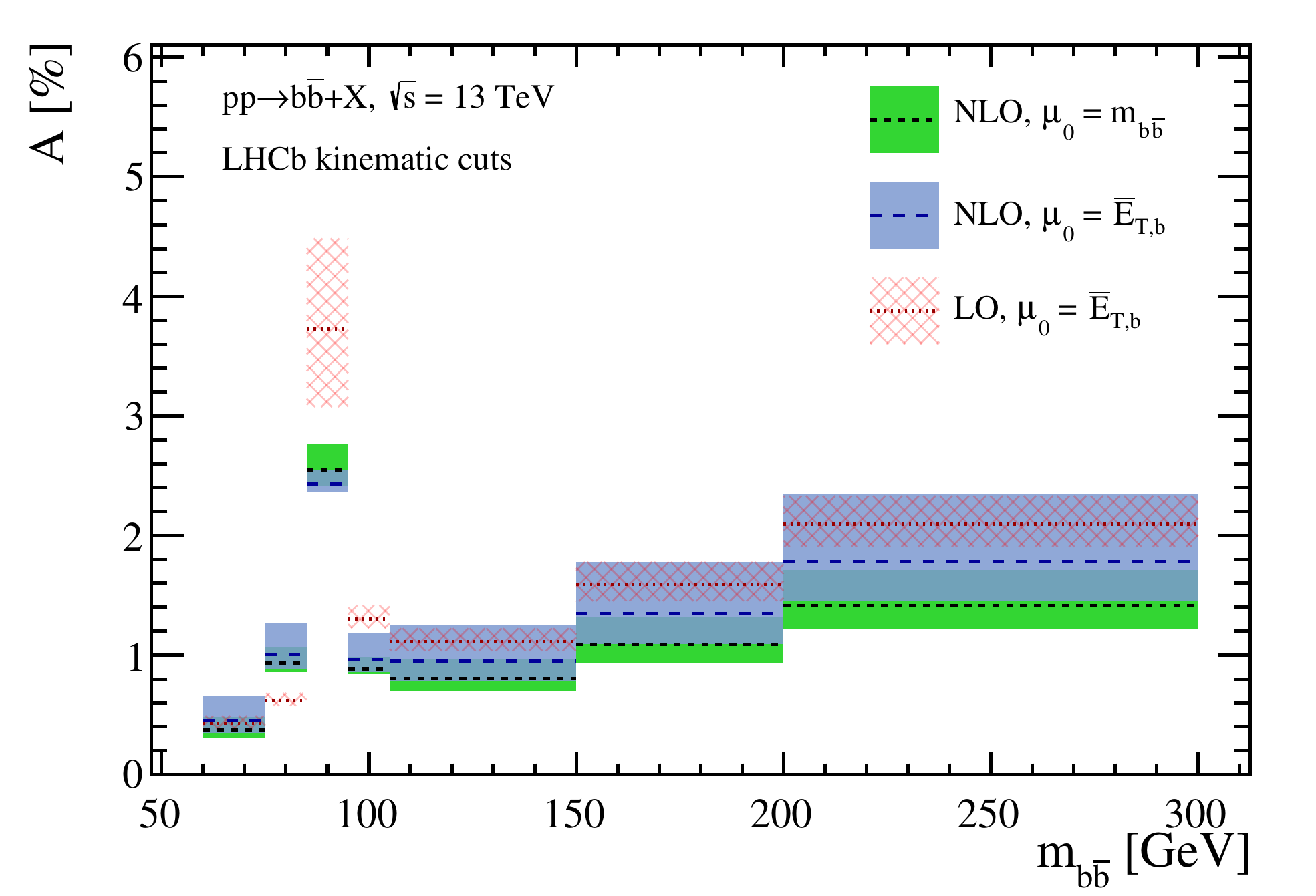} \hfill
\includegraphics[width=.49\linewidth]{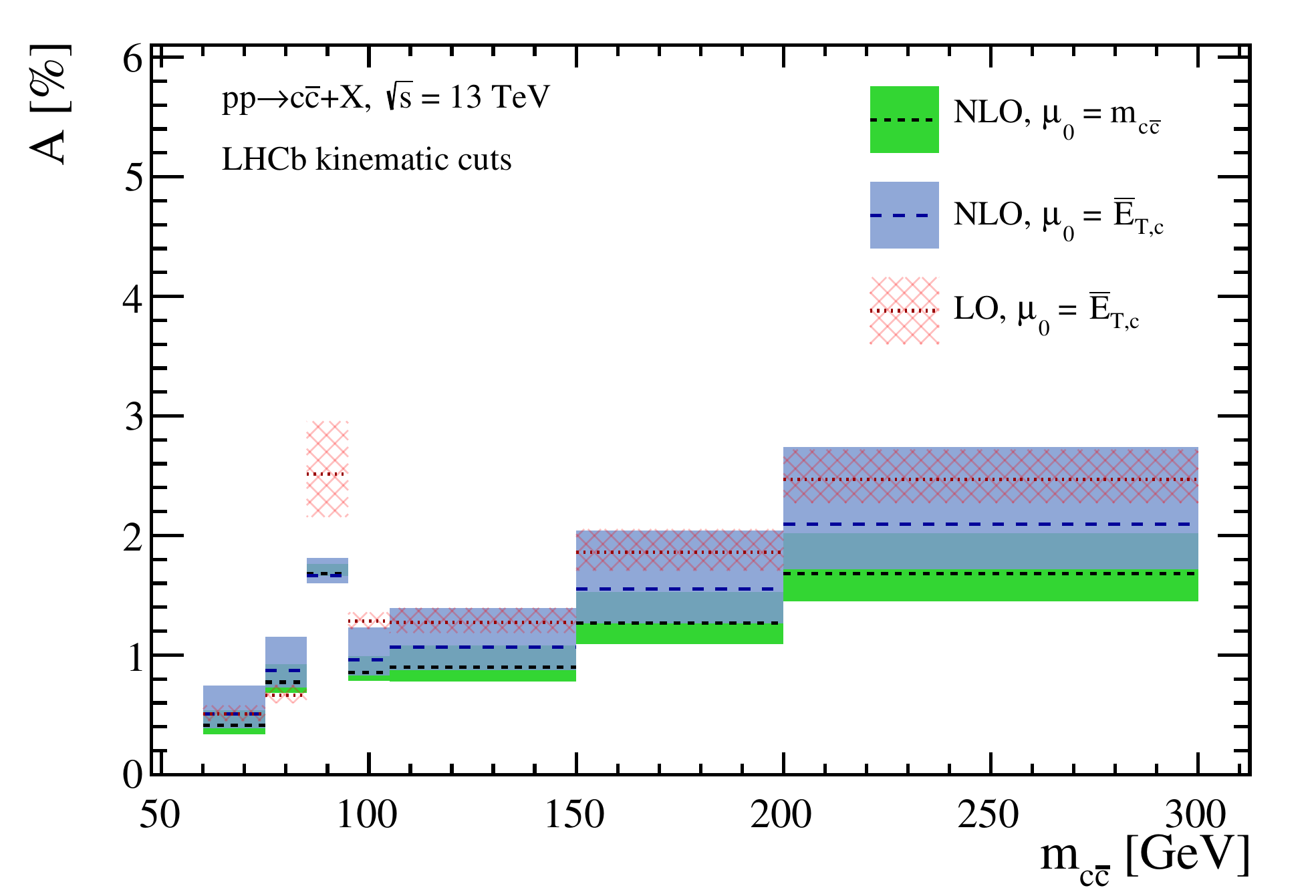} 
\vspace{2mm}
\caption{
Differential asymmetries for $b$- (left) and $c$-jet (right) pairs within the LHCb acceptance~\eqref{eq:fiducial} at $\sqrt{s} = 13 \, {\rm TeV}$.  The NLO (LO) results have been  obtained for $\mu_0 = m_{Q\bar{Q}}$ and $\mu_0  = \overline{E}_{T,Q}$ ($\mu_0 = \overline{E}_{T,Q}$), and the shown error bands correspond to scale variations. 
}
\label{fig:asym}
\end{figure}

In Figure~\ref{fig:asym} our results for the $b \bar b$ (left) and $c \bar c$ (right) asymmetries are presented. The NLO predictions corresponding to the two different scales choices~(\ref{eq:scalechoices})  lead to consistent results across the considered mass range. Close to the $Z$ peak it is found that the NLO corrections have an important impact on the absolute value of the asymmetries as well as the uncertainty estimates. In both the low- and high-mass regions, the uncertainty estimate from the LO prediction is artificially small and should not be considered robust. 

%------------------------------------------------
%------------------------------------------------

Compared to the results presented in~\cite{Gauld:2015qha}, our current predictions include the following improvements. First, a jet definition consistent with the LHCb flavour-tagging algorithm~\cite{Aaij:2015yqa} is used throughout. Second,  based on the recent measurement of $b\bar b$ production in the vicinity of the $Z$ peak~\cite{Aaij:2017eru}, numerical predictions for both the cross section and asymmetry are provided in fine bins in this invariant-mass region. Third,  more precise theoretical predictions are obtained by including a number of subleading NLO corrections, which were previously absent. Fourth, the numerical predictions are computed with updated PDFs which include LHC data, and have been obtained with two reference scale settings which lead to a more reliable uncertainty estimate. These improvements should allow for a more precise comparison to data, which can in turn be used to perform more stringent tests of the SM as well as new physics. Two  applications along these lines are discussed in Section~\ref{sec:applications}.

Before discussing these applications, it is important to estimate the potential sensitivity of future experimental measurements. The original measurement of the $b$-jet pair asymmetry at LHCb was performed with an integrated luminosity of $1 \, {\rm fb}^{-1}$ collected at~$7 \, {\rm TeV}$~\cite{Aaij:2014ywa}. Results were presented in the $m_{b \bar b}$ bins of $[40,75] \, {\rm GeV}$, $[75,105] \, {\rm GeV}$ and $[105,300] \, {\rm GeV}$, and the measurement was statistically limited in each bin. To estimate the statistical sensitivity expected at $13 \, {\rm TeV}$, we compute the corresponding uncertainties via 
\begin{align} \label{eq:stat}
\delta A_{\rm stat}^2 = \frac{(1-A^2)}{N} \,,
\end{align}
where $A$ denotes the central value of the NLO asymmetry obtained with  $\mu_0 = m_{Q\bar{Q}}$, and~$N$~is the expected number of events within the data set. To calculate $N$, we use  our  cross-section predictions, assume a data set of $ 5 \, {\rm fb}^{-1}$, and further apply experimental efficiencies for the reconstruction of a pair of charged- and flavour-tagged jets of $\epsilon_{b\bar b} = 0.6\%$ and $\epsilon_{c\bar c} = 0.3\%$. The values of these efficiencies are obtained by inverting~\eqref{eq:stat} for the $7 \, {\rm TeV}$ measurement~\cite{Aaij:2014ywa} using the corresponding central NLO prediction at $7 \, {\rm TeV}$. We note that the value of $\epsilon_{b\bar b} = 0.6\%$ corresponds to a  factor of two improvement compared to what has been achieved in the original measurement. 

\begin{figure}[t!]
\centering
\includegraphics[width=.49\linewidth]{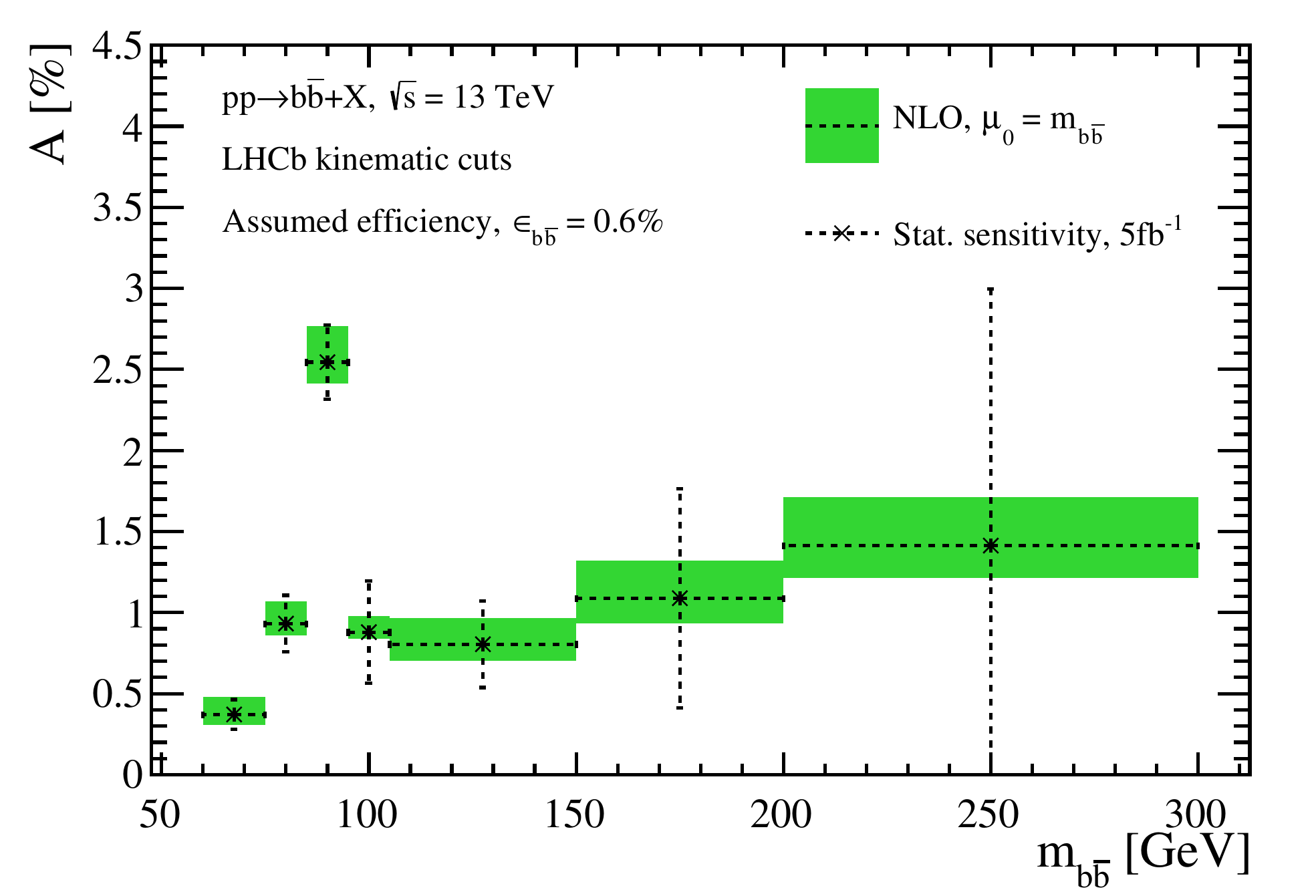} \hfill
\includegraphics[width=.49\linewidth]{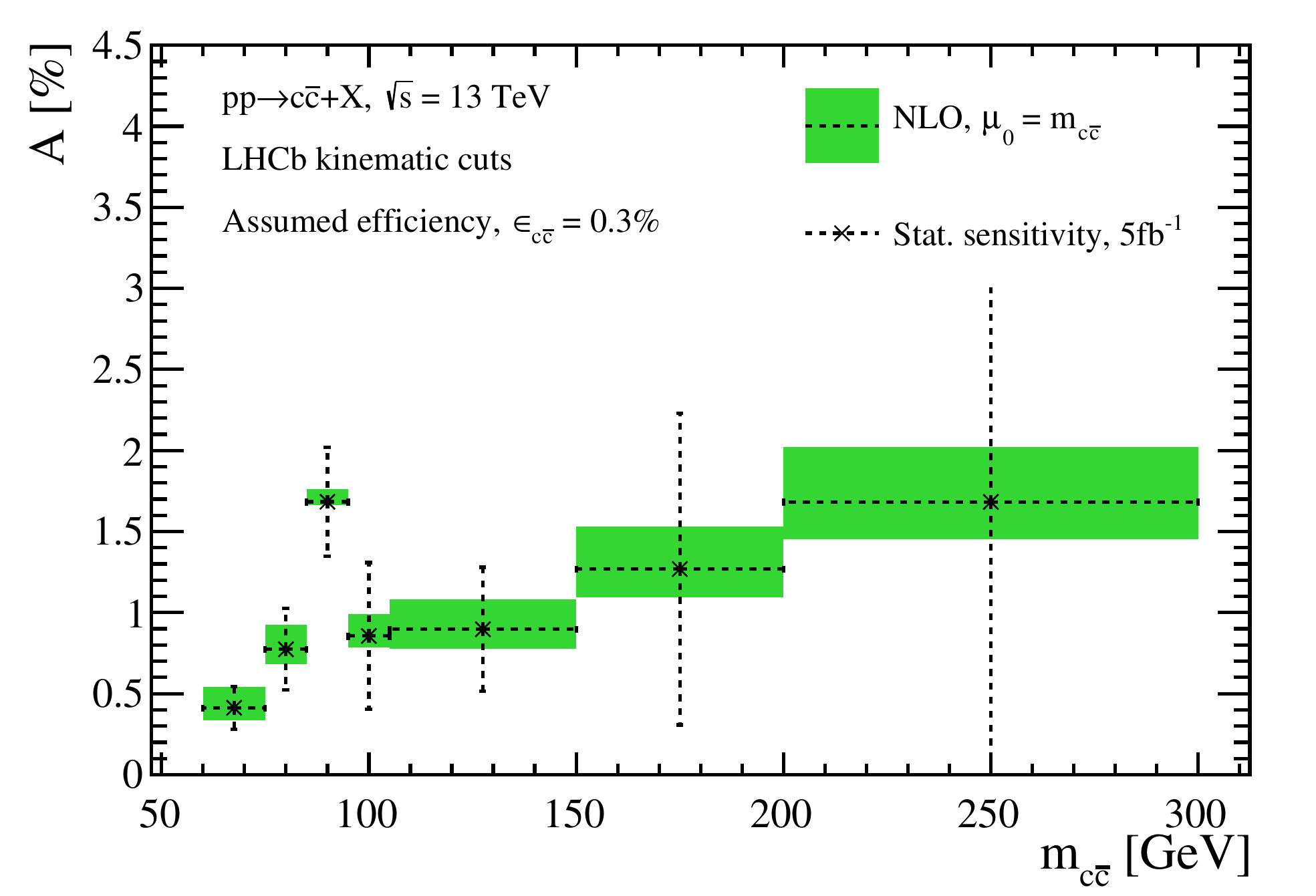} 
\vspace{2mm}
\caption{
Differential asymmetry for $b$- (left) and $c$-jet (right) pairs within the LHCb fiducial region~(\ref{eq:fiducial}) at $\sqrt{s} = 13 \, {\rm TeV}$. The shown NLO distributions are obtained with the scale choice $\mu_0 = m_{Q\bar{Q}}$, and  the estimated statistical sensitivity of a future measurement at LHCb with $5 \, {\rm fb}^{-1}$ of integrated luminosity has been indicated.
}
\label{fig:stat}
\end{figure}

The results of our sensitivity study are shown in Figure~\ref{fig:stat}, where the projections for the statistical uncertainties~(\ref{eq:stat}) are overlaid on the predictions for the $b$- (left) and $c$-jet~(right) pair asymmetries. This study indicates that a significant improvement in statistical precision will be achievable with future data sets, and that  finely binned measurements close to the $Z$ pole should be possible. This is a consequence of the higher cross sections, the increased data sample size, and the assumption about the improved signal efficiency. In the event that a data sample of $50 \, {\rm fb}^{-1}$ is collected at LHCb~\cite{Bediaga:2018lhg} (such as in the high-luminosity phase of the LHC), it is likely that measurements of the heavy-quark asymmetries will be systematically and not statistically limited.

%------------------------------------------------
%------------------------------------------------

\begin{figure}[t!]
\centering
\includegraphics[width=.49\linewidth]{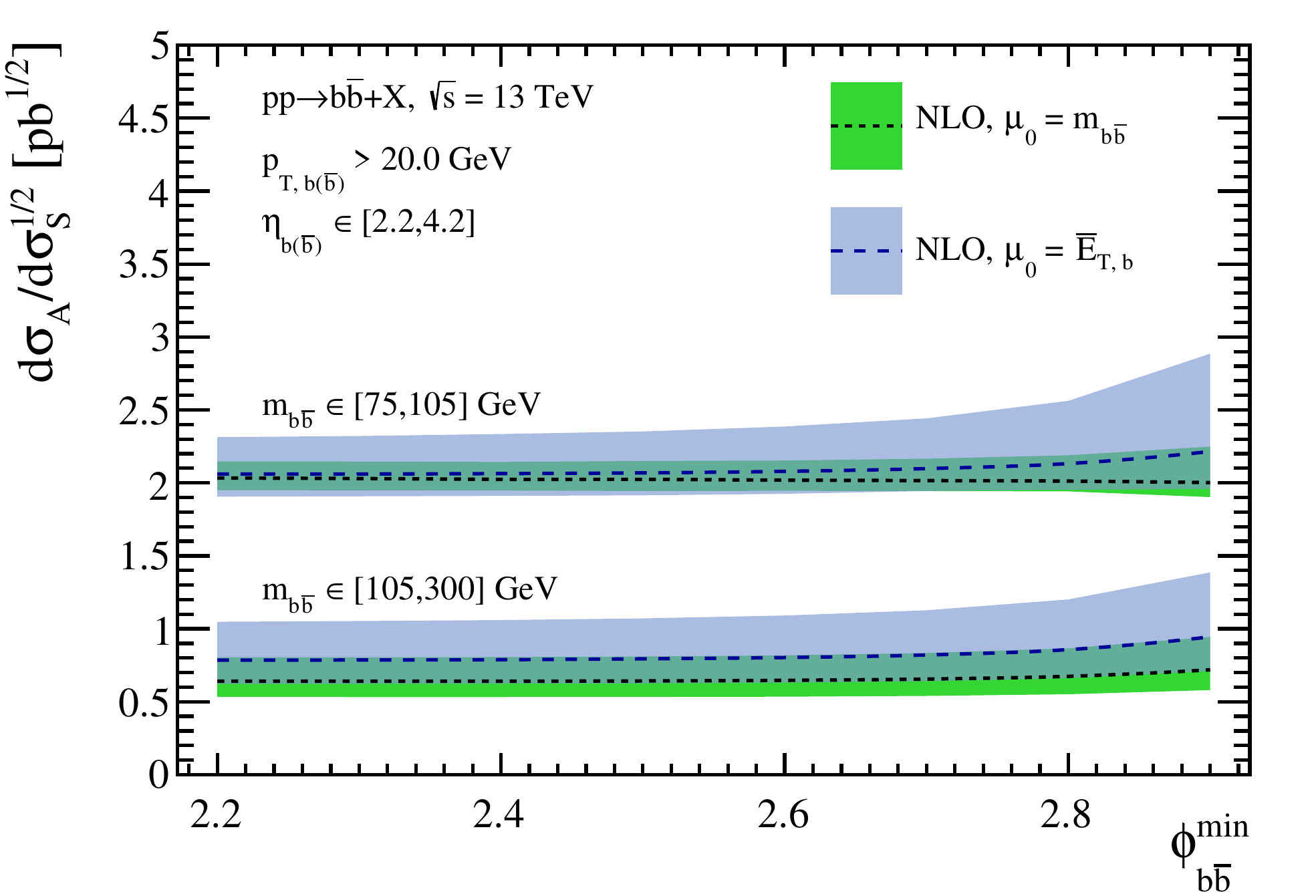} 
\caption{
Asymmetric $b \bar b$ cross section within $m_{b\bar b} \in [75,105] \, {\rm GeV}$ and $m_{b\bar b} \in [105,300] \, {\rm GeV}$ normalised to the square root of the symmetric cross section as a function of~$\phi_{b\bar b}^{\rm min}$. The shown NLO predictions correspond to the LHCb fiducial region~(\ref{eq:fiducial}) at $\sqrt{s} = 13 \, {\rm TeV}$, and  employ the two scale choices introduced in~(\ref{eq:scalechoices}). The depicted error bands are due to scale variations. }
\label{fig:dphi_asym}
\end{figure}

We conclude this Section by returning to the choice of the angular  cut $\phi_{Q\bar Q}$ used in defining the fiducial region~\eqref{eq:fiducial}. As mentioned in Section~\ref{sigma}, the motivation for introducing this cut is to increase the sensitivity to the asymmetry by enhancing ``non-$gg$ production mechanisms". To assess this statement, we study the impact of the choice of $\phi^{\rm min}_{Q\bar Q}$ on the observable $\sigma_A/\sigma_S^{1/2}$, where $\sigma_{S \, (A)}$ is the (a)symmetric production cross section. The motivation behind this definition is that the significance of a statistically limited measurement of the asymmetry  is approximately $A/\delta A_{\rm stat}$. Our definition is therefore useful as it estimates the overall statistical sensitivity to the asymmetry measurement itself, rather than just the asymmetry. This is relevant because, while the value of the asymmetry may increase as the value of the cut $\phi_{Q\bar Q}^{\rm min}$ is increased, the number of analysed events simultaneously decreases. Our predictions for $\sigma_A/\sigma_S^{1/2}$  as a function of $\phi_{b \bar b}^{\rm min}$ are shown in Figure~\ref{fig:dphi_asym}. The two different sets of predictions correspond to the results restricted to the invariant mass bins $m_{b\bar b} \in [75,105] \, {\rm GeV}$ and $m_{b\bar b} \in [105,300] \, {\rm GeV}$. The obtained  distributions are close to flat as $\phi_{b\bar b}^{\rm min}$ increases, indicating that from a statistical point of view the sensitivity to the asymmetry is not improved by requiring larger~$\phi_{b\bar b}^{\rm min}$ values. We have therefore chosen to provide predictions for $\phi_{Q\bar Q}^{\rm min} = 2.6$, which matches the original value advocated in~\cite{Aaij:2014ywa}. It is worth noting that the choice of this cut may also be important for background rejection~(i.e.~from light jets). In the far future, if the asymmetry measurements becomes systematically limited, it may be worthwhile to perform a dedicated experimental study of this~issue. 

%------------------------------------------------
%------------------------------------------------

\section{Applications} 
\label{sec:applications}

In this Section we present two applications of our calculations of heavy-quark production. We will first discuss the model-independent constraints that future LHCb measurements of the ratio of the $b \bar b$ and $c \bar c$ asymmetry may allow to set  on the couplings of the $Z$ boson~to bottom- and charm-quark pairs. We will compare the results of our sensitivity studies to~the constraints on the $Zb \bar b$ and $Z c \bar c$ couplings that arise from the $Z$-pole measurements performed at LEP~\cite{ALEPH:2005ab}. Our second application consists in using the recent LHCb measurement of $Z \to b \bar b$ production~\cite{Aaij:2017eru} to put constraints on the new-physics model proposed in~\cite{Liu:2017xmc} which aims at explaining the long-standing LEP anomaly of the forward-backward asymmetry of the bottom quark.  

\subsection[Constraints on $Zb \bar b$ and $Z c \bar c$ couplings]{Constraints on $\bm{Zb \bar b}$ and $\bm{Z c \bar c}$ couplings}

Similarly to the top-quark asymmetry, the dominant contribution to the asymmetry of bottom- and charm-quark arises from  QCD for most values of the invariant mass of the heavy-quark pair.  An important exception is the mass region close to the $Z$-pole, where the double-resonant contribution from $Z$-$Z$ interference becomes dominant, and accounts for the bulk of the total asymmetry~\cite{Grinstein:2013iws,Murphy:2015cha,Gauld:2015qha}. Measurements of the bottom- and charm-quark asymmetry can therefore be used to set limits on the $Zb \bar b$ and $Z c \bar c$ couplings~\cite{Murphy:2015cha}.

In order to put model-independent constraints on the $Z$-boson couplings to bottom and charm quarks, we consider in the following the ratio $R_{b/c} = A_{b}/A_{c}$ of the asymmetry in $b \bar b$ and $c \bar c$ production restricted to the mass bin $[75, 105] \, {\rm GeV}$. This ratio can be predicted to high accuracy in the SM~\cite{Gauld:2015qha}, since many uncertainties cancel between numerator and denominator. For $pp$ collisions at $\sqrt{s} = 13 \, {\rm TeV}$ and employing the standard LHCb cuts~\eqref{eq:fiducial}, we find within the SM the result 
\beq \label{eq:RbbccSM}
R_{b/c}^{\rm SM}  = 1.33 \pm 0.07 \,. 
\eeq
The given central value corresponds to the reference scale choice $\mu_0 = m_{Q\bar Q}$ and the quoted uncertainty of around $5\%$ includes scale variations as described in Section~\ref{sec:PDFsEWSCAL} and~PDF uncertainties.  The dominant source of uncertainty arises from missing higher-order QCD corrections, meaning that the stated total uncertainty  is in principle improvable by including NNLO corrections. We add that using instead the scale setting $\mu_0 = \overline{E}_{T,Q}$ leads to a central value of $R_{b/c}^{\rm SM}$ that agrees within errors with that in~(\ref{eq:RbbccSM}) and to a comparable total uncertainty. 

The experimental measurements of the EW $Z$-pole observables obtained at~LEP can be used to precisely extract the $Z$-boson couplings to all SM quarks but the top quark. In the case of the $Zb \bar b$ and $Z c \bar c$ couplings the combined results are~\cite{ALEPH:2005ab}
\beq \label{eq:gbcexp}
\begin{split}
g_L^b & =  -0.4182 \pm 0.0015 \,, \qquad g_R^b = 0.0962 \pm  0.0063 \,, \\[1mm]
g_L^c  & =  0.3453 \pm 0.0036 \,, \qquad \phantom{-} g_R^c = -0.1580 \pm 0.0051 \,,
\end{split}
\eeq	
and the corresponding correlation matrix is given by~\cite{ALEPH:2005ab} 
\beq \label{eq:correlation}
\rho = \begin{pmatrix}
1.00 & 0.88 & -0.09 & -0.17 \\
0.88 & 1.00 & -0.14 & -0.13  \\
-0.09  & -0.14 & 1.00 & 0.30  \\
 -0.17 & -0.13 & 0.30 & 1.00
\end{pmatrix} \,.
\eeq
The SM predictions for the  $Zb \bar b$ and $Z c \bar c$ couplings can be extracted with the help of  {\tt ZFITTER}~\cite{Arbuzov:2005ma} and read 
\beq \label{eq:gbcSM}
\begin{split}
(g_L^b)^{\rm SM} & =  -0.42114  \,, \qquad (g_R^b)^{\rm SM} = 0.077420 \,, \\[1mm]
(g_L^c)^{\rm SM} & = 0.34674 \,, \qquad \phantom{-} (g_R^c)^{\rm SM} = -0.15470  \,.
\end{split}
\eeq
Since the uncertainties associated with~(\ref{eq:gbcSM}) are negligible compared to the uncertainties quoted in~(\ref{eq:gbcexp})  only the central values of the SM expectations have been given here. Compared to the experimental measurements~(\ref{eq:gbcexp}) and~(\ref{eq:correlation}) the SM values~(\ref{eq:gbcSM}) have a $\chi^2$ per degree of freedom ($\chi^2/{\rm dof}$) of 2.8.  

%------------------------------------------------
\begin{figure}
\centering
\includegraphics[width=.95\linewidth]{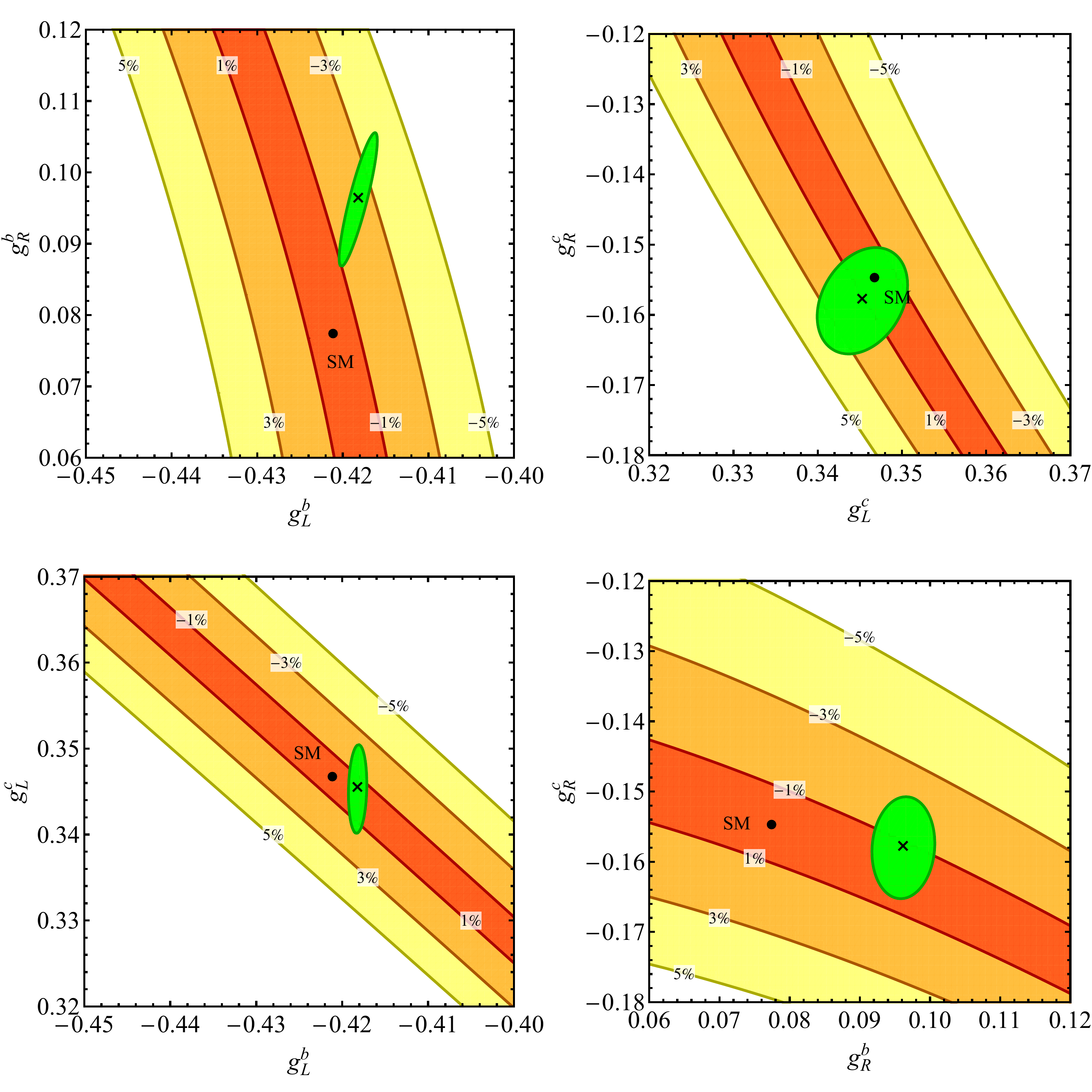} 
\caption{Relative deviations of the ratio  $R_{b/c}$ in the $g_L^b$--$\hspace{0.25mm} g_R^b$ (upper left), $g_L^c$--$\hspace{0.25mm} g_R^c$ (upper right), $g_L^b$--$\hspace{0.25mm} g_L^c$ (lower left) and $g_R^b$--$\hspace{0.25mm} g_R^c$ (lower right) plane.  For comparison also the 68\%~confidence level~(CL) regions favoured by the LEP measurements of the EW $Z$-pole observables are shown. The SM and best-fit points are indicated by black dots and black crosses, respectively. }
\label{fig:gZQQ}
\end{figure}
%------------------------------------------------

In Figure~\ref{fig:gZQQ} we show the relative deviations of   $R_{b/c}$ in four different planes of $Zb \bar b$ and $Z c \bar c$ couplings. Overlaid in green are the 68\%~CL regions that follow from the LEP measurements~(\ref{eq:gbcexp}) and~(\ref{eq:correlation}). The SM and best-fit points are indicated by black dots and black crosses in the figure. Numerically, we find that the best fits lead to relative deviations of $-2.4\%$, $1.1\%$, $-0.4\%$ and $-0.8\%$  in $R_{b/c}$ in the  $g_L^b$--$\hspace{0.25mm} g_R^b$ (upper left), $g_L^c$--$\hspace{0.25mm} g_R^c$ (upper right), $g_L^b$--$\hspace{0.25mm} g_L^c$ (lower left) and $g_R^b$--$\hspace{0.25mm} g_R^c$ (lower right) plane, respectively. These numbers indicate that for future LHCb measurements of the $b \bar b$ and $c \bar c$ asymmetries to reach the sensitivity of the existing LEP constraints on the $Zb \bar b$ and $Z c \bar c$ couplings, determinations of the ratio $R_{b/c}$ at the percent level are needed.  Notice that to reach this goal not only the experimental precision but also the theoretical  accuracy of the SM predicition~(\ref{eq:RbbccSM}) needs to be improved. Such a theoretical improvement would require the inclusion of NNLO QCD corrections in the prediction of  $R_{b/c}$, which is technically viable in view of~\cite{Czakon:2014xsa}.

\subsection{Constraints on new light gauge bosons}

Precision measurements of the gauge sector have shown agreement with expected SM properties at the permille level.  Among the many observables, the bottom-quark forward-backward asymmetry $A_{\rm FB}^{b}$ measured at LEP however presents a $3\sigma$ deviation with respect to the values expected in the SM~\cite{ALEPH:2005ab}.  While  this deviation could be a result of  statistical fluctuations, it is intriguing since it also could be associated with a large modification of the right-handed $Zb \bar b$ coupling $\big($cf.~(\ref{eq:gbcexp}) and~(\ref{eq:gbcSM})$\big)$, which can for instance arise  if the   $Z$~boson is mixed with additional neutral gauge bosons. 

A recently proposed model~\cite{Liu:2017xmc} that aims at explaining the  long-standing LEP anomaly of $A_{\rm FB}^{b}$ contains a new $U(1)_D$ boson (the corresponding mass eigenstate will be called $Z^\prime$ in what follows), which couples with opposite charges to the right-handed components of the bottom and charm quarks. The low-energy spectrum of the model also includes two Higgs doublets, a singlet, and a charged and a neutral vector-like singlet, the latter being a good dark matter candidate. The phenomenology of the $Z$ and the $Z^\prime$ bosons is fully described by the masses $M_Z$ and $M_{Z^\prime}$ of the two gauge bosons, the new coupling constant~$g_D$, the sine $s_\alpha$ of the mixing angle $\alpha$  of the neutral weak eigenstates and the mixings of the bottom~(charm) quark with a heavy  vector-like bottom~(charm) partner, parameterised by the four variables $s_{b,L}$, $s_{b,R}$, $s_{c,L}$ and~$s_{c,R}$. 

The following values of the~$U(1)_D$  gauge coupling and the mixing parameters   
\beq \label{eq:benchmark}
g_D = 0.36 \,, \quad s_\alpha = -0.03 \,, \quad s_{b,L} = -0.07 \,,  \quad s_{c,L} = -0.1 \,, \quad  s_{b,R} =s_{c,R} = -0.001 \,, 
\eeq
have been used in the article~\cite{Liu:2017xmc} as a benchmark. In fact, for these choices the  $Zb \bar b$ and $Z c \bar c$  couplings  in the $U(1)_D$ model take the values  
\beq \label{eq:gbcU(1)D}
\begin{split}
(g_L^b)^{U(1)_D} & =  -0.4185  \,, \qquad (g_R^b)^{U(1)_D} = 0.0920 \,, \\[1mm]
(g_L^c)^{U(1)_D} & =  0.3416 \,, \qquad \phantom{-} (g_R^c)^{U(1)_D} =  -0.1693 \,.
\end{split}
\eeq
These couplings lead to a $\chi^2/{\rm dof}$ of $1.6$, which represents a visible improvement compared to the  $\chi^2/{\rm dof}$  value quoted after~(\ref{eq:gbcSM}).  

%------------------------------------------------
\begin{figure}
\centering
\includegraphics[width=.975\linewidth]{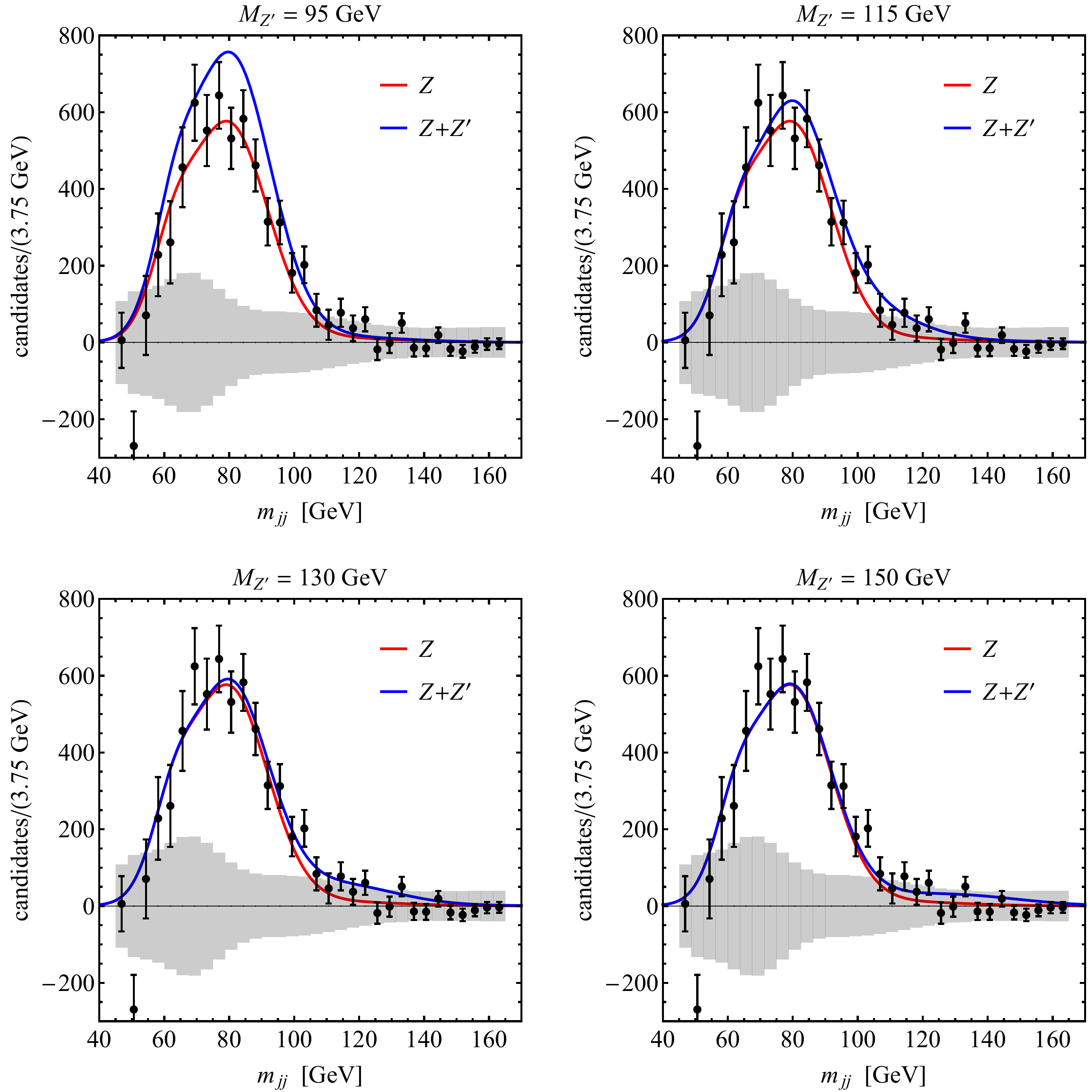} 
\caption{Dijet mass distributions in $Z \to b \bar b$ predicted in the $U(1)_D$ model for different $Z^\prime$ masses~(blue curves). All predictions employ the benchmark parameters~(\ref{eq:gbcU(1)D}). The background-subtracted dijet mass distribution  as measured by LHCb in~\cite{Aaij:2017eru}~(black error bars), the $Z \to b \bar b$ mass model within the SM~(red curves) and the one standard deviation total uncertainty band~(grey bands) is also shown. The uncertainty band includes statistical  and systematic uncertainties. See text for further details. }
\label{fig:mjjexamples}
\end{figure}
%------------------------------------------------

Constraints on the Peskin-Takeuchi parameter $T$ now put a bound on the size of the allowed mass splitting $M_{Z^\prime} - M_Z$~\cite{Liu:2017xmc}. For the  $s_\alpha$ value given in~(\ref{eq:gbcSM}), one finds that the constraint  $T = 0.07 \pm 0.12$~\cite{Tanabashi:2018oca} translates into the following 95\%~CL limit on the mass of the new gauge boson 
\beq \label{eq:MZpbound}
M_{Z^\prime} \in  [91.2, 174] \, {\rm GeV} \,.
\eeq
As pointed out in~\cite{Liu:2017xmc}, the presence of a $Z^\prime$ boson in this mass range is subject to several constraints. The first constraint comes from the CMS search~\cite{Sirunyan:2017nvi} for  narrow spin-1 resonances decaying to a $q \bar q$ pair in association with a high-transverse momentum  jet from ISR. Other relevant constraints arise from the $Z^\prime \to \ell^+ \ell^-$ searches~\cite{Aad:2012cfr,Aaboud:2017buh}. For the benchmark parameters~(\ref{eq:benchmark}) the combination of the constraints~\cite{Sirunyan:2017nvi,Aad:2012cfr,Aaboud:2017buh} can however be shown to be fulfilled for most of the $Z^\prime$-boson masses in~(\ref{eq:MZpbound}). In fact, the search~\cite{Sirunyan:2017nvi} features an $2.9 \sigma$ local excess for dijet invariant masses around $115 \, {\rm GeV}$, which has been interpreted in~\cite{Liu:2017xmc}   as a $Z^\prime$ boson in the $U(1)_D$ model with $M_{Z^\prime} \simeq 115 \, {\rm GeV}$ and~(\ref{eq:gbcU(1)D}). 

\begin{figure}
\centering
\includegraphics[width=.45\linewidth]{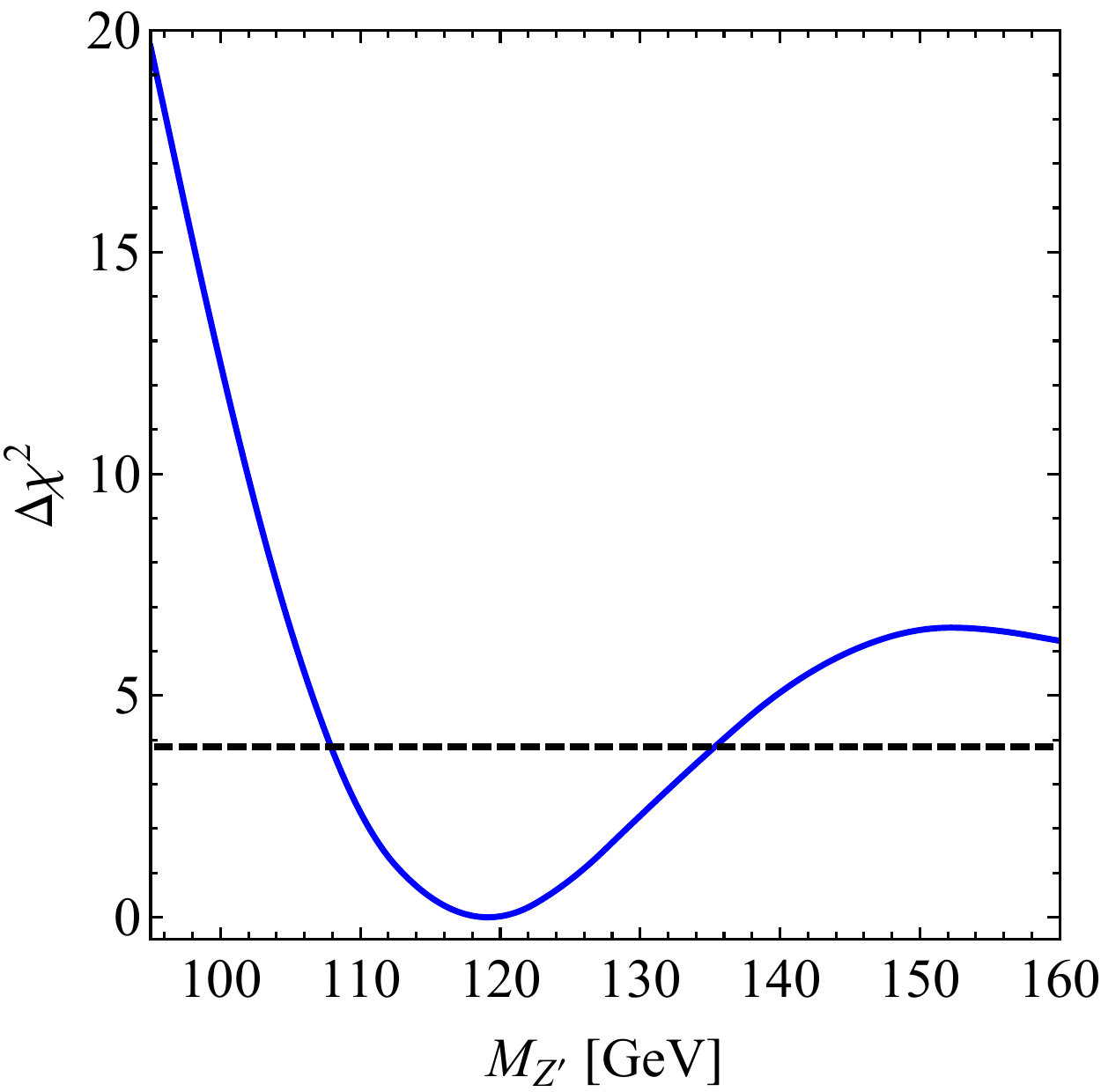} 
\caption{$\Delta \chi^2$ distribution in the $U(1)_D$ model as a function of  the $Z^\prime$-boson mass following from the LHCb measurement~\cite{Aaij:2017eru}. The benchmark parameters~(\ref{eq:gbcU(1)D}) have been used to obtain the shown predictions. The dashed black line corresponds to $\Delta \chi^2 = 3.84$,~i.e.~the 95\%~CL for a Gaussian distribution. }
\vspace{2mm}
\label{fig:deltachi2}
\end{figure}

In the following, we point out that $Z^\prime$ boson with the properties~(\ref{eq:gbcU(1)D}) and~(\ref{eq:MZpbound}), can also be probed by the LHCb measurement of $Z \to b \bar b$ production in the forward direction~\cite{Aaij:2017eru}. To this purpose, we show in Figure~\ref{fig:mjjexamples} four different dijet mass  ($m_{jj}$) distributions predicted in the $U(1)_D$ model (blue curves). The chosen parameters are given in~(\ref{eq:benchmark}). The background-subtracted dijet mass distribution\footnote{We thank Lorenzo~Sestini for providing the LHCb $Z \to b \bar b$ mass model to us.}   as measured by the LHCb collaboration in~\cite{Aaij:2017eru} (black error bars), the SM prediction (red curves) and the one standard deviation total uncertainty band (grey bands) is also displayed. From the upper left panel (lower right panel)  one observes that a $Z^\prime$ boson with mass $M_Z^\prime = 95 \, {\rm GeV}$ ($M_Z^\prime = 150 \, {\rm GeV}$) is disfavoured by the data since it leads to an excess in the peak region (the tail) of the~$m_{jj}$ distribution. This statement is quantified in Figure~\ref{fig:deltachi2} which shows the $\Delta \chi^2$ in the $U(1)_D$ model as a function of $M_{Z^\prime}$. One sees that for the parameters~(\ref{eq:gbcU(1)D}) only $Z^\prime$ bosons with masses in the range 
\beq \label{eq:MZpboundLHCb}
M_{Z^\prime} \in  [108, 135] \, {\rm GeV} \,,
\eeq
are compatible with the LHCb measurements of $Z \to b \bar b$ production  at the 95\%~CL. In fact, the minimum of $\Delta \chi^2$ is located at $M_{Z^\prime} \simeq 120 \, {\rm GeV}$ in close proximity to the local excess in the CMS measurement~\cite{Sirunyan:2017nvi} observed for dijet masses around~$115 \, {\rm GeV}$. The corresponding $\chi^2/{\rm dof}$ is $1.9$ and thus slightly better than the SM fit which leads to $\chi^2/{\rm dof} = 2.0$. While the finding that both the CMS and LHCb measurement may indicate that  a light new gauge boson is hiding in the  data is probably accidental, we emphasise that the two-sided bound~(\ref{eq:MZpboundLHCb}) on the mass of the $Z^\prime$ boson is stronger than the limit that derives from a combination of the searches~\cite{Sirunyan:2017nvi,Aad:2012cfr,Aaboud:2017buh}. \\

%------------------------------------------------
%------------------------------------------------

The above applications of our SM results presented in Sections~\ref{sigma} and~\ref{asym} show that  the LHCb experiment can provide unique probes of new physics in heavy-quark production due to its efficient triggering, excellent vertexing and accurate event reconstruction.  While already a handful of similar proposals of such ``exotic"  new-physics searches  exist that specifically exploit the remarkable capabilities of LHCb (see~e.g.~\cite{Kahawala:2011sm,Murphy:2015cha,Schwaller:2015gea,Haisch:2016hzu,Ilten:2016tkc,Aaij:2017rft,Ilten:2018crw,Haisch:2018kqx,Aaij:2018xpt,Bediaga:2018lhg}), we believe that further research in this rich and developing field can turn out to be potentially very rewarding.   

\acknowledgments We thank Valentin Hirschi, Alexander Huss, Mitesh Patel and Lorenzo Sestini for useful communications and discussions. RG acknowledges support by the Dutch Organization for Scientific Research (NWO) through a VENI grant. UH acknowledges the  hospitality and support of the Particle Theory Group at the University of Oxford  and the CERN Theoretical Physics Department at various stages of this project.  We acknowledge the computing resources provided to us by the Swiss National Supercomputing Centre (CSCS) under the
project ID p501b.

\appendix \label{appendix}

\section{Analytic results}
\label{calculation}

In this appendix, we provide predictions for the hadronic process $pp\to Q\bar Q X$ assuming the factorisation theorem of the form
\beq
\rd\sigma \left ({pp  \to Q\bar Q X} \right )  = \sum_{i,j} \int \rd x_1 \,\rd x_2 \, f_{i/p}(x_1,\mu_F^2)\,f_{j/p}(x_2,\mu_F^2) \, \rd \hat{\sigma}_{ij}  \,,
\eeq
where the universal  PDFs $f_{i/p} (x, \mu_F)$ describe the probability of finding the parton $i$ in the proton with longitudinal momentum fraction $x$ and $\mu_F$ is the factorisation scale. In analogy to~\eqref{eq:expansion},  the  differential  partonic cross sections $\rd \hat{\sigma}_{ij}$ may be written as an expansion in terms of the electromagnetic and strong coupling constant according to
\beq \label{eq:expansionappendix}
\rd \hat{\sigma}_{ij} = \sum_{n,\,m} \alpha^n \hspace{0.25mm} \alpha_s^m \hspace{0.5mm} \rd \hat{\sigma}_{ij}^{(n,\,m)} \,,
\eeq
where $\rd  \hat{\sigma}_{ij}^{(n,\,m)}$ denotes the differential partonic cross sections for the initial state $ij$ with the $\alpha$ and $\alpha_s$ factors stripped off. 

The differential cross sections can be further decomposed to isolate the contributions which are asymmetric under interchange of the the final state heavy quark and antiquark according to
\beq \label{eq:dsigmaA}
\rd\hat{\sigma}_{ij, \hspace{0.25mm} A} = \frac{1}{2} \, \Big [ \rd \hat{ \sigma } \left (ij \to Q \bar Q X \right ) -  \rd \hat{\sigma} \left (ij \to \bar Q Q X \right ) \Big  ] \,.
\eeq 
Here the notation indicates that in the process labelled by  $ij \to Q \bar Q X$ ($ij \to \bar Q  Q X$) the angle $\theta$ corresponds to the scattering angle of the heavy quark (heavy antiquark) in the partonic centre-of-mass (CM) frame. The benefit of this decomposition is that the symmetric and asymmetric contributions to the  cross sections can be numerically integrated separately, which substantially improves the efficiency of the numerical evaluation.

The results in this work have been obtained at NLO accuracy ($n + m = 3$) for both symmetric and asymmetric contributions. Relevant results for the differential cross sections to this accuracy have previously been obtained in~\cite{Jersak:1981sp,Nason:1987xz,Nason:1989zy,Mangano:1991jk,Beenakker:1990maa,Beenakker:1988bq,Kuhn:2005it,Kuhn:2006vh,Bernreuther:2006vg,Hollik:2007sw,Kuhn:2009nf}, and in many cases the analytic results have been given. The purpose of this appendix is to collect the analytic results for asymmetric bottom- and charm-quark pair production, suitable for direct numerical integration. We will provide  analytic expressions for both the (renormalised) virtual and real emission contributions to the partonic cross section for various subprocesses, which contain explicit and implicit divergences, respectively. A number of these contributions contain only soft divergences, and so we have also included a soft function which describes the radiation of soft gluons integrated in phase space up to a cut~$E_{\rm cut}$ in the gluon energy. This function is suitable for applying the technique of phase-space slicing~\cite{Harris:2001sx}, which is found to be stable for these types of processes. In cases where both soft and collinear divergences are present, we have also regularised the numerical integration using dipole subtraction~\cite{Catani:1996vz}. In the next subsection, we  introduce our notation for describing the kinematics of two- and three-body partonic final states, and then list the relevant formulas  ordered by their powers in the expansion~(\ref{eq:expansionappendix}). 

\subsection{Kinematics and notation}

The partonic cross section for heavy-quark pair-production receives Born-level contributions from gluon fusion, quark annihilation as well as  gluon-photon and photon-photon scattering. As an example of a partonic $2 \to 2$ subprocess, we consider quark annihilation 
\beq \label{eq:treelevelprocess}
q(p_1) + \bar{q}(p_2) \rightarrow Q(p_3) + \bar{Q}(p_4) \, ,
\eeq
where the four-momenta $p_{1,2}$ of the initial-state partons can be expressed as the fractions $x_{1,2}$ of the four-momenta $P_{1,2}$ of the colliding protons. The partonic cross section is a function of the kinematic invariants
\beq \label{eq:stu}
\sh = (p_1+p_2)^2 \, ,  \qquad \hat t_Q = (p_1-p_3)^2 -m_Q^2 \, , \qquad \hat u_Q = (p_2-p_3)^2 -m_Q^2 \, ,
\eeq
and momentum conservation implies that $\sh + \hat t_Q + \hat u_Q = 0$. In addition to $\sh$, $\hat t_Q$ and $\hat u_Q$, we also use the velocity of the heavy quark and scattering angle
\beq \label{eq:rhobeta}
 \beta = \sqrt{1-  \frac{4m_Q^2}{\hat s}} \,, \qquad c = \beta \cos \theta \,,
\eeq
to write our results,  where $\theta$ denotes the angle between $\vec{p}_1$ and $\vec{p}_3$  in the partonic CM frame. Notice that 
\beq
\hat t_Q = -\frac{\hat s}{2} \left ( 1 - c  \right ) \,, \qquad \hat u_Q = -\frac{\hat s}{2} \left ( 1 + c  \right ) \,,
\eeq
which implies that $c = (\hat t_Q - \hat u_Q)/\hat s$, and as a result the variable $c$ is strictly speaking not needed when writing our results. In some cases we will however use $c$, because the  obtained expressions turn out to be more compact than those written in terms of $\sh$, $\hat t_Q$ and~$\hat u_Q$.
In addition to these kinematic variables, it also useful to introduce the following mass variables
\beq \label{eq:notdef}
y_Q  = \frac{m_Q^2}{\hat s} \,, \qquad y_W = \frac{M_W^2}{\hat s} \,, \qquad \mu_Z = M_Z^2 - i \Gamma_Z M_Z \,.
\eeq
The complex squared-mass $\mu_Z$ is introduced as the $Z$ boson is treated as an unstable particle throughout our calculation. This is necessary when describing  bottom- and charm-quark pair production in the vicinity of the $Z$-boson resonance.

The NLO corrections also involves the evaluation of  $2 \to 3$ real emission processes of the form
\beq \label{eq:realemission}
\begin{aligned}
q(p_1) + \bar{q}(p_2) &\rightarrow Q(p_3) + \bar{Q}(p_4) + g(p_5) \, ,
\end{aligned}
\eeq
where again we have used the quark-annihilation subprocess as an example. The analytic formula for these processes are provided in terms of the following dimensionless variables 
\beq \label{eq:yij}
y_{ij} = \frac{2\, p_i \cdot p_j}{\hat{s}} \, . 
\eeq
All $2 \to 3$ processes can be characterised by five independent scalar quantities~\cite{Mangano:1991jk}. For instance, choosing $y_{14}$, $y_{23}$, $y_{34}$, $y_{35}$ and $y_{45}$,  the remaining five $y_{ij}$ variables are  related (by momentum conservation) to the made choices by the following equalities
\beq \label{eq:yijrelations}
\begin{aligned}
y_{14} &= y_{12} - y_{13} - y_{15}  \, , \\[1mm]
y_{23} &= y_{12} - y_{24} - y_{25}  \, , \\[1mm]
y_{34} &= y_{12} - y_{15} - y_{25} - 2 y_Q \, , \\[1mm]
y_{35} &= y_{13} + y_{15} - y_{24}  \, , \\[1mm]
y_{45} &=-y_{13} + y_{24} + y_{25} \, .
\end{aligned}
\eeq
We emphasise that below we will write the $2 \to 3$ results such that the obtained expressions become as short as possible, and as a result our  formulas will  involve more than five independent   $y_{ij}$  parameters.

\subsection[{${\mathcal{O}(\alpha^2)}$ contributions}] {$\bm{\mathcal{O}(\alpha^2)}$ contributions} \label{pureQCD}

\begin{figure}[!t]
\begin{center}
\vspace{-5mm}
\includegraphics[width=0.95\textwidth]{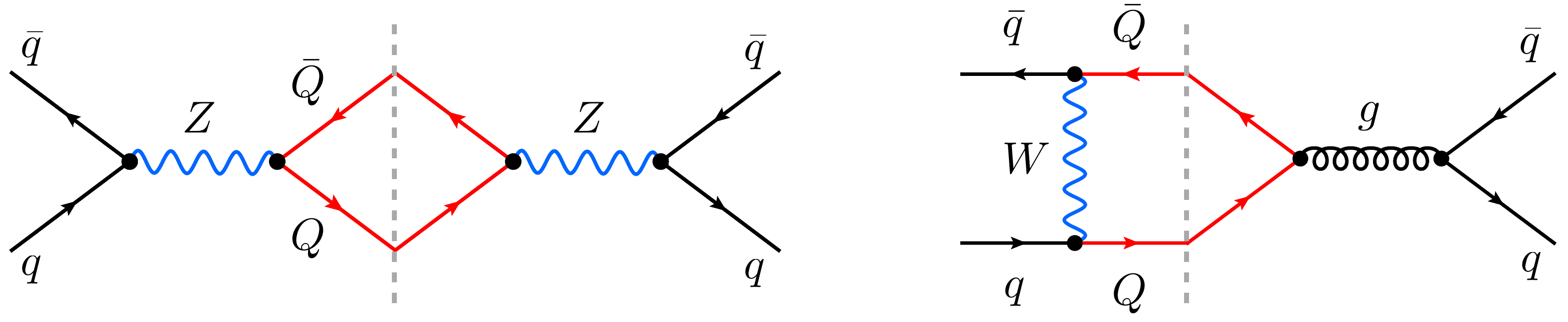} 
\vspace{2mm}
\caption{\label{fig:EW} Left: Tree-level $s$-channel $Z$-boson exchange contribution to  asymmetric heavy-quark production  at ${\cal O} (\alpha^2)$.  Right: Interference contribution between $t$-channel $W$-boson exchange and tree-level $s$-channel gluon exchange.  In both diagrams the relevant particle cuts are represented by a dashed line. See text for further details. }
\end{center}
\end{figure}

The asymmetric ${\cal O} (\alpha^2)$  effects  arise from the interference between the partonic processes $q \bar q \to Z/\gamma \to Q \bar Q$. The relevant diagram with $s$-channel $Z$-boson exchange is shown on the left-hand side in Figure~\ref{fig:EW}. In agreement with \cite{Hollik:2011ps}, we obtain for the corresponding asymmetric differential cross section the following compact expression
\beq \label{eq:dsigmaAEW} % Uli: 6/1/19: Rechecked! See Appendix.nb
\left ( \frac{\rd\sigma_{q \bar q, \hspace{0.25mm} A}}{\rd \cos \theta} \right)_{{\mathcal{O}(\alpha^2)}}  = \frac{\pi \alpha^2}{4} \, \beta c \;
  \frac{a_q a_Q}{(\sh - M_Z^2)^2 + \Gamma_Z^2 M_Z^2} \,  \Big [ v_q v_Q \, \sh  + 2 e_q e_Q \hspace{0.25mm} \left ( \sh - M_Z^2 \right) \Big ] \,.
\eeq
where
\beq \label{eq:va}
a_{f} = \frac{T_3^{f}}{s_w\hspace{0.25mm}c_w}  \,, \qquad v_{f} = \frac{T_3^{f} - 2 e_fs_w^2}{s_w\hspace{0.25mm}c_w} \,,
\eeq
are the axial-vector and vector coupling of the $Z$ boson to a fermion $f$. These couplings depend on the third component~$T_3^{f}=\pm1/2$ of the weak isospin, the electric charge $e_f$, and  the sine~$s_w$ and cosine~$c_w$ of the weak mixing angle. 

\subsection[{${\mathcal{O}(\alpha \alpha_s)}$ contributions}] {$\bm{\mathcal{O}(\alpha \alpha_s)}$ contributions}

In order to obtain the ${\cal O} (\alpha \alpha_s)$ contributions one has to consider interference contributions between $t$-channel $W$-boson (and would-be Goldstone boson) exchange and $s$-channel gluon exchange. A diagram of this kind is given on the right in~Figure~\ref{fig:EW}. In the case of $c \bar c$ production from a $d \bar d$ initial state, we arrive at 
\beq \label{eq:dsigmaAEWQCD1} % Uli: 20/12/18: Checked! See Appendix.nb
\begin{split}
\left ( \frac{\rd\sigma_{d \bar d, \hspace{0.25mm} A}}{\rd \cos \theta} \right)_{{\mathcal{O}(\alpha \alpha_s)}} & = \frac{\pi \alpha \alpha_s |V_{cd}|^2 }{18 s_w^2} \frac{\beta c}{\sh}  \\[1mm] & \phantom{xx} \times \frac{8 y_c^2  + y_c \left ( c^2 - 4 y_W -1\right )  + 2 y_W \left ( c^2 + 4 y_W + 3 \right )}{y_W \left (c^2 - \left ( 1 - 2 y_c + 2 y_W \right )^2 \right )} \,,
\end{split}
\eeq
where $V_{cd}$ denotes the relevant CKM matrix element. Our result (\ref{eq:dsigmaAEWQCD1}) agrees with the expression given in \cite{Murphy:2015cha}. In the case of asymmetric $b \bar b$ production, the $W$-boson mediated $t$-channel contributions are strongly suppressed either by the small CKM element $V_{ub}$  or by a bottom-quark PDF. Consequently, we  include~(\ref{eq:dsigmaAEWQCD1}) in our numerical analysis only in the case of charm-quark pair production. As these corrections are numerically small, we have included neither the QCD nor the QED/weak correction to this process in our predictions.

\subsection[{$\mathcal{O}(\alpha_s^3)$ contributions}]{$\bm{\mathcal{O}(\alpha_s^3)}$ contributions}
\label{sec:appendixQCD}

There is no asymmetric contribution to the production of heavy-quark pairs at $\mathcal{O}(\alpha_s^2)$. Starting at ${\cal O} (\alpha_s^3)$, however, quark annihilation $q \bar q \to Q \bar Q (g)$ as well as flavour excitation $q g \to Q \bar Q q$ receive charge-asymmetric contributions. The gluon-fusion $gg \to Q \bar Q X$ subprocess must be convoluted with a symmetric initial state to provide a hadronic cross-section prediction, and so does not lead to an observable asymmetry.

\begin{figure}[!t]
\begin{center}
\vspace{-5mm}
\includegraphics[width=0.9 \textwidth]{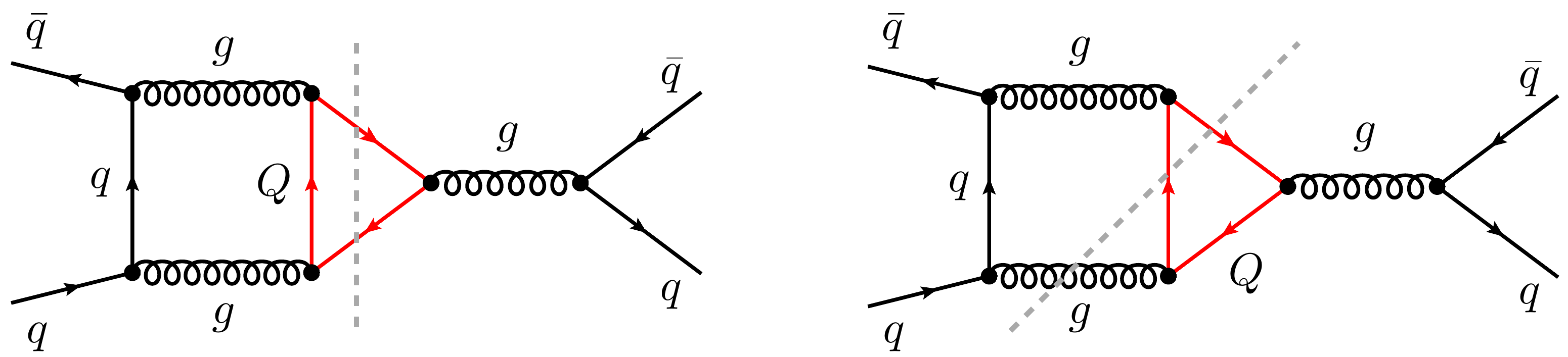} 

\vspace{1cm}

\includegraphics[width=0.9 \textwidth]{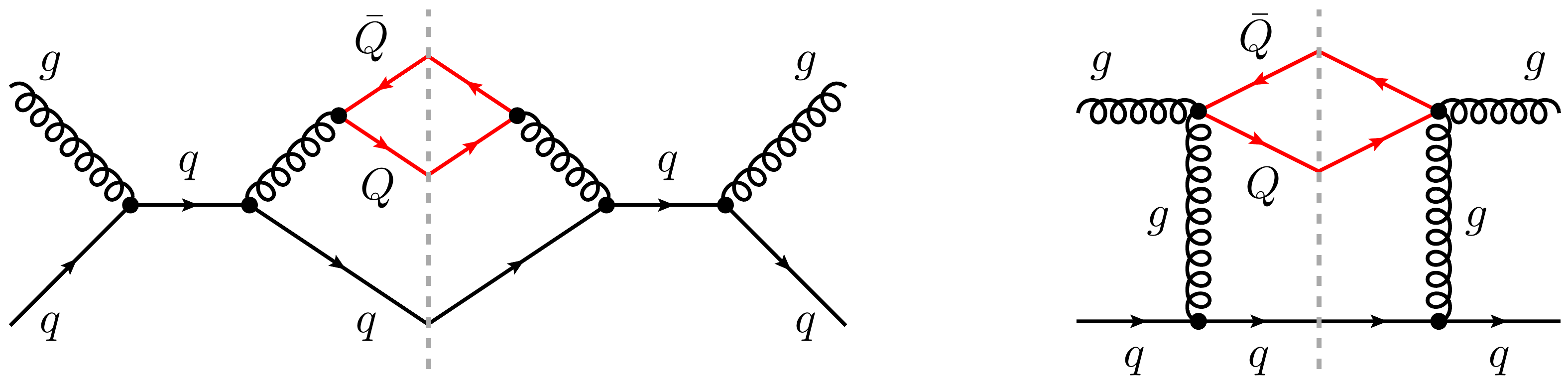} 
\vspace{4mm}
\caption{\label{fig:QCD1} Representative Feynman diagrams that contribute to the asymmetric production cross section of heavy-quark pairs at ${\cal O} (\alpha_s^3)$.  
Upper row: The two-particle cut (right) describes the interference of the one-loop box diagram with the tree-level graph, while the three-particle cut~(left) corresponds to the interference of final-state with initial-state gluon corrections. Lower row:~Three-particle cuts that represent a production of heavy quarks via flavour excitation.}
\end{center}
\end{figure}

Charge conjugation invariance can be invoked to show that, as far as the virtual corrections to $q \bar q \to Q \bar Q$ are concerned, only the interference between the lowest-order and the QCD box graphs contributes to the asymmetry at ${\cal O} (\alpha_s^3)$. An example of a Feynman diagram that furnishes a contribution is shown on the left-hand side in the upper row of~Figure~\ref{fig:QCD1}.  The corresponding virtual corrections can be written as 
\beq \label{eq:QCDvirt} % Uli: 20/12/18: Checked! See Appendix.nb
\left ( \frac{\rd \sigma_{q\bar q, \hspace{0.25mm} A}}{\rd \cos \theta} \right )_{{\cal O} (\alpha_s^3)}^{ \rm{virt}} =  \frac{\alpha_s^3 \hspace{0.25mm} d^{\hspace{0.25mm} 2}_{abc}}{16 N_c^2} \, \frac{\beta}{\hat s}  \; \frac{\Aone (\hat t_Q, \hat u_Q) - \Aone (\hat u_Q, \hat t_Q)}{\sh} \,,
\eeq
with $N_c = 3$ and  $d_{abc}^{\hspace{0.25mm} 2} = 40/3$ colour factors. Here $d_{abc} = 2 \hspace{0.25mm} {\rm Tr} \left ( \left \{ T^a, T^b \right \} T^c \right )$, while $T^a$ are the colour generators normalised such that ${\rm Tr} \left ( T^a  T^b \right ) =  \delta_{ab}/2$. The one-loop function appearing in (\ref{eq:QCDvirt}) is given by 
\bea \label{eq:Aone} % Uli: 20/12/18: Checked! See Appendix.nb
\begin{aligned}
\Aone( v, w) & =   \frac{v}{1- 4 y_Q} \left [ B_0 ( \sh, 0,0 ) - 4 y_Q \left( 2 + \frac{v^2+w^2 - 4 \sh^2 y_Q}{2 v w} \right) B_0 (m_Q^2, 0, m_Q^2)  \right.  \\[2mm]
	 & \phantom{xx} - \sh \left (1- 4 y_Q \right ) C_0 (\sh,0,0,0,0,0) \\
	 & \phantom{xx}  - \sh \, \big (1 - 8 y_Q  \left (1 - y_Q \right ) \! \big ) \, C_0 (\sh,m_Q^2, m_Q^2, 0, 0, m_Q^2)\Bigg ]  \hspace{10mm} \\ 
	 &  \phantom{xx} + \left( w - \frac{2 \hspace{0.25mm} \sh^2 y_Q}{v} \right) B_0 (v+m_Q^2,0,m_Q^2) \\[2mm]
	 &  \phantom{xx} -v \left(  v - w + 2  \hspace{0.25mm} \sh \hspace{0.25mm} y_Q \right) C_0 (0,m_Q^2, x+m_Q^2,0,0,m_Q^2) \\[2mm]
	 &  \phantom{xx} -\frac{v}{2} \left( 3 v^2+w^2 + 2 \hspace{0.25mm} \sh^2 y_Q\right) D_0(0,0,m_Q^2,m_Q^2, \sh,v+m_Q^2,0,0,0,m_Q^2)  \,,
\end{aligned}
\eea
where our definition of the Passarino-Veltman scalar integrals $B_0$, $C_0$ and $D_0$ follows that used in {\tt FormCalc}. We have verified that the formulas (\ref{eq:QCDvirt}) and (\ref{eq:Aone}) recover the analytic results given in~\cite{Kuhn:1998kw}.

As in the case of the virtual contributions also for the real corrections, only the interference between the amplitudes that are odd under the exchange of $Q$ and $\bar Q$ leads to a non-zero correction of the form (\ref{eq:dsigmaA}).  A relevant Feynman graph is displayed on the right in the upper row of Figure~\ref{fig:QCD1}. For the real gluon corrections to the asymmetric cross section, we find the result 
\bea \label{KuhnHard} % Uli: 20/12/18: Checked! See Appendix.nb
\begin{aligned}
\left ( \frac{\rd \sigma_{q\bar{q}, \hspace{0.25mm} A}}{\rd y_{35} \hspace{0.25mm} \rd y_{45} \hspace{0.25mm}  \rd\Omega} \right)_{\rm {\cal O} (\alpha_s^3)}^{{\rm real}} &= \frac{\alpha_s^3 \hspace{0.25mm} d^{\hspace{0.25mm} 2}_{abc}}{64 \pi N_c^2}   \, \frac{1}{ \hat{s} \hspace{0.25mm} y_{12}  \hspace{0.25mm} y_{35} \left( y_{34} + 2 y_Q \right) } \\[2mm]
	& \hspace{-1cm} \times \left\{ \frac{y_{13}}{y_{15}} \left[ y_{13}^2+y_{14}^2+y_{23}^2+y_{24}^2+2 \left(y_{12} +y_{34} +2 y_Q \right) y_Q \right] + 4 y_{24}  \hspace{0.25mm} y_Q  \right\} \hspace{10mm} \\[2mm]
	&  \phantom{xx}-\left(1\leftrightarrow2\right)-\left(3\leftrightarrow4\right)+\left(1\leftrightarrow2,3\leftrightarrow4\right) \, ,
\end{aligned}
\eea
where $d\Omega = d \cos \theta \hspace{0.25mm} d\varphi$ is the differential solid angle with $\varphi$ the azimuthal angle. The~expression given in~(\ref{KuhnHard}) agrees with the results provided in~\cite{Kuhn:1998kw}.  

Soft gluon radiation in the process $ q \bar q \to Q \bar Q g$ integrated in phase space up to a cut~$E_{\rm cut}$ in the gluon energy leads to the expression 
\beq \label{eq:QCDsoft} % Uli: 20/12/18: Changed! Rhorry, please check!
\left ( \frac{\rd \sigma_{q \bar q, \hspace{0.25mm} A}}{\rd \cos \theta} \right )_{{\cal O} (\alpha_s^3)}^{  \rm{soft} } = \frac{\alpha_s^3 \hspace{0.25mm} d^{\hspace{0.25mm} 2}_{abc}}{32 N_c^2}  \, \frac{\beta}{\hat s} \,  S \,,
\eeq
where 
\beq \label{eq:Sfunction} % Uli: 20/12/18: Changed! Rhorry, please check!
S = \left ( c^2 +1+ 4 y_Q \right) \Bigg\{  \ln \left(\frac{\hat t_Q}{\hat u_Q}\right)
	\left[ -\frac{2}{\epsilon} + 2 \ln B + 4\ln \left( \frac{E_{\rm cut}}{\mu}\right) \right] + \Atwo (\hat t_Q) - \Atwo (\hat u_Q) \Bigg\} \,,
\eeq
is the relevant soft function. Here $\epsilon = (4 - d)/2$ arises from dimensional regularisation in $d$ dimensions, $\mu$~denotes the corresponding renormalisation scale, and we have furthermore introduced 
\bea \label{eq:BA} % Uli: 20/12/18: Changed! Rhorry, please check!
 \Atwo(v) = \ln^2 \left(\frac{-v}{\sh \hspace{0.25mm}  \sqrt{y_Q}} \right) + 2\hspace{0.25mm} {\rm Li}_2\left(1 - \frac{B \hspace{0.25mm}  v}{\sh \hspace{0.25mm}   \sqrt{y_Q}}\right) - 2\hspace{0.25mm} {\rm Li}_2\left(1 - \frac{B \hspace{0.25mm}  \sh \hspace{0.25mm}   \sqrt{y_Q}}{v}\right)  , \hspace{3mm}  B = \sqrt{\frac{1+\beta}{1-\beta}} \,, \hspace{9mm}
\eea
with ${\rm Li}_2 (z) = \int_z^0 dt \ln (1 - t)/t$ denoting the usual dilogarithm. Our results (\ref{eq:QCDsoft}), (\ref{eq:Sfunction}) and~(\ref{eq:BA}) can be shown to agree with the expressions presented in~\cite{Kuhn:1998kw}.  Note that the IR~$1/\epsilon$ pole in~(\ref{eq:Sfunction}) cancels against that in the virtual corrections~(\ref{eq:QCDvirt}) so that the sum of the virtual and soft contributions is IR finite and can be numerically integrated in four dimensions.

The asymmetric ${\cal O} (\alpha_s^3)$ contribution to the heavy-quark production cross section that is associated to the flavour excitation process can be obtained from the result (\ref{KuhnHard}) by crossing,~i.e.~interchanging the indices $2\leftrightarrow5$ in the variables $y_{ij}$ as defined in (\ref{eq:yij}). Examples of relevant Feynman diagrams are given in the lower row of Figure~\ref{fig:QCD1}. Noting a difference in the colour factor for averaging over the initial-state gluon with respect to~\cite{Kuhn:1998kw}, we find the expression
\bea \label{qg} % Uli: 23/12/18: Changed! Crossing does not agree with Kuehn! Direct calculation does agree with Kuehn!  See Appendix.nb
\begin{aligned}
\left ( \frac{\rd \sigma_{qg, \hspace{0.25mm} A}}{\rd y_{35} \hspace{0.25mm} \rd y_{45}  \hspace{0.25mm} \rd\Omega} \right)_{{\cal O} (\alpha_s^3)} &= \frac{\alpha_s^3 \hspace{0.25mm} d^{\hspace{0.25mm} 2}_{abc}}{64 \pi N_c \left (N_c^2-1 \right )}  \, \frac{1}{\hat s  \hspace{0.25mm} y_{15} \hspace{0.25mm} y_{23}  \left( y_{34} + 2 y_Q \right) } \\[2mm]
	& \hspace{-2cm} \phantom{xx} \times \Bigg\{ \! \left(\frac{y_{13}}{y_{12}}-\frac{y_{35}}{y_{25}}\right) \left[y_{13}^2+y_{14}^2+y_{35}^2+y_{45}^2+2 \left(y_{34}-y_{15} + 2 y_Q\right) y_Q \right] \hspace{4mm} \\[2mm]
	& \hspace{-2cm} \phantom{xxxxx} + 4  \left (y_{14}+y_{45} \right ) y_Q  \Bigg\} -\left(3\leftrightarrow4\right) \, .
\end{aligned}
\eea
The same result also holds in the case of the partonic reaction $\bar q g \to Q \bar Q \bar q$. In contrast to the asymmetric contribution from $q \bar q \to Q \bar Q g$, the flavour excitation processes  $ q g \to Q \bar Q q$ and $\bar q g \to Q \bar Q \bar q$ are IR finite.  

\subsection[{${\mathcal{O}(\alpha \alpha_s^2)}$ contributions}] {$\bm{\mathcal{O}(\alpha \alpha_s^2)}$ contributions}

The structure of the ${\cal O} (\alpha \alpha_s^2)$  contributions to the asymmetric production cross section of heavy-quark pairs that involve a photon is very similar to that of the pure QCD corrections. In fact, all subprocesses that contribute at ${\cal O} (\alpha \alpha_s^2)$ can be obtained from the~${\cal O} (\alpha_s^3)$ corrections presented in Appendix~\ref{sec:appendixQCD} by rescaling with 
\beq \label{eq:QEDQCDrescaling} % Uli: 23/12/18: Changed! Checked! See Appendix.nb
R_{{\cal O} (\alpha \alpha_s^2)}^\gamma  = \frac{12 \alpha \hspace{0.25mm} e_q \hspace{0.25mm} e_Q}{5 \alpha_s} \, .
\eeq
Here the factor $12/5$ arises from the ratio of QED and QCD colour factors $(N_c^2-1)/4 = 2$ and $d_{abc}^2/16 = 5/6$. 

\begin{figure}[!t]
\begin{center}
\vspace{-5mm}
\includegraphics[width=0.875 \textwidth]{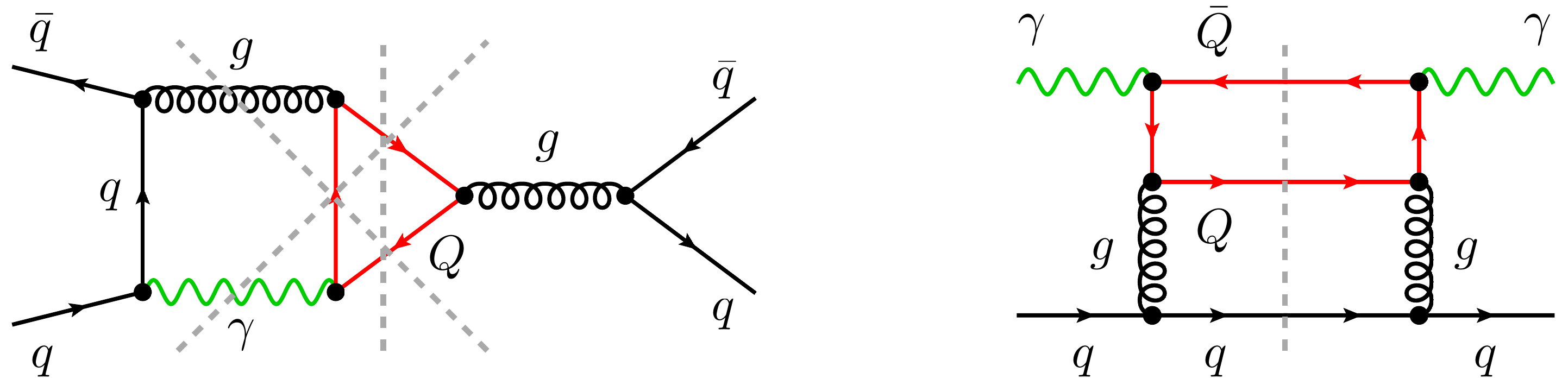} 
\vspace{2mm}
\caption{\label{fig:QED} Left: Possible two- and three-particle cuts that  contribute  via $q \bar q \to Q \bar Q$, $q \bar q \to Q \bar Q g$ and $q \bar q \to Q \bar Q \gamma$  to the asymmetric cross section in heavy-quark production  at ${\cal O} (\alpha \alpha_s^2)$.  Right: Example of a Feynman diagram that leads to a photon-initiated production asymmetry of heavy quarks at  ${\cal O} (\alpha \alpha_s^2)$.}
\end{center}
\end{figure}

In the case of quark annihilation, we find the following relation between the corrections of  ${\cal O} (\alpha \alpha_s^2)$ and ${\cal O} ( \alpha_s^3)$  to the differential asymmetric cross sections
\beq \label{eq:qqQED} % Uli: 23/12/18: Changed! Checked! See Appendix.nb
(\rd\sigma_{q \bar q, \hspace{0.25mm} A} )_{{\cal O} (\alpha \alpha_s^2)}^\gamma = 3 R_{{\cal O} (\alpha \alpha_s^2)}^\gamma \, (\rd\sigma_{q \bar q, \hspace{0.25mm} A} )_{{\cal O} ( \alpha_s^3)} \,,
\eeq
irrespectively of whether the contributions are virtual, real or soft. The additional overall factor of 3 reflects the three possible attachments of the photon in diagrams like the one shown on the left-hand side in Figure~\ref{fig:QED}. In  the case of the $qg$-initiated transition only two different photon attachments are possible so that 
\beq \label{eq:qgQED} % Uli: 21/12/18: Changed! Checked! See Appendix.nb
(\rd\sigma_{q g, \hspace{0.25mm} A} )_{{\cal O} (\alpha \alpha_s^2)}^\gamma  = 2 R_{{\cal O} (\alpha \alpha_s^2)}^\gamma \, (\rd\sigma_{q g, \hspace{0.25mm} A} )_{{\cal O} ( \alpha_s^3)} \,,
\eeq
and similarly for $\bar q g \to Q \bar Q \bar q$. Our formulas~(\ref{eq:qqQED}) and (\ref{eq:qgQED}) agree with the findings of the article~\cite{Kuhn:1998kw}. Finally, in the case of the photon-initiated process $q  \gamma \to Q \bar Q  q$, which receives contributions from Feynman diagrams such as the one displayed on the right in Figure~\ref{fig:QED}, we obtain 
\beq \label{eq:qaQED} % Uli: 23/12/18: Changed! Checked! Rhorry, please double-check! RG: 04/12/19: Done
(\rd\sigma_{q \gamma, \hspace{0.25mm} A} )_{{\cal O} (\alpha \alpha_s^2)}^\gamma  = 8 R_{{\cal O} (\alpha \alpha_s^2)}^\gamma \, (\rd\sigma_{q g, \hspace{0.25mm} A} )_{{\cal O} ( \alpha_s^3)} \,,
\eeq
and the same result applies to the $\bar q \gamma$ initial state. Notice that the factor of $N_c^2 -1=  8$ arises from averaging over the photon in the initial state  rather than the gluon. 

\begin{figure}[!t]
\begin{center}
\vspace{-5mm}
\includegraphics[width=0.95 \textwidth]{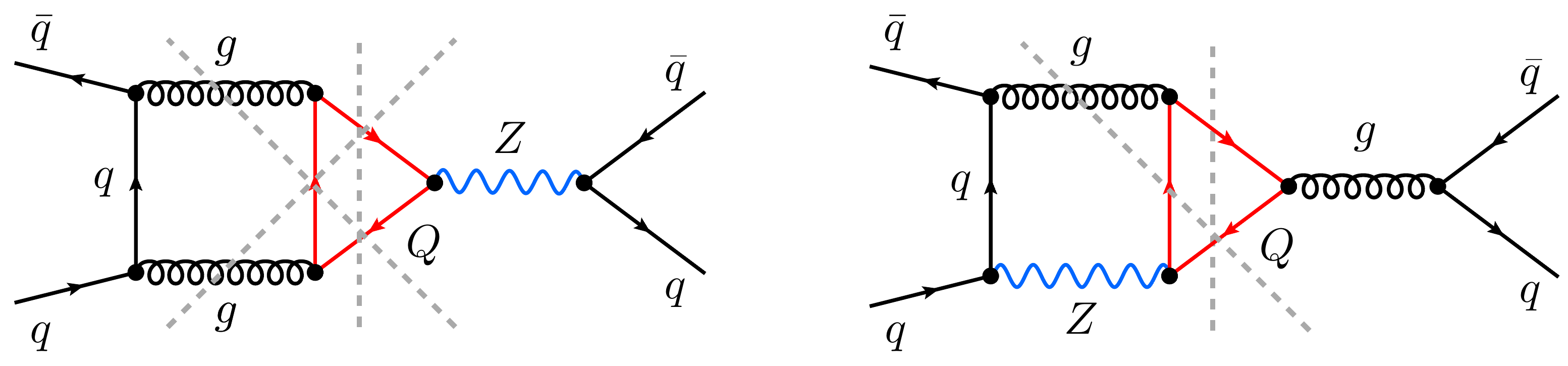} 
\vspace{2mm}
\caption{\label{fig:EWQCD1} Examples of graphs that affect the production asymmetry of heavy quarks at  ${\cal O} (\alpha \alpha_s^2)$ and involve the exchange of a $Z$ boson. Consult the main text for additional explanations.}
\end{center}
\end{figure}

As in the case of pure QCD, the ${\cal O} (\alpha \alpha_s^2)$ corrections associated to $Z$-boson exchange receive contributions from both virtual and real corrections. For the interference contributions of box diagrams with tree-level $s$-channel exchange graphs, we obtain the following expression
\beq \label{eq:EWQCDvirt} % Uli: 24/12/18: Changed! See Appendix.nb. Rhorry, please double-check! RG: 4/1/19: Changed slightly to make more self consistent. Matches normalisation in code. Uli: 6/1/19: Rechecked! 
\begin{aligned}
\left ( \frac{\rd \sigma_{q\bar q, \hspace{0.25mm} A}}{\rd \cos \theta} \right )_{{\cal O} (\alpha \alpha_s^2)}^{{\rm virt}, Z} =  \frac{\alpha \alpha_s^2}{2 N_c^2} \, \frac{\beta}{\sh}   \, \Re \Bigg \{ & \left( \frac{v_q v_Q}{\sh   - \mu_Z } \right)^{\ast} \,\left(\Aone (\hat{t}_Q, \hat{u}_Q) - \Aone (\hat{u}_Q, \hat{t}_Q)\right)  \\[1mm] &  + \frac{v_q v_Q}{\sh}\left(\Athree (\hat{t}_Q, \hat{u}_Q) - \Athree(\hat{u}_Q, \hat{t}_Q)\right) \Bigg \} \,,
\end{aligned}
\eeq
for the asymmetric heavy-quark pair production cross section. The loop function $\Aone (v,w)$ has already been given in~(\ref{eq:Aone}).  It arises in the context of~(\ref{eq:EWQCDvirt}) from two-particle cut diagrams like the one depicted on the left in Figure~\ref{fig:EWQCD1}. The function $\Athree(v,w)$ is instead related to two-particle cuts in graphs of the type shown on the right-hand side of the same figure. In terms of standard Passarino-Veltman scalar integrals $B_0$, $C_0$ and $D_0$, we arrive at the result 
\bea \label{eq:Athree} % Uli: 24/12/18: Changed! See Appendix.nb. RG: 04/12/19, introduced complex Z-boson mass. Uli: 7/1/19: Rechecked! 
\begin{aligned}
\Athree( v, w) & =   \frac{v}{1-4y_Q} \, \Bigg [ \, 2 \hspace{0.25mm} B_0 ( \sh, 0,\mu_Z ) - 2 \hspace{0.25mm} \sh \left (1- 4 y_Q \right ) C_0 (\sh,0,0,0,0,\mu_Z) \\[1mm] 
         &  \phantom{xx}  - 4 y_Q \left( 2 + \frac{v^2+w^2 - 4 \sh^2 y_Q}{2 \hspace{0.25mm} v w} \right) \Big [ B_0 (m_Q^2, 0, m_Q^2) + B_0 (m_Q^2, \mu_Z, m_Q^2) \Big  ]   \\[1mm]
	 & \phantom{xx}  - \sh \left (2 - 16  \hspace{0.25mm}  y_Q  \left (1 -  y_Q \right )  -\frac{4 \mu_Z \hspace{0.25mm}y_Q}{v + w} \right ) C_0 (m_Q^2, m_Q^2, \sh, 0, m_Q^2,\mu_Z)\Bigg ]  \hspace{10mm} \\ 
	 &  \phantom{xx} + 2 \left( w - \frac{2 \sh^2 y_Q}{ v} \right) B_0 (v+m_Q^2,0,m_Q^2) \\[1mm]
	 &  \phantom{xx} -v \, \Big (  v - w +2 \hspace{0.25mm} \sh y_Q  + \mu_Z \Big )  \, C_0 (0,m_Q^2, v+m_Q^2,0,0,m_Q^2) \\[1mm]
	 &  \phantom{xx} -v \left [  v - w + 2 \hspace{0.25mm} \sh y_Q + \mu_Z \left ( 1 + \frac{2 \hspace{0.25mm} \sh^2 y_Q}{v^2} \right ) \right] C_0 (v+m_Q^2,m_Q^2,0,0,m_Q^2,\mu_Z) \\[1mm]
	 &  \phantom{xx} -v \, \Big (  3 v^2+w^2 + 2 \hspace{0.25mm} \sh^2 y_Q  +2 \mu_Z \hspace{0.5mm} \big ( v \left ( 1 - y_Q  \right) - w y_Q \big )  + \mu_Z^2 \Big) \\[1mm] 
	 & \hspace{1.2cm} \times D_0(0,m_Q^2,m_Q^2, 0,v+m_Q^2, \sh, 0,0,m_Q^2,\mu_Z)  \,.
\end{aligned}
\eea
Notice that the these expressions are a function of the complex squared-mass $\mu_Z$, which is necessary to account for the width effects in region close to the $Z$-boson resonance (this is not required in the case of top-quark pair production since $2 m_{t} > m_{Z}$). If the calculation is performed in the complex-mass scheme, the couplings $v_f$ become complex as discussed for instance in~\cite{Denner:2005fg}.

The real emission and soft contributions of ${\cal O} (\alpha \alpha_s^2)$ can again be obtained by rescaling the corresponding QCD results. We define 
\beq \label{eq:Raas2} % Uli: 28/12/18: Changed! Checked! See Appendix.nb. RG: 04/12/19: Double checked/fixed s2w normalisation. Uli: 7/1/19: Rechecked! 
R_{{\cal O} (\alpha \alpha_s^2)}^Z=  \frac{12 \alpha \hspace{0.25mm} v_q v_Q}{5 \alpha_s}  \,.
\eeq
In terms of (\ref{KuhnHard}) and (\ref{eq:Raas2}), we find for the real corrections
\bea \label{eq:EWQCDhardreal} % Uli: 28/12/18: Changed! Checked! See Appendix.nb. RG: 04/12/19: Moved vector couplings into Re(part), to get the complex width parts
\big ( \rd \sigma_{q\bar{q}, \hspace{0.25mm} A} \big)_{{\cal O} (\alpha \alpha_s^2)}^{{\rm real}, Z} =   {\rm Re} \left [R_{{\cal O} (\alpha \alpha_s^2)}^Z  \left(\frac{y_{12}  \hspace{0.25mm} \hat s}{y_{12} \hspace{0.25mm} \hat s-\mu_Z} + \frac{\left ( y_{34}+2 y_Q \right )  \hat s}{\left ( y_{34}+2y_Q \right )  \hat s-\mu_Z} \right) \right ]  \big ( \rd \sigma_{q\bar{q}, \hspace{0.25mm} A} \big)_{{\cal O} ( \alpha_s^3)}^{{\rm real}}  \,. \hspace{8mm}
\eea
We emphasise that this contribution corresponds to the three-particle cuts displayed in Figure~\ref{fig:EWQCD1}, while real $Z$-boson emission of  is not included in~(\ref{eq:EWQCDhardreal}) as the $Z$ boson is considered unstable in our calculation.

The corresponding soft function is given by
\beq \label{eq:EWQCDsoftreal} % Uli: 28/12/18: Changed! RG: 04/12/19: checked
\big ( \rd \sigma_{q\bar{q}, \hspace{0.25mm} A} \big)_{{\cal O} (\alpha \alpha_s^2)}^{{\rm soft}, Z} =   {\rm Re} \left [ R_{{\cal O} (\alpha \alpha_s^2)}^Z  \left(\frac{2 \hat s}{\hat s -\mu_Z} \right) \right ] \big ( \rd \sigma_{q\bar{q}, \hspace{0.25mm} A} \big)_{{\cal O} ( \alpha_s^3)}^{{\rm soft}}  \,,
\eeq
where the relevant QCD results can be found  in (\ref{eq:QCDsoft}) and (\ref{eq:Sfunction}). 

The ${\cal O} (\alpha \alpha_s^2)$ contribution to asymmetric heavy-quark production arising from the flavour excitation process can be obtained from (\ref{eq:EWQCDhardreal}) by crossing. Explicitly, we have  
\bea \label{eq:EWQCDqg} % Uli: 28/12/18: Changed! Checked! See Appendix.nb. Rhorry, please double-check! RG: 04/12/19: checked  Uli: 7/1/19: Rechecked! 
\big ( \rd \sigma_{q g, \hspace{0.25mm} A} \big)_{{\cal O} (\alpha \alpha_s^2)}^{Z} =   {\rm Re} \left [ R_{{\cal O} (\alpha \alpha_s^2)}^Z  \left( \frac{y_{15} \hspace{0.25mm} \hat s}{y_{15}  \hspace{0.25mm} \hat s +\mu_Z} + \frac{\left ( y_{34}+2 y_Q \right )  \hat s}{\left ( y_{34}+2y_Q \right )  \hat s -\mu_Z} \right) \right ]   \big ( \rd \sigma_{q g, \hspace{0.25mm} A} \big)_{{\cal O} ( \alpha_s^3)}  \,, \hspace{8mm}
\eea
where the expression for QCD term has already been given in (\ref{qg}). 

\subsection[{$\mathcal{O}(\alpha^2 \alpha_s)$ contributions}]{$\bm{\mathcal{O}(\alpha^2 \alpha_s)}$ contributions}
\label{QCDDY}

\begin{figure}[!t]
\begin{center}
\vspace{-5mm}
\includegraphics[width=0.975 \textwidth]{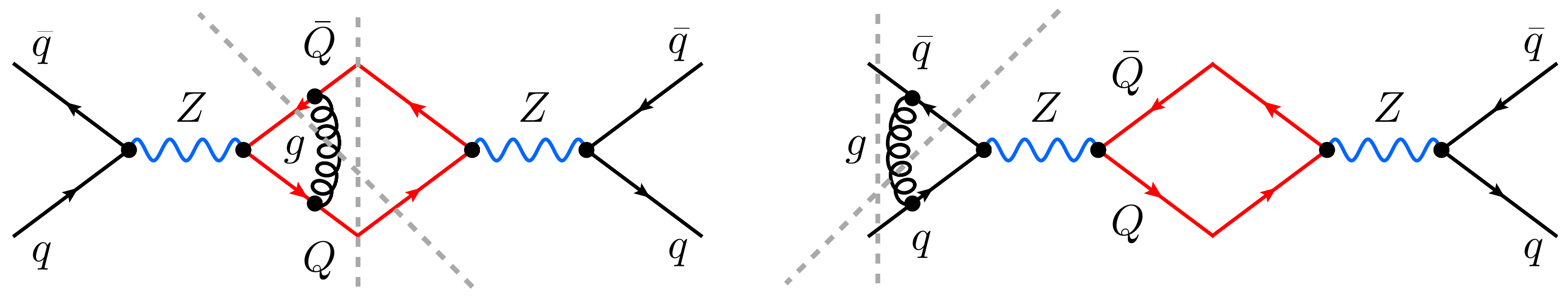} 
\vspace{2mm}
\caption{\label{fig:EWQCD2} Possible two- and three-particle cuts that contribute to asymmetric heavy-quark production at ${\cal O} (\alpha^2 \alpha_s)$. Graphs with photon exchange in the $s$-channel also give a contribution but are not explicitly shown. See text for additional details. }
\end{center}
\end{figure}

We finally consider the corrections of $\mathcal{O}(\alpha^2 \alpha_s)$, which correspond to QCD corrections to the contributions of $\mathcal{O}(\alpha^2)$ provided in~(\ref{eq:dsigmaAEW}). Due to the colour structure of these corrections, they can be separated into those to either the massive final-state quarks or the massless initial-state quarks. Example diagrams are depicted in Figure~\ref{fig:EWQCD2}. 

The relevant results for the  corrections associated to final-state radiation~(FSR) have been provided in~\cite{Jersak:1981sp}, and may be written as
\bea \label{eq:DYvirt}  %Uli: 11/1/19: Checked first bit! %Uli: 14/1/19: Checked second bit!
\begin{split}
\big ( \rd \sigma_{q\bar{q}, \hspace{0.25mm} A} \big)_{{\cal O} (\alpha^2 \alpha_s)}^{{\rm virt}, {\rm FSR}} & =  
\left(F^{{\rm virt}}_{{\cal O} (\alpha_s)} + 2 \delta Z^Q_{\mathcal{O}(\alpha_s)} \right) \big(\rd\sigma_{q \bar q, \hspace{0.25mm} A}\big)_{{\mathcal{O}(\alpha^2)}}  \\[1mm]
& \hspace{-1cm}+ \alpha^2 \alpha_s \hspace{0.25mm} C_F\,\pi\,  \frac{\left ( \beta^2-1 \right ) c}{4} \,
  \frac{e_q e_Q \hspace{0.25mm} a_q a_Q  }{(\sh - M_Z^2)^2 + \Gamma_Z^2 M_Z^2} \; \Gamma_Z M_Z  \,,
\end{split}  
\eea
with $C_F = 4/3$. Here $\delta Z^Q_{\mathcal{O}(\alpha_s)}$ denotes the one-loop vector  QCD wave-function renormalisation constant for the heavy quark, while $F^{{\rm virt}}_{{\cal O} (\alpha_s)}$ is a form factor applied to the Born-level cross section. These quantities are given  in the on-shell renormalisation scheme for the heavy quark by
\beq % Uli: 11/1/19: Checked!
\begin{split}
\delta Z^Q_{\mathcal{O}(\alpha_s)} & 
= -\frac{\alpha_s C_F}{2 \pi} \left \{  \frac{3}{2} \left [ \frac{1}{\epsilon} + \ln \left( \frac{\mu^2}{m_{Q}^2}  \right ) \right ] + 2 \right \} \,,
\\[3mm]
F^{{\rm virt}}_{{\cal O} (\alpha^2 \alpha_s)}  = \frac{\alpha_s C_F}{2 \pi} \, \Re \, \Bigg\{  & - 2
	 +  \left(3 +\frac{1}{\beta^2}\right) \, B_0(m_Q^2,0,m_Q^2)  \\[1mm]
	& - \left(2 +\frac{1}{\beta^2}\right)\, B_0(\sh,m_Q^2,m_Q^2)  \\[1mm]
	&   - \hat{s} \left (1+\beta^2 \right ) \, C_0 (m_Q^2, \sh, m_Q^2, 0, m_Q^2,m_Q^2) \Bigg \} \,.
\end{split}
\eeq
The additional term appearing in~\eqref{eq:DYvirt} arises from the interference of amplitudes with $Z$ boson and photon exchange, and is proportional to the imaginary part of the $Z$-boson propagator. This contribution is numerically unimportant.

The soft contribution to the FSR process takes the form 
\beq
\big ( \rd \sigma_{q\bar{q}, \hspace{0.25mm} A} \big)_{{\cal O} (\alpha^2 \alpha_s)}^{{\rm soft}, {\rm FSR}} =  
F^{{\rm soft}}_{{\cal O} (\alpha^2 \alpha_s)} \big(\rd\sigma_{q \bar q, \hspace{0.25mm} A}\big)_{{\mathcal{O}(\alpha^2)}}  \,,
\eeq
where
\bea % Uli: 11/1/19: Checked that UV & IR poles cancel between (A.32) to (A.35)!
\begin{split}
F^{{\rm soft}}_{{\cal O} (\alpha^2 \alpha_s)} &=   \frac{\alpha_s C_F}{2 \pi} \, \Bigg \{
	\left[  2 + \frac{1+\beta^2}{\beta} \, \ln \left( \frac{1-\beta}{1+\beta} \right) \right] \left[ \frac{1}{\epsilon} + 2 \ln \left( \frac{2 E_{\rm cut}}{\mu} \right) \right] \\[2mm]
	& \hspace{-1cm} -\frac{1}{2 \beta}\,  \ln \left( \frac{1-\beta}{1+\beta} \right) \left[ 4 +  \left(1+\beta^2\right) \ln \left( \frac{1-\beta}{1+\beta} \right) \right]
	- \frac{2 \left ( 1 + \beta^2 \right)}{\beta} \; {\rm Li}_2\left( 1-\frac{1-\beta}{1+\beta} \right)
	  \Bigg\} \,, \hspace{1cm}
\end{split}	  
\eea
with $E_{\rm cut}$ denoting the upper limit on the gluon energy. 

For the real emission corrections from the heavy quark lines, the differential cross section reads
\beq \label{eq:FSRreal}% RG: introduced 07/01/19 UH: 14/01/19: Checked!
\begin{aligned}
\big ( \rd \sigma_{q\bar{q}, \hspace{0.25mm} A} \big)_{{\cal O} (\alpha^2 \alpha_s)}^{{\rm real}, {\rm FSR}} & = \frac{9 \alpha^2 \alpha_s  C_F}{16  \pi N_c^2}\frac{a_q a_Q \, \sh  \hspace{0.25mm} y_{12} }{ \left( \sh  \hspace{0.25mm} y_{12} - M_Z^2  \right )^2 +  \Gamma_Z^2 M_Z^2 }  \\[1mm] 
& \phantom{xx}  \times \left [  v_q v_Q + 2 e_q e_Q \hspace{0.25mm} \left ( 1 - \frac{M_Z^2}{\sh  \hspace{0.25mm} y_{12}} \right ) \right ] f_{Q} \,,  
\end{aligned}
\eeq
where the kinematic function $f_Q$ is defined as
\bea \label{eq:fheavy}%RG: re-introduced 07/01/19. UH: 14/01/19: Changed signs! Please recheck! 
\begin{split}
f_{Q} &= \frac{2 \left ( y_{12} - y_{13} - 2 y_Q  \right ) - y_{45}}{y_{35}} + \frac{2 \left  (  y_{12} - y_{24} - 2 y_Q  \right ) - y_{35}}{y_{45}} \\[1mm]
& \phantom{xx}  + \frac{2 \left  (y_{12} - 2 y_{13} \right ) y_Q}{y_{45}^2} + \frac{2  \left (y_{12} - 2 y_{24} \right ) y_Q}{y_{35}^2} - \frac{2 \left  ( y_{12}-y_{13}-y_{24}  \right ) \left  ( y_{12}  -2 y_Q  \right )}{y_{35} y_{45} } \,. \hspace{8mm}
\end{split}
\eea

\begin{figure}[t!]
\centering
\includegraphics[width=.49\linewidth]{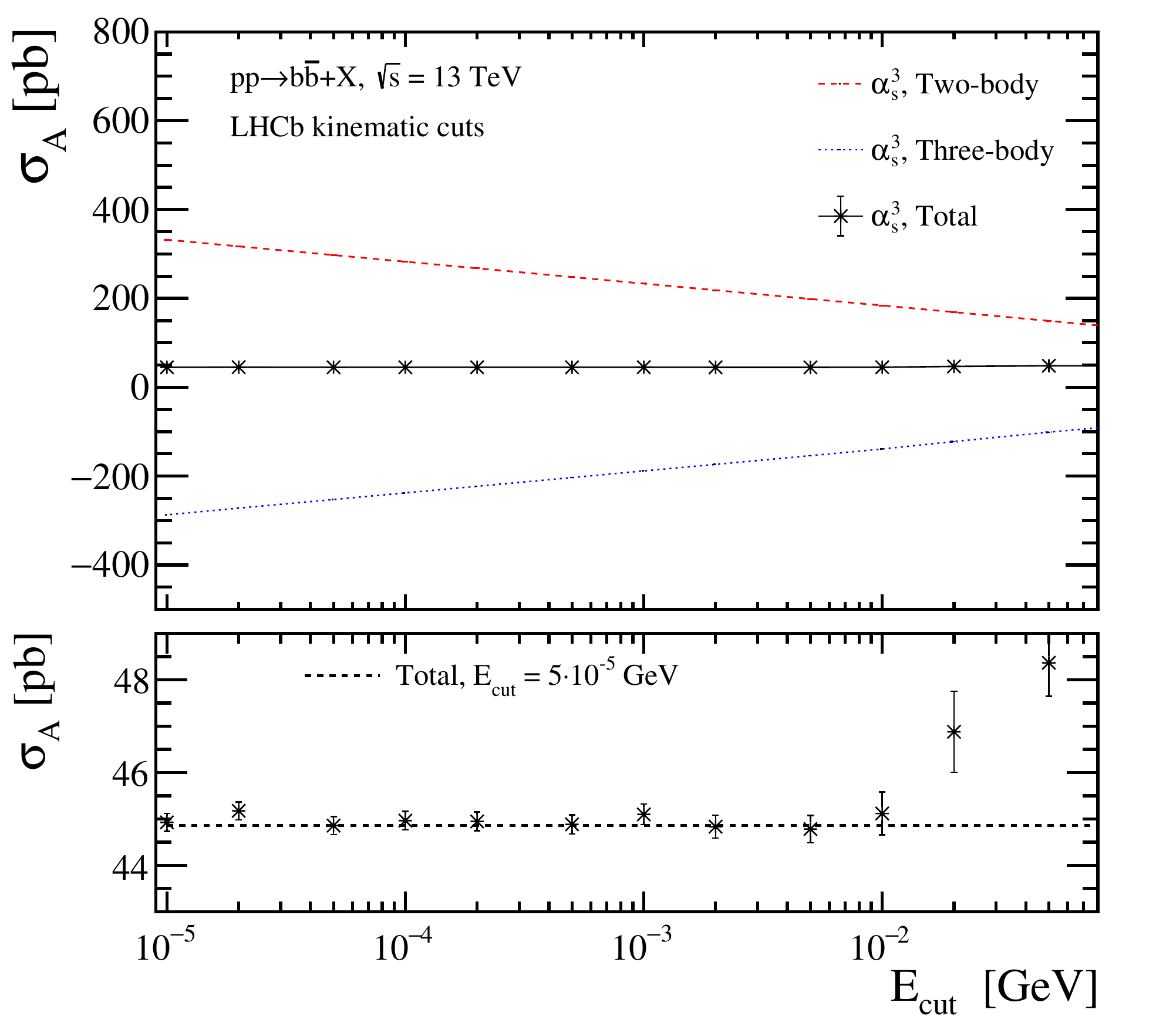} \hfill
\includegraphics[width=.49\linewidth]{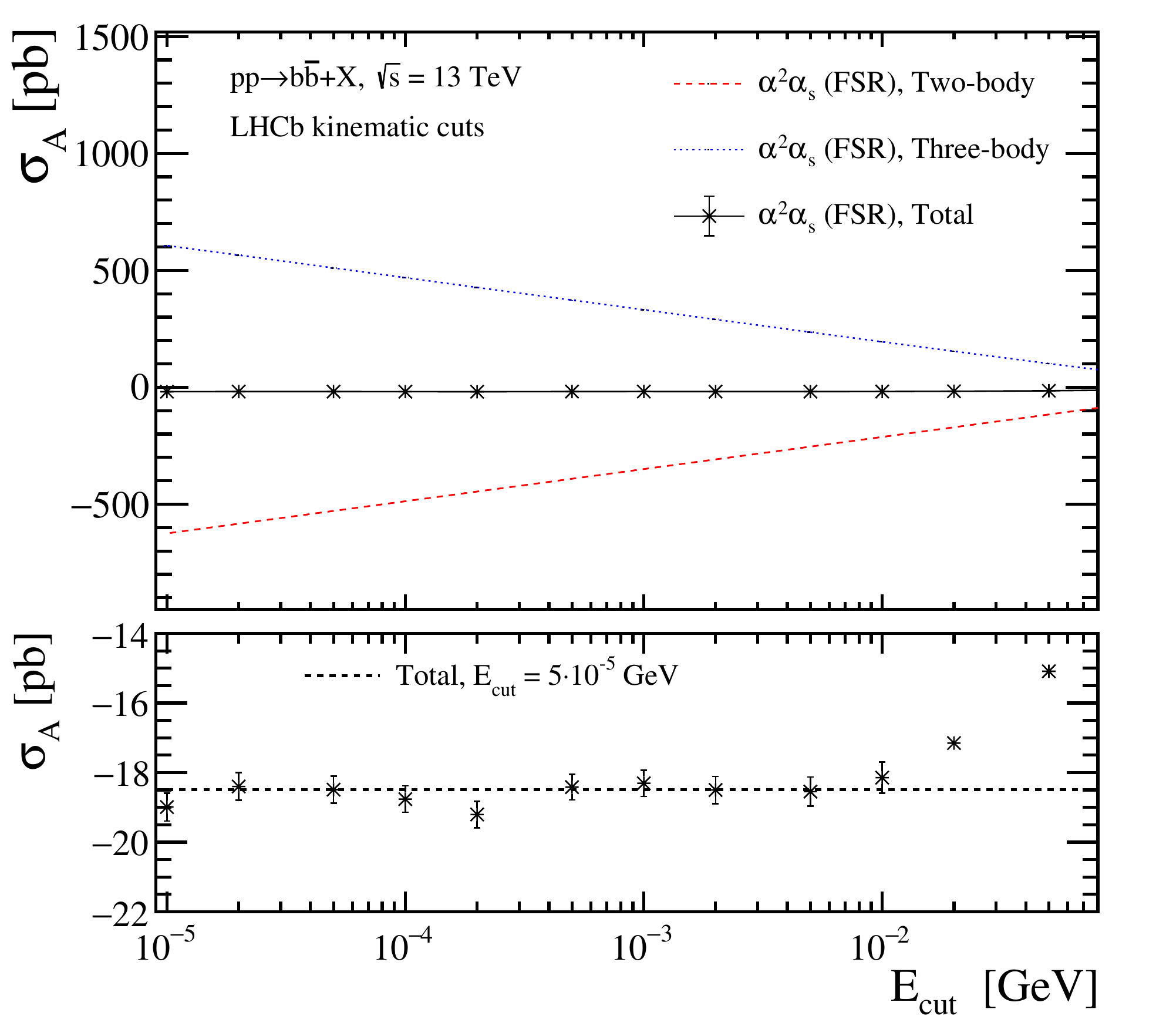} 
\vspace{2mm}
\caption{
Dependence on the slicing parameter $E_{\rm cut}$ which defines the soft region of the three-body corrections. Results are shown for the asymmetric $b\bar b$ cross section within $m_{b\bar b} \in [75,105] \, {\rm GeV}$, adopting the experimental selections~\eqref{eq:fiducial}.  The left panel illustrates the $\mathcal{O}(\alpha_s^3)$ corrections, while the right panel depicts the $\mathcal{O}(\alpha^2 \alpha_s)$ corrections to the massive final-state quark lines.}
\label{fig:slicing}
\end{figure}

The QCD corrections to the massless initial-state quark lines contain soft and/or collinear divergences. In this case we have chosen to provide an implementation of the ${\cal O} (\alpha^2 \alpha_s)$ corrections using the technique of phase-space slicing~\cite{Harris:2001sx}, and performed a cross-check using dipole subtraction~\cite{Catani:1996vz}. Rather than repeating the necessary details of both techniques, we instead refer the reader to Section D of~\cite{Harris:2001sx} for phase-space slicing, and Appendix D of~\cite{Catani:1996vz} for dipole subtraction. In the latter case, the relevant formula for the virtual correction is given in~(D.9), while the general formula for the operator insertions are collected in Appendix~C. These relative $\mathcal{O}(\alpha_s)$ corrections can be applied to our result for the Born-level cross section of $\mathcal{O}(\alpha^2)$ provided in~\eqref{eq:dsigmaAEW}. 

In addition to this, we provide the result for the real emission contributions to the differential cross section. They read
\beq \label{eq:ISRreal}% RG: introduced 07/01/19 UH: 14/01/19: Checked! Agrees!
\begin{aligned}
\big ( \rd \sigma_{q\bar{q}, \hspace{0.25mm} A} \big)_{{\cal O} (\alpha^2 \alpha_s)}^{{\rm real},  {\rm ISR}} & = \frac{9 \alpha^2 \alpha_s  C_F}{16 \pi N_c^2}
\frac{a_q a_Q \, \sh}{ \left( \sh \left(y_{34}+2 y_Q\right) - M_Z^2 \right)^2  + \Gamma_Z^2 M_Z^2 } \\[1mm] 
& \phantom{xx}  \times \, \bigg \{ v_q v_Q \hspace{0.25mm} \left(y_{34}+2 y_Q\right) + 2 e_q e_Q \hspace{0.25mm} \left[y_{34}+2 y_Q -    \frac{M_Z^2}{\sh} \right] \bigg \} \; f_{q}   \,, 
\end{aligned}
\eeq
with the kinematic function $f_{q}$ given by 
\beq \label{eq:flight}% RG: introduced 07/01/19 UH: 14/01/19: Checked! Agrees!
f_{q} = \frac{1}{y_{15} y_{25}} \, \Big [ \left (y_{12} - y_{15} \right ) \left ( y_{12} - 2 y_{13} -y_{15} \right ) + \left(1\leftrightarrow2,3\leftrightarrow4\right)  \Big ] \,. 
\eeq
The cross section for the $qg$-initiated contributions can be obtained from crossing, and by additionally adjusting the colour averaging over the initial state.

\subsection{Slicing parameter dependence}

To conclude this appendix, we perform a numerical study of the dependence on the slicing parameter $E_{\rm cut}$ used in the phase-space slicing technique. We~do this by computing the asymmetric cross section for $b$-quark pair production within the LHCb acceptance~\eqref{eq:fiducial} for $pp$ collisions at $\sqrt{s} = 13 \, {\rm TeV}$. The calculation is performed with the input parameters given in Section~\ref{sec:PDFsEWSCAL}, with factorisation and renormalisation scales set to $\mu_F = \mu_R = m_{b\bar b}$. The invariant mass of the $b$-jet pair is furthermore restricted to $m_{b\bar b} \in [75,105] \, {\rm GeV}$. In the left panel of Figure~\ref{fig:slicing} the contribution to the asymmetric cross section is shown for both the two- and three-body $\mathcal{O}(\alpha_s^3)$ contributions as well as their sum. This type of correction has been discussed in Appendix~\ref{pureQCD}. In the lower panel, the $y$-axis is zoomed into the region around the sum of the $\mathcal{O}(\alpha_s^3)$ contributions, where the shown uncertainties are due to the accuracy of the numerical integration. A similar study is presented for the $\mathcal{O}(\alpha^2 \alpha_s)$ FSR corrections discussed in Appendix~\ref{QCDDY}. The corresponding results are given on the right-hand side in Figure~\ref{fig:slicing}. The numerical results of this work employ the choice $E_{\rm cut} = 5 \cdot 10^{-5} \, {\rm GeV}$, and the NLO coefficients are  obtained with a relative precision of around $1\%$ to $2\%$ in this case. 

%\bibliography{HVQasym}

\providecommand{\href}[2]{#2}\begingroup\raggedright\begin{thebibliography}{}

\end{thebibliography}\endgroup


\begin{thebibliography}{10}

\bibitem{ALEPH:2005ab}
{\bf SLD Electroweak Group, DELPHI, ALEPH, SLD, SLD Heavy Flavour Group, OPAL,
  LEP Electroweak Working Group, L3} Collaboration, S.~Schael et~al., {\it
  {Precision electroweak measurements on the $Z$ resonance}},  {\em Phys.
  Rept.} {\bf 427} (2006) 257--454,
  [\href{http://arxiv.org/abs/hep-ex/0509008}{{\tt hep-ex/0509008}}].

\bibitem{Abazov:2014ysa}
{\bf D0} Collaboration, V.~M. Abazov et~al., {\it {Measurement of the
  Forward-Backward Asymmetry in the Production of $B^{\pm}$ Mesons in
  $p\bar{p}$ Collisions at $\sqrt{s}$~=~1.96~TeV}},  {\em Phys. Rev. Lett.}
  {\bf 114} (2015) 051803, [\href{http://arxiv.org/abs/1411.3021}{{\tt
  arXiv:1411.3021}}].

\bibitem{Aaltonen:2016azt}
{\bf CDF} Collaboration, T.~A. Aaltonen et~al., {\it {Measurement of the
  forward-backward asymmetry in low-mass bottom-quark pairs produced in
  proton-antiproton collisions}},  {\em Phys. Rev.} {\bf D93} (2016), no.~11
  112003, [\href{http://arxiv.org/abs/1601.06526}{{\tt arXiv:1601.06526}}].

\bibitem{Aaij:2014ywa}
{\bf LHCb} Collaboration, R.~Aaij et~al., {\it {First measurement of the charge
  asymmetry in beauty-quark pair production}},  {\em Phys. Rev. Lett.} {\bf
  113} (2014), no.~8 082003, [\href{http://arxiv.org/abs/1406.4789}{{\tt
  arXiv:1406.4789}}].

\bibitem{Bai:2011ed}
Y.~Bai, J.~L. Hewett, J.~Kaplan, and T.~G. Rizzo, {\it {LHC Predictions from a
  Tevatron Anomaly in the Top Quark Forward-Backward Asymmetry}},  {\em JHEP}
  {\bf 1103} (2011) 003, [\href{http://arxiv.org/abs/1101.5203}{{\tt
  arXiv:1101.5203}}].

\bibitem{Kahawala:2011sm}
D.~Kahawala, D.~Krohn, and M.~J. Strassler, {\it {Measuring the Bottom-Quark
  Forward-Central Asymmetry at the LHC}},  {\em JHEP} {\bf 1201} (2012) 069,
  [\href{http://arxiv.org/abs/1108.3301}{{\tt arXiv:1108.3301}}].

\bibitem{Saha:2011wr}
P.~Saha, {\it {Bottom Pair Production and Search for Heavy Resonances}},  {\em
  Phys. Lett.} {\bf B700} (2011) 221--228,
  [\href{http://arxiv.org/abs/1101.5797}{{\tt arXiv:1101.5797}}].

\bibitem{Manohar:2012rs}
A.~V. Manohar and M.~Trott, {\it {Electroweak Sudakov Corrections and the Top
  Quark Forward-Backward Asymmetry}},  {\em Phys. Lett.} {\bf B711} (2012)
  313--316, [\href{http://arxiv.org/abs/1201.3926}{{\tt arXiv:1201.3926}}].

\bibitem{Drobnak:2012cz}
J.~Drobnak, J.~F. Kamenik, and J.~Zupan, {\it {Flipping $t \bar t$ Asymmetries
  at the Tevatron and the LHC}},  {\em Phys. Rev.} {\bf D86} (2012) 054022,
  [\href{http://arxiv.org/abs/1205.4721}{{\tt arXiv:1205.4721}}].

\bibitem{Delaunay:2012kf}
C.~Delaunay, O.~Gedalia, Y.~Hochberg, and Y.~Soreq, {\it {Predictions from
  Heavy New Physics Interpretation of the Top Forward-Backward Asymmetry}},
  {\em JHEP} {\bf 12} (2012) 053, [\href{http://arxiv.org/abs/1207.0740}{{\tt
  arXiv:1207.0740}}].

\bibitem{Ipek:2013zi}
S.~Ipek, {\it {Light Axigluon Contributions to $b\bar{b}$ and $c\bar{c}$
  Asymmetry and Constraints on Flavor Changing Axigluon Currents}},  {\em Phys.
  Rev.} {\bf D87} (2013), no.~11 116010,
  [\href{http://arxiv.org/abs/1301.3990}{{\tt arXiv:1301.3990}}].

\bibitem{Grinstein:2013iws}
B.~Grinstein and C.~W. Murphy, {\it {Bottom-Quark Forward-Backward Asymmetry in
  the Standard Model and Beyond}},  {\em Phys. Rev. Lett.} {\bf 111} (2013)
  062003, [\href{http://arxiv.org/abs/1302.6995}{{\tt arXiv:1302.6995}}].
  [Erratum: {\it Phys. Rev. Lett.} {\bf 112} (2014) 239901].

\bibitem{Murphy:2015cha}
C.~W. Murphy, {\it {Bottom-Quark Forward-Backward and Charge Asymmetries at
  Hadron Colliders}},  {\em Phys. Rev.} {\bf D92} (2015), no.~5 054003,
  [\href{http://arxiv.org/abs/1504.02493}{{\tt arXiv:1504.02493}}].

\bibitem{Gauld:2015qha}
R.~Gauld, U.~Haisch, B.~D. Pecjak, and E.~Re, {\it {Beauty-quark and
  charm-quark pair production asymmetries at LHCb}},  {\em Phys. Rev.} {\bf
  D92} (2015) 034007, [\href{http://arxiv.org/abs/1505.02429}{{\tt
  arXiv:1505.02429}}].

\bibitem{Alves:2008zz}
{\bf LHCb} Collaboration, J.~Alves et~al., {\it {The LHCb Detector
  at the LHC}},  {\em JINST} {\bf 3} (2008) S08005.

\bibitem{Aaij:2015yqa}
{\bf LHCb} Collaboration, R.~Aaij et~al., {\it {Identification of beauty and
  charm quark jets at LHCb}},  {\em JINST} {\bf 10} (2015), no.~06 P06013,
  [\href{http://arxiv.org/abs/1504.07670}{{\tt arXiv:1504.07670}}].

\bibitem{Aaij:2017eru}
{\bf LHCb} Collaboration, R.~Aaij et~al., {\it {First observation of forward $Z
  \rightarrow b \bar{b}$ production in $pp$ collisions at $\sqrt{s}=8$ TeV}},
  {\em Phys. Lett.} {\bf B776} (2018) 430--439,
  [\href{http://arxiv.org/abs/1709.03458}{{\tt arXiv:1709.03458}}].

\bibitem{Cacciari:2008gp}
M.~Cacciari, G.~P. Salam, and G.~Soyez, {\it {The anti-$k_t$ jet clustering
  algorithm}},  {\em JHEP} {\bf 0804} (2008) 063,
  [\href{http://arxiv.org/abs/0802.1189}{{\tt arXiv:0802.1189}}].

\bibitem{Collins:1989gx}
J.~C. Collins, D.~E. Soper, and G.~F. Sterman, {\it {Factorization of Hard
  Processes in QCD}},  {\em Adv. Ser. Direct. High Energy Phys.} {\bf 5} (1989)
  1--91, [\href{http://arxiv.org/abs/hep-ph/0409313}{{\tt hep-ph/0409313}}].

\bibitem{Nason:1987xz}
P.~Nason, S.~Dawson, and R.~K. Ellis, {\it {The Total Cross-Section for the
  Production of Heavy Quarks in Hadronic Collisions}},  {\em Nucl.Phys.} {\bf
  B303} (1988) 607.

\bibitem{Nason:1989zy}
P.~Nason, S.~Dawson, and R.~K. Ellis, {\it {The One Particle Inclusive
  Differential Cross-Section for Heavy Quark Production in Hadronic
  Collisions}},  {\em Nucl.Phys.} {\bf B327} (1989) 49--92.

\bibitem{Mangano:1991jk}
M.~L. Mangano, P.~Nason, and G.~Ridolfi, {\it {Heavy quark correlations in
  hadron collisions at next-to-leading order}},  {\em Nucl.Phys.} {\bf B373}
  (1992) 295--345.

\bibitem{Beenakker:1990maa}
W.~Beenakker, W.~van Neerven, R.~Meng, G.~Schuler, and J.~Smith, {\it {QCD
  corrections to heavy quark production in hadron hadron collisions}},  {\em
  Nucl.Phys.} {\bf B351} (1991) 507--560.

\bibitem{Beenakker:1988bq}
W.~Beenakker, H.~Kuijf, W.~van Neerven, and J.~Smith, {\it {QCD Corrections to
  Heavy Quark Production in $p \bar p$ Collisions}},  {\em Phys.Rev.} {\bf D40}
  (1989) 54--82.

\bibitem{Kuhn:2005it}
J.~H. K{\"u}hn, A.~Scharf, and P.~Uwer, {\it {Electroweak corrections to
  top-quark pair production in quark-antiquark annihilation}},  {\em Eur. Phys.
  J.} {\bf C45} (2006) 139--150,
  [\href{http://arxiv.org/abs/hep-ph/0508092}{{\tt hep-ph/0508092}}].

\bibitem{Kuhn:2006vh}
J.~H. K{\"u}hn, A.~Scharf, and P.~Uwer, {\it {Electroweak effects in top-quark
  pair production at hadron colliders}},  {\em Eur. Phys. J.} {\bf C51} (2007)
  37--53, [\href{http://arxiv.org/abs/hep-ph/0610335}{{\tt hep-ph/0610335}}].

\bibitem{Bernreuther:2006vg}
W.~Bernreuther, M.~F{\"u}cker, and Z.-G. Si, {\it {Weak interaction corrections
  to hadronic top quark pair production}},  {\em Phys. Rev.} {\bf D74} (2006)
  113005, [\href{http://arxiv.org/abs/hep-ph/0610334}{{\tt hep-ph/0610334}}].

\bibitem{Hollik:2007sw}
W.~Hollik and M.~Kollar, {\it {NLO QED contributions to top-pair production at
  hadron collider}},  {\em Phys. Rev.} {\bf D77} (2008) 014008,
  [\href{http://arxiv.org/abs/0708.1697}{{\tt arXiv:0708.1697}}].

\bibitem{Kuhn:2009nf}
J.~H. K{\"u}hn, A.~Scharf, and P.~Uwer, {\it {Weak effects in b-jet production
  at hadron colliders}},  {\em Phys. Rev.} {\bf D82} (2010) 013007,
  [\href{http://arxiv.org/abs/0909.0059}{{\tt arXiv:0909.0059}}].

\bibitem{Buckley:2014ana}
A.~Buckley, J.~Ferrando, S.~Lloyd, K.~Nordstr{\"o}m, B.~Page, M.~R{\"u}fenacht,
  M.~Sch{\"o}nherr, and G.~Watt, {\it {LHAPDF6: parton density access in the
  LHC precision era}},  {\em Eur. Phys. J.} {\bf C75} (2015) 132,
  [\href{http://arxiv.org/abs/1412.7420}{{\tt arXiv:1412.7420}}].

\bibitem{Hahn:2004fe}
T.~Hahn, {\it {CUBA: A Library for multidimensional numerical integration}},
  {\em Comput. Phys. Commun.} {\bf 168} (2005) 78--95,
  [\href{http://arxiv.org/abs/hep-ph/0404043}{{\tt hep-ph/0404043}}].

\bibitem{vanHameren:2009dr}
A.~van Hameren, C.~G. Papadopoulos, and R.~Pittau, {\it {Automated one-loop
  calculations: A Proof of concept}},  {\em JHEP} {\bf 09} (2009) 106,
  [\href{http://arxiv.org/abs/0903.4665}{{\tt arXiv:0903.4665}}].

\bibitem{vanHameren:2010cp}
A.~van Hameren, {\it {OneLOop: For the evaluation of one-loop scalar
  functions}},  {\em Comput.Phys.Commun.} {\bf 182} (2011) 2427--2438,
  [\href{http://arxiv.org/abs/1007.4716}{{\tt arXiv:1007.4716}}].

\bibitem{Frixione:2007nw}
S.~Frixione, P.~Nason, and G.~Ridolfi, {\it {A Positive-weight
  next-to-leading-order Monte Carlo for heavy flavour hadroproduction}},  {\em
  JHEP} {\bf 0709} (2007) 126, [\href{http://arxiv.org/abs/0707.3088}{{\tt
  arXiv:0707.3088}}].

\bibitem{Hirschi:2011pa}
V.~Hirschi, R.~Frederix, S.~Frixione, M.~V. Garzelli, F.~Maltoni, and
  R.~Pittau, {\it {Automation of one-loop QCD corrections}},  {\em JHEP} {\bf
  05} (2011) 044, [\href{http://arxiv.org/abs/1103.0621}{{\tt
  arXiv:1103.0621}}].

\bibitem{Hirschi:2015iia}
V.~Hirschi and O.~Mattelaer, {\it {Automated event generation for loop-induced
  processes}},  {\em JHEP} {\bf 10} (2015) 146,
  [\href{http://arxiv.org/abs/1507.00020}{{\tt arXiv:1507.00020}}].

\bibitem{Alwall:2014hca}
J.~Alwall, R.~Frederix, S.~Frixione, V.~Hirschi, F.~Maltoni, O.~Mattelaer,
  H.~S. Shao, T.~Stelzer, P.~Torrielli, and M.~Zaro, {\it {The automated
  computation of tree-level and next-to-leading order differential cross
  sections, and their matching to parton shower simulations}},  {\em JHEP} {\bf
  07} (2014) 079, [\href{http://arxiv.org/abs/1405.0301}{{\tt
  arXiv:1405.0301}}].

\bibitem{mg5online}
\url{https://cp3.irmp.ucl.ac.be/projects/madgraph/wiki/LoopInducedTimesTree}.

\bibitem{Hahn:2000kx}
T.~Hahn, {\it {Generating Feynman diagrams and amplitudes with FeynArts 3}},
  {\em Comput.Phys.Commun.} {\bf 140} (2001) 418--431,
  [\href{http://arxiv.org/abs/hep-ph/0012260}{{\tt hep-ph/0012260}}].

\bibitem{Hahn:1998yk}
T.~Hahn and M.~Perez-Victoria, {\it {Automatized one loop calculations in
  four-dimensions and D-dimensions}},  {\em Comput.Phys.Commun.} {\bf 118}
  (1999) 153--165, [\href{http://arxiv.org/abs/hep-ph/9807565}{{\tt
  hep-ph/9807565}}].

\bibitem{Harris:2001sx}
B.~W. Harris and J.~F. Owens, {\it {The Two cutoff phase space slicing
  method}},  {\em Phys. Rev.} {\bf D65} (2002) 094032,
  [\href{http://arxiv.org/abs/hep-ph/0102128}{{\tt hep-ph/0102128}}].

\bibitem{Catani:1996vz}
S.~Catani and M.~H. Seymour, {\it {A General algorithm for calculating jet
  cross-sections in NLO QCD}},  {\em Nucl. Phys.} {\bf B485} (1997) 291--419,
  [\href{http://arxiv.org/abs/hep-ph/9605323}{{\tt hep-ph/9605323}}]. [Erratum:
  Nucl. Phys.B510,503(1998)].

\bibitem{Czakon:2014xsa}
M.~Czakon, P.~Fiedler, and A.~Mitov, {\it {Resolving the Tevatron Top Quark
  Forward-Backward Asymmetry Puzzle: Fully Differential
  Next-to-Next-to-Leading-Order Calculation}},  {\em Phys. Rev. Lett.} {\bf
  115} (2015), no.~5 052001, [\href{http://arxiv.org/abs/1411.3007}{{\tt
  arXiv:1411.3007}}].

\bibitem{Currie:2016bfm}
J.~Currie, E.~W.~N. Glover, and J.~Pires, {\it {Next-to-Next-to Leading Order
  QCD Predictions for Single Jet Inclusive Production at the LHC}},  {\em Phys.
  Rev. Lett.} {\bf 118} (2017), no.~7 072002,
  [\href{http://arxiv.org/abs/1611.01460}{{\tt arXiv:1611.01460}}].

\bibitem{deFlorian:2016spz}
{\bf LHC Higgs Cross Section Working Group} Collaboration, D.~de~Florian
  et~al., {\it {Handbook of LHC Higgs Cross Sections: 4. Deciphering the Nature
  of the Higgs Sector}},  \href{http://arxiv.org/abs/1610.07922}{{\tt
  arXiv:1610.07922}}.

\bibitem{Manohar:2016nzj}
A.~Manohar, P.~Nason, G.~P. Salam, and G.~Zanderighi, {\it {How bright is the
  proton? A precise determination of the photon parton distribution function}},
   {\em Phys. Rev. Lett.} {\bf 117} (2016), no.~24 242002,
  [\href{http://arxiv.org/abs/1607.04266}{{\tt arXiv:1607.04266}}].

\bibitem{Banfi:2006hf}
A.~Banfi, G.~P. Salam, and G.~Zanderighi, {\it {Infrared safe definition of jet
  flavor}},  {\em Eur. Phys. J.} {\bf C47} (2006) 113--124,
  [\href{http://arxiv.org/abs/hep-ph/0601139}{{\tt hep-ph/0601139}}].

\bibitem{Czakon:2013goa}
M.~Czakon, P.~Fiedler, and A.~Mitov, {\it {The total top quark pair production
  cross-section at hadron colliders through $O(\alpha_S^4)$}},  {\em Phys. Rev.
  Lett.} {\bf 110} (2013) 252004, [\href{http://arxiv.org/abs/1303.6254}{{\tt
  arXiv:1303.6254}}].

\bibitem{Ball:2017nwa}
{\bf NNPDF} Collaboration, R.~D. Ball et~al., {\it {Parton distributions from
  high-precision collider data}},  {\em Eur. Phys. J.} {\bf C77} (2017), no.~10
  663, [\href{http://arxiv.org/abs/1706.00428}{{\tt arXiv:1706.00428}}].

\bibitem{Cacciari:1998it}
M.~Cacciari, M.~Greco, and P.~Nason, {\it {The $p_T$ spectrum in heavy flavor
  hadroproduction}},  {\em JHEP} {\bf 05} (1998) 007,
  [\href{http://arxiv.org/abs/hep-ph/9803400}{{\tt hep-ph/9803400}}].

\bibitem{deFlorian:2015ujt}
D.~de~Florian, G.~F.~R. Sborlini, and G.~Rodrigo, {\it {QED corrections to the
  Altarelli-Parisi splitting functions}},  {\em Eur. Phys. J.} {\bf C76}
  (2016), no.~5 282, [\href{http://arxiv.org/abs/1512.00612}{{\tt
  arXiv:1512.00612}}].

\bibitem{Harland-Lang:2016apc}
L.~A. Harland-Lang, V.~A. Khoze, and M.~G. Ryskin, {\it {The photon PDF in
  events with rapidity gaps}},  {\em Eur. Phys. J.} {\bf C76} (2016), no.~5
  255, [\href{http://arxiv.org/abs/1601.03772}{{\tt arXiv:1601.03772}}].

\bibitem{Harland-Lang:2016kog}
L.~A. Harland-Lang, V.~A. Khoze, and M.~G. Ryskin, {\it {Photon-initiated
  processes at high mass}},  {\em Phys. Rev.} {\bf D94} (2016), no.~7 074008,
  [\href{http://arxiv.org/abs/1607.04635}{{\tt arXiv:1607.04635}}].

\bibitem{Manohar:2017eqh}
A.~V. Manohar, P.~Nason, G.~P. Salam, and G.~Zanderighi, {\it {The Photon
  Content of the Proton}},  {\em JHEP} {\bf 12} (2017) 046,
  [\href{http://arxiv.org/abs/1708.01256}{{\tt arXiv:1708.01256}}].

\bibitem{Bertone:2017bme}
{\bf NNPDF} Collaboration, V.~Bertone, S.~Carrazza, N.~P. Hartland, and
  J.~Rojo, {\it {Illuminating the photon content of the proton within a global
  PDF analysis}},  {\em SciPost Phys.} {\bf 5} (2018), no.~1 008,
  [\href{http://arxiv.org/abs/1712.07053}{{\tt arXiv:1712.07053}}].

\bibitem{Pagani:2016caq}
D.~Pagani, I.~Tsinikos, and M.~Zaro, {\it {The impact of the photon PDF and
  electroweak corrections on $t \bar{t}$ distributions}},  {\em Eur. Phys. J.}
  {\bf C76} (2016), no.~9 479, [\href{http://arxiv.org/abs/1606.01915}{{\tt
  arXiv:1606.01915}}].

\bibitem{Czakon:2017wor}
M.~Czakon, D.~Heymes, A.~Mitov, D.~Pagani, I.~Tsinikos, and M.~Zaro, {\it
  {Top-pair production at the LHC through NNLO QCD and NLO EW}},  {\em JHEP}
  {\bf 10} (2017) 186, [\href{http://arxiv.org/abs/1705.04105}{{\tt
  arXiv:1705.04105}}].

\bibitem{Czakon:2017lgo}
M.~Czakon, D.~Heymes, A.~Mitov, D.~Pagani, I.~Tsinikos, and M.~Zaro, {\it
  {Top-quark charge asymmetry at the LHC and Tevatron through NNLO QCD and NLO
  EW}},  {\em Phys. Rev.} {\bf D98} (2018), no.~1 014003,
  [\href{http://arxiv.org/abs/1711.03945}{{\tt arXiv:1711.03945}}].

\bibitem{Fleischer:1993ub}
J.~Fleischer, O.~V. Tarasov, and F.~Jegerlehner, {\it {Two loop heavy top
  corrections to the rho parameter: A Simple formula valid for arbitrary Higgs
  mass}},  {\em Phys. Lett.} {\bf B319} (1993) 249--256.

\bibitem{Denner:2005fg}
A.~Denner, S.~Dittmaier, M.~Roth, and L.~H. Wieders, {\it {Electroweak
  corrections to charged-current $e^+ e^- \to 4$~fermion processes: Technical
  details and further results}},  {\em Nucl. Phys.} {\bf B724} (2005) 247--294,
  [\href{http://arxiv.org/abs/hep-ph/0505042}{{\tt hep-ph/0505042}}]. [Erratum:
  {\it Nucl. Phys.} {\bf B854} (2012) 504].

\bibitem{Czakon:2018nun}
M.~Czakon, A.~Ferroglia, D.~Heymes, A.~Mitov, B.~D. Pecjak, D.~J. Scott,
  X.~Wang, and L.~L. Yang, {\it {Resummation for (boosted) top-quark pair
  production at NNLO+NNLL$^\prime$ in QCD}},  {\em JHEP} {\bf 05} (2018) 149,
  [\href{http://arxiv.org/abs/1803.07623}{{\tt arXiv:1803.07623}}].

\bibitem{Bediaga:2018lhg}
{\bf LHCb} Collaboration, R.~Aaij et~al., {\it {Physics case for an LHCb
  Upgrade II - Opportunities in flavour physics, and beyond, in the HL-LHC
  era}},  \href{http://arxiv.org/abs/1808.08865}{{\tt arXiv:1808.08865}}.

\bibitem{Liu:2017xmc}
D.~Liu, J.~Liu, C.~E.~M. Wagner, and X.-P. Wang, {\it {Bottom-quark
  Forward-Backward Asymmetry, Dark Matter and the LHC}},  {\em Phys. Rev.} {\bf
  D97} (2018), no.~5 055021, [\href{http://arxiv.org/abs/1712.05802}{{\tt
  arXiv:1712.05802}}].

\bibitem{Arbuzov:2005ma}
A.~B. Arbuzov, M.~Awramik, M.~Czakon, A.~Freitas, M.~W. Gr{\"u}newald,
  K.~M{\"o}nig, S.~Riemann, and T.~Riemann, {\it {ZFITTER: A Semi-analytical
  program for fermion pair production in $e^+ e^-$ annihilation, from version
  6.21 to version 6.42}},  {\em Comput. Phys. Commun.} {\bf 174} (2006)
  728--758, [\href{http://arxiv.org/abs/hep-ph/0507146}{{\tt hep-ph/0507146}}].

\bibitem{Tanabashi:2018oca}
{\bf Particle Data Group} Collaboration, M.~Tanabashi et~al., {\it {Review of
  Particle Physics}},  {\em Phys. Rev.} {\bf D98} (2018), no.~3 030001.

\bibitem{Sirunyan:2017nvi}
{\bf CMS} Collaboration, A.~M. Sirunyan et~al., {\it {Search for low mass
  vector resonances decaying into quark-antiquark pairs in proton-proton
  collisions at $\sqrt{s}$~=~13~TeV}},  {\em JHEP} {\bf 01} (2018) 097,
  [\href{http://arxiv.org/abs/1710.00159}{{\tt arXiv:1710.00159}}].

\bibitem{Aad:2012cfr}
{\bf ATLAS} Collaboration, G.~Aad et~al., {\it {Search for the neutral Higgs
  bosons of the Minimal Supersymmetric Standard Model in $pp$ collisions at
  $\sqrt{s}$~=~7~TeV with the ATLAS detector}},  {\em JHEP} {\bf 02} (2013)
  095, [\href{http://arxiv.org/abs/1211.6956}{{\tt arXiv:1211.6956}}].

\bibitem{Aaboud:2017buh}
{\bf ATLAS} Collaboration, M.~Aaboud et~al., {\it {Search for new high-mass
  phenomena in the dilepton final state using 36~fb$^{-1}$ of proton-proton
  collision data at $ \sqrt{s}$~=~13~TeV with the ATLAS detector}},  {\em JHEP}
  {\bf 10} (2017) 182, [\href{http://arxiv.org/abs/1707.02424}{{\tt
  arXiv:1707.02424}}].

\bibitem{Schwaller:2015gea}
P.~Schwaller, D.~Stolarski, and A.~Weiler, {\it {Emerging Jets}},  {\em JHEP}
  {\bf 05} (2015) 059, [\href{http://arxiv.org/abs/1502.05409}{{\tt
  arXiv:1502.05409}}].

\bibitem{Haisch:2016hzu}
U.~Haisch and J.~F. Kamenik, {\it {Searching for new spin-0 resonances at
  LHCb}},  {\em Phys. Rev.} {\bf D93} (2016), no.~5 055047,
  [\href{http://arxiv.org/abs/1601.05110}{{\tt arXiv:1601.05110}}].

\bibitem{Ilten:2016tkc}
P.~Ilten, Y.~Soreq, J.~Thaler, M.~Williams, and W.~Xue, {\it {Proposed
  Inclusive Dark Photon Search at LHCb}},  {\em Phys. Rev. Lett.} {\bf 116}
  (2016), no.~25 251803, [\href{http://arxiv.org/abs/1603.08926}{{\tt
  arXiv:1603.08926}}].

\bibitem{Aaij:2017rft}
{\bf LHCb} Collaboration, R.~Aaij et~al., {\it {Search for Dark Photons
  Produced in 13 TeV $pp$ Collisions}},  {\em Phys. Rev. Lett.} {\bf 120}
  (2018), no.~6 061801, [\href{http://arxiv.org/abs/1710.02867}{{\tt
  arXiv:1710.02867}}].

\bibitem{Ilten:2018crw}
P.~Ilten, Y.~Soreq, M.~Williams, and W.~Xue, {\it {Serendipity in dark photon
  searches}},  {\em JHEP} {\bf 06} (2018) 004,
  [\href{http://arxiv.org/abs/1801.04847}{{\tt arXiv:1801.04847}}].

\bibitem{Haisch:2018kqx}
U.~Haisch, J.~F. Kamenik, A.~Malinauskas, and M.~Spira, {\it {Collider
  constraints on light pseudoscalars}},  {\em JHEP} {\bf 03} (2018) 178,
  [\href{http://arxiv.org/abs/1802.02156}{{\tt arXiv:1802.02156}}].

\bibitem{Aaij:2018xpt}
{\bf LHCb} Collaboration, R.~Aaij et~al., {\it {Search for a dimuon resonance
  in the $\Upsilon$ mass region}},  {\em JHEP} {\bf 09} (2018) 147,
  [\href{http://arxiv.org/abs/1805.09820}{{\tt arXiv:1805.09820}}].

\bibitem{Jersak:1981sp}
J.~Jersak, E.~Laermann, and P.~M. Zerwas, {\it {Electroweak Production of Heavy
  Quarks in $e^+ e^-$ Annihilation}},  {\em Phys. Rev.} {\bf D25} (1982) 1218.
  [Erratum: {\it Phys. Rev.} {\bf D36} (1987) 310].

\bibitem{Hollik:2011ps}
W.~Hollik and D.~Pagani, {\it {The electroweak contribution to the top quark
  forward-backward asymmetry at the Tevatron}},  {\em Phys.Rev.} {\bf D84}
  (2011) 093003, [\href{http://arxiv.org/abs/1107.2606}{{\tt
  arXiv:1107.2606}}].

\bibitem{Kuhn:1998kw}
J.~H. K{\"u}hn and G.~Rodrigo, {\it {Charge asymmetry of heavy quarks at hadron
  colliders}},  {\em Phys.Rev.} {\bf D59} (1999) 054017,
  [\href{http://arxiv.org/abs/hep-ph/9807420}{{\tt hep-ph/9807420}}].

\end{thebibliography}

\providecommand{\href}[2]{#2}\begingroup\raggedright\endgroup

\end{document}